\documentclass[12pt]{article}
\usepackage{epsf}

\usepackage{epsfig,graphics}
\usepackage {graphicx}
\usepackage {epsfig}
\usepackage {subfigure}
\usepackage {tabularx} 
\usepackage{rotate}	
\usepackage{slashed}

\usepackage{amsmath}
\usepackage{amsfonts}
\usepackage{amssymb}
\usepackage{graphicx}
\usepackage{cite}
\usepackage{color}
\usepackage{bbm}

\usepackage{fancyhdr}
\usepackage{hyperref}

\newcommand{\op}{\hspace{1pt}}
\newcolumntype{M}[1]{>{\centering\arraybackslash}m{#1}}
\newcolumntype{N}{@{}m{0pt}@{}}

\newcommand{\bmat}{\left(\begin{array}}
\newcommand{\emat}{\end{array}\right)}

\def\preal{{\rm Re\,}}
\def\pim{{\rm Im\,}}
\def\ds{\displaystyle}
\def\yzero{\smash{\hbox{$y\kern-4pt\raise1pt\hbox{${}^\circ$}$}}}

\def\a{\alpha}

\def\d{\delta}
\def\beq{\begin{equation}}
\def\eeq{\end{equation}}
\def\beqa{\begin{eqnarray}}
\def\eeqa{\end{eqnarray}}

\def\-{\hphantom{-}}

\def\s2{\frac{1}{\sqrt2}}

\def\oh{\frac{1}{2}}
\def\beq{\begin{equation}}
\def\eeq{\end{equation}}
\def\beqa{\begin{eqnarray}}
\def\eeqa{\end{eqnarray}}

\def\IF{\relax{\rm I\kern-.18em F}}
\def\II{\relax{\rm I\kern-.18em I}}
\def\IP{\relax{\rm I\kern-.18em P}}
\def\IC{\relax\hbox{\kern.25em$\inbar\kern-.3em{\rm C}$}}
\def\IR{\relax{\rm I\kern-.18em R}}

\def\cc{{\mathcal C}}

\def\CK{{\kappa}}
\def\ci{{\mathcal I}}
\def\cj{{\mathcal J}}

\def\cn{{\mathcal N}}
\def\cam{{\mathcal M}}
\def\cad{{\mathcal D}}

\def\cv{{\mathcal V}}

\def\cf{{\mathcal F}}

\def\Dsl{\,\raise.15ex\hbox{/}\mkern-13.5mu D} 
\def\IZ{{\mathbb Z}}
\def\RR{{\mathbb R}}

\def\tD{{\text D}}

\def\R{{\mathcal R}}
\def\ca{{\mathcal A}}

\catcode`\@=11   
\newdimen\@rotdimen
\newbox\@rotbox  

\def\@vspec#1{\special{ps:#1}}
\def\@rotstart#1{\@vspec{gsave currentpoint currentpoint translate
   #1 neg exch neg exch translate}}
\def\@rotfinish{\@vspec{currentpoint grestore moveto}}
\def\@rotr#1{\@rotdimen=\ht#1\advance\@rotdimen by\dp#1%
   \hbox to\@rotdimen{\hskip\ht#1\vbox to\wd#1{\@rotstart{90 rotate}%
   \box#1\vss}\hss}\@rotfinish}
\def\@rotl#1{\@rotdimen=\ht#1\advance\@rotdimen by\dp#1%
   \hbox to\@rotdimen{\vbox to\wd#1{\vskip\wd#1\@rotstart{270 rotate}%
   \box#1\vss}\hss}\@rotfinish}%
\def\@rotu#1{\@rotdimen=\ht#1\advance\@rotdimen by\dp#1%
   \hbox to\wd#1{\hskip\wd#1\vbox to\@rotdimen{\vskip\@rotdimen
   \@rotstart{-1 dup scale}\box#1\vss}\hss}\@rotfinish}%
\def\@rotf#1{\hbox to\wd#1{\hskip\wd#1\@rotstart{-1 1 scale}%
   \box#1\hss}\@rotfinish}%
\def\rotate{\@ifnextchar[{\@rotate}{\@rotate[l]}}
\def\@rotate[#1]#2{\setbox\@rotbox=\hbox{#2}\@nameuse{@rot#1}\@rotbox}

\catcode`\@=12

\topmargin
-1.5cm
\textwidth
15.5cm
\textheight
23.5cm
\oddsidemargin
0.7cm
\evensidemargin
0.7cm

\begin{document}

\makeatletter
\@addtoreset{equation}{section}
\makeatother
\renewcommand{\theequation}{\thesection.\arabic{equation}}
\pagestyle{empty}
\vspace{-0.2cm}
\rightline{ IFT-UAM/CSIC-19-50}
\vspace{1.2cm}
\begin{center}


\LARGE{The Swampland Distance  Conjecture \\ and Towers of Tensionless  Branes\\ [13mm]}
  \large{ A. Font$^1$, A. Herr\'aez$^2$ and L.E. Ib\'a\~nez$^2$,
   \\[6mm]}
\small{
$^1$  {\em Facultad de Ciencias, Universidad Central de Venezuela, A.P.20513, Caracas 1020-A,  Venezuela
}  \\[0pt]
  $^2$ {\em Departamento de F\'{\i}sica Te\'orica
and Instituto de F\'{\i}sica Te\'orica UAM/CSIC,\\[-0.3em]
Universidad Aut\'onoma de Madrid,
Cantoblanco, 28049 Madrid, Spain} \\[0pt]
}
\small{\bf Abstract} \\[6mm]
\end{center}
\begin{center}
\begin{minipage}[h]{15.22cm}
The Swampland Distance Conjecture states that at infinite distance in the scalar moduli space an infinite  tower of 
particles become exponentially massless. We study this issue in the context of 4d type IIA and type IIB Calabi-Yau compactifications. 
We find that for large moduli not only towers of particles but also  domain walls  and strings become tensionless. 
We study in detail the case of  type IIA and IIB ${\cal N}=1$ CY orientifolds and show how for infinite K\"ahler  and/or 
complex structure moduli towers of domain walls and strings become tensionless, depending on the
particular direction in moduli space. For the type IIA case we construct the monodromy orbits of domain walls in detail.
 We study the structure of mass scales in these  limits and find that
 these towers may occur at the same scale as the fundamental string scale or the KK scale making sometimes difficult
 an effective field theory description. The structure of IIA and IIB towers are consistent with mirror symmetry,
 as long as towers of exotic domain walls 
 associated to non-geometric fluxes also appear.
 We briefly discuss the
issue of emergence within this context and the possible implications for 4d vacua.

\end{minipage}
\end{center}
\newpage
\setcounter{page}{1}
\pagestyle{plain}
\renewcommand{\thefootnote}{\arabic{footnote}}
\setcounter{footnote}{0}

\tableofcontents

\section{Introduction}

One of the most striking proposals in the Swampland program \cite{swampland,WGC,distance} (see \cite{review,vafafederico} for reviews)
is the Swampland Distance Conjecture (SDC) \cite{distance}.
In simple terms it states 
that starting from a point $p_0$ in moduli space, 
and moving to a point $p$ an infinite distance $d(p_0,p) \to \infty$ away, there appears a tower of states which becomes exponentially massless according to
\beq
m \sim  m_0 e^{-\alpha d(p_0,p)} \ .
\eeq
This proposal has been tested in different situations in string theory 
 \cite{Klaewer:2016kiy,Palti,Hebecker:2017lxm,Harlow:2016,emergence1,emergence2,irene,Grimm2,Corvilain,andriolo,blumenrefined,timo1,timo2,timo3,Krishnan,dS1,dS3,Baume:2016,Valenzuela:2017,Blumenhagen:2017,Reecephotons,Hebecker:2019,BlumenhaggenKKLT,Buratti:2018xjt,Klemm:2018,SSWGC,infinitons}.
 In particular, in \cite{irene} it was shown how in the large complex structure 
limit of type IIB Calabi-Yau (CY) compactifications, towers of states indeed become exponentially massless. In this example, further studied in \cite{Grimm2}, 
the towers  are provided by states formed by D3-branes wrapping 3-cycles in the compact space. The type IIA mirror situation,
in which the towers come from bound states of D0 and D2-branes wrapping 2-cycles in the CY, was studied in \cite{Corvilain}. 
In these references it was shown that
points at infinite distance are characterized by an infinite order monodromy matrix.

Intuitively  one can argue that in the above examples the towers appear in order to fulfill the 
Weak Gravity Conjecture (WGC) \cite{WGC} (see also \cite{WGC16,WGC21,WGC22,WGC23,WGC24,WGC25,WGC26,WGC27,WGC28,WGC31,WGC32,WGC33,imuv} for some recent work on the WGC and applications) in the following sense. There are in general massless 
RR gauge bosons coming from the $C_4$ RR potential, and the towers of particles from the wrapped D3-branes are charged under these $U(1)$'s.
The charge dependence on the complex structure is such that they go exponentially to zero in the large complex structure limit. The WGC applied to these $U(1)$'s forces the towers of particles to become exponentially light, to avoid global $U(1)$ symmetries to develop. The magnetic WGC already  tells us that it will not only be one particle but that a threshold of new states opens, corresponding to a full tower.

The case of towers of light particles arising for large moduli is just a particular example of a more general phenomenon, which we address in this paper. Indeed we find that not only particles but also domain walls and strings become exponentially tensionless for large moduli. 
This could be expected on general grounds, but is less obvious how the different scales of particles, strings and domain walls appear and which ones become lighter in different directions in moduli space. It is also not evident
what happens when the various types of moduli (K\"ahler, complex structure and complex dilaton)
go to infinite distance along different trajectories.

The presence of towers of domain walls is easy to understand.
In  type II CY compactifications there are massless 3-forms. For instance, one can very explicitly write down the scalar potential 
for type IIA ${\cal N}=1$ orientifolds in terms of 4-form field strengths coupling to Chern-Simons polynomials depending  only on the axions, fluxes and intersection numbers, bot not on the saxions\cite{Bielleman:2015ina,Herraez:2018vae} (see also \cite{Martucci1,Martucci2,Martucci3,Escobar:2018tiu,Escobar:2018rna}). The 3-forms couple to  RR and NS domain walls, which separate regions with different values of flux vacua.  
The associated charges are again moduli dependent and vanish for large moduli. The generalization of the WGC for 3-forms coupling to
domain walls forces the latter to become exponentially tensionless. Something similar happens with strings, which couple to  RR or NS massless 2-forms (or their 4d dual axions). 

In this  paper we analyse in detail how towers of tensionless membranes and strings appear as we move to infinite distance in 
K\"ahler, complex structure and complex dilaton moduli space for both type IIA and type IIB  4d CY  compactifications.
We concentrate in the case of ${\cal N}=1$ orientifolds in which the structure of 3-form couplings is simpler.
However, most of our results apply to the parent  ${\cal N}=2$ CY compactification before the orientifold projection,
as we describe in the text.
We do the analysis at fixed Planck scale $M_P$ (instead of
string scale) which is arguably more appropriate in the context of Swampland conjectures. 
In type IIA, domain walls arise from D2, D4, D6 and D8-branes  wrapping even cycles and NS5-branes wrapping 3-cycles.
We deduce the tensions of these domain walls from the DBI action and find their moduli dependence. 
We also compute the tensions
using the field theoretical formula for BPS states, involving the flux-induced superpotential, and check that both results agree, as expected.

Depending on the particular direction in moduli space, some or all of the different domain walls may become tensionless, and we give a
detailed description of the various posiblities. In the limit of large K\"ahler moduli in type IIA, the domain walls from D2 and D4-branes
become tensionless. We study the associated towers in some detail, in terms of the monodromies present at infinite distance points. 
The results turn out to be analogous to 
the towers of massless particles in ${\cal N}=2$ coming from D0 and D2-branes wrapping 2-cycles studied in \cite{Corvilain}.  
We also study the corresponding type IIB mirror which features tensionless domain walls formed by wrapping D5-branes on 3-cycles. 
Other towers of domain walls appear in other infinite directions in moduli space such as large complex structure.

To illustrate our general results, we compare the masses and tensions of 
particles, strings and domain walls in a type IIA $T^6/\IZ_2 \times \IZ_2^\prime$ orbifold setting. 
This simple orientifold has $h_+^{1,1}=0$ so that there are no massless RR vector bosons and,
consistently, no towers of particles from D0 and D2-banes wrapping 2-cycles. However both are present in the
parent ${\cal N}=2$ compactification.
A few general conclusions may be drawn. In practically all cases the KK scale is the lightest
scale in the problem. An exception occurs in the ${\cal N}=2$ parent model where as we said D0 towers appear
when at least one K\"ahler moduli goes to infinity,
corresponding to an M-theory limit.  This is in fact the case which is dual to the large complex structure case for IIB studied in
\cite{irene,Grimm2,Corvilain}. We find however that in the type IIA mirror domain walls from D2-branes and  
NS5-branes wrapping 3-cycles as well as
strings from NS5-branes wrapping 4-cycles appear at a mass scale similar to that of the tower of massless particles 
both in the ${\cal N}=1$ and ${\cal N}=2$ models.  This 
implies that the cut-off scale is close to the scale of the towers of particles, making difficult an analysis of the effective field theory.
There are other limits though in which there are no towers of particles coming from wrapped branes that become light but rather domain walls, both in the ${\cal N}=1$ and 
${\cal N}=2$ cases.
This is for example the case of large complex structure for type IIA, with K\"ahler moduli fixed. Yet in other moduli directions 
the KK scale is of order of the string scale, making the 10d action unreliable. 

The spectra of particles that we find is consistent with mirror symmetry. However, full consistency requires the
existence of new classes of exotic extended objects coupling to new gauge forms. In particular, as mentioned, 
in the type IIA side for large K\"ahler moduli there are towers of domain walls and strings coming from
NS5-branes wrapping 3-cycles and 4-cycles with no apparent type IIB dual for large complex structure.
They should come from exotic extended objects, coupling to 3-forms associated to non-geometric fluxes \cite{Shelton:2005cf,acfi}.
This is general, objects coming from wrapping NS5-branes often do not have mirrors unless exotic 
fluxes and branes are included.  This means that the large moduli limit of type II string compactifications
will give rise to towers of tensionless branes corresponding to exotic branes. Going to points at infinite distance 
is a way to probe the exotic extended objects in string theory. In some cases one may obtain information about 
these exotic towers (e.g. the tensions)  in terms of the flux superpotential, using the BPS formula.

Another interesting problem is the effect of the towers of domain walls and strings in the low-energy effective action.
One logical question is whether the presence of these towers of states may invalidate some of the
moduli fixing scenarios (such as KKLT \cite{KKLT} or LSV \cite{LVS,Conlon:2005ki} or type IIA with fluxes
\cite{Louis:2002ny,Villadoro:2005cu,DeWolfe:2005uu,Camara:2005dc,Palti:2008mg}) in some limit of moduli space. We do not 
analyse this point in detail but find that for instance in the KKLT scenario or type IIA toroidal orbifolds with fluxes the
towers of branes do not seem to endanger the region of the effective action relevant for the minima. 

More generally, we discuss whether towers of branes may bear on the question of emergence 
\cite{review,Harlow:2016,emergence1,emergence2,irene,Klaewer:2016kiy}
of couplings in
string vacua. The counting of number of species, which is relevant for emergence, would be in 
principle affected by the nearby domain and string towers. We discuss the case of the towers of domain walls which
could be related to the emergence of flux-dependent potentials in the effective theory.

The structure of this paper is as follows. In the next section we present a review of type IIA CY orientifolds, their moduli
and K\"ahler potential. In section 3 we discuss the towers of tensionless domain walls in type IIA orientifolds.
After discussing the limit of infinite moduli and the associated monodromy, we discuss how the towers are constructed and
populated by states. We then specialize to the case of a toroidal orientifold to discuss the different scales of extended objects
which arise in different infinite directions in moduli space. We also compute the charges of the 3-forms coupling to the
domain walls and check how they verify the WGC, and in fact saturate the BPS bounds. In subsection (3.6) we discuss the
${\cal N}=2$ CY case in which towers of tensionless strings and massless particles  are present,
including a discussion of the toroidal orbifold.
In section 4 we do a similar analysis for the
case of type IIB orientifolds and check how they are consistent with mirror symmetry. In section 5 we discuss several 
consequences of our findings, including how exotic branes and emergence may appear in the effective action in the
infinite limit in moduli space. Several appendices contain material complementary to the main text.

\section{Review of type IIA orientifolds}
\label{ss:2a}

In this section we briefly describe type IIA orientifold compactifications with fluxes,
mostly following \cite{Grimm:2004ua}. 
The purpose is to collect some basic results and to establish notation. 
The construction will be illustrated in a $T^6/\IZ_2 \times \IZ_2^\prime$ example \cite{Camara:2005dc}.

We consider the standard compactification of IIA strings on an orientifold of
$\RR^{1,3} \times \cam$, where $\cam$ is a compact Calabi-Yau 3-fold. 
The orientifold projection is generated by $\Omega_p (-1)^{F_L} \R$.  
It involves the world-sheet parity operator $\Omega_p$, the
left-moving fermion number $F_L$ and an anti-holomorphic involution $\R$ of $\cam$.
The latter acts on the K\"ahler 2-form $J$ and the holomorphic 3-form $\Omega$ of $\cam$
as  $\R J=-J$ and  $\R \Omega=\bar\Omega$. The fixed loci of $\R$ are 3-cycles
supporting the internal part of orientifold O6-planes. The RR tadpoles induced by the O6-planes
can be cancelled by a combination of D6-branes and background fluxes.

The matter content resulting upon compactification depends on the cohomology of $\cam$.
The groups $H^n(\cam)$ are divided into  $H_+^n(\cam)$ and $H_-^n(\cam)$
according to whether the elements are even or odd under the involution $\R$.
For 2- and 4-forms we use the notation
\beq
\left\{\omega_a\right\} \in H^2_-(\cam) \, , \qquad \left\{\tilde\omega^a\right\} \in H^4_+(\cam) \, ,
\qquad a=1,\ldots, h^{1,1}_- \, .
\label{forms24}
\eeq
The Hodge number $h^{1,1}_-$ counts the odd $(1,1)$ forms. Hodge duality requires $h^{2,2}_+=h^{1,1}_-$. 
For simplicity, we restrict to internal manifolds with $h_+^{1,1}=0$ but comment on the relaxation of this condition later. 
On the other hand, the numbers of even and odd 3-forms are both equal to $1+h^{1,2}$. The bases are denoted
\beq
\left\{\alpha_K\right\} \in H^3_+(\cam) \, , \qquad \left\{\beta^K\right\} \in H^3_-(\cam) \, ,
\qquad K=0,\ldots, h^{1,2} \, .
\label{forms3}
\eeq
More generically,  $H^3_+(\cam)$ and $H^3_-(\cam)$ are spanned respectively by $\left\{\alpha_k, \beta^\lambda\right\}$ 
and $\left\{\alpha_\lambda, \beta^k\right\}$. For simplicity we assume that the pairs $(\alpha_\lambda, \beta^\lambda)$ 
are absent in $\cam$. For the forms that are kept the non-trivial intersections are
\beq
\int_\cam \omega_a \wedge \tilde\omega^b = \delta_a^b \, , \qquad
\int_\cam \alpha_K\wedge \beta^L = \delta_K^L \, .
\label{int243}
\eeq
To unclutter expressions, explicit factors of the string length \mbox{$\ell_s=2\pi \sqrt{\alpha^\prime}$} are
not included above nor in the following. Such factors can be reinserted later to account for the proper dimensions.

The orientifold projection gives rise to a theory with $\cn=1$ supersymmetry in four dimensions.
Besides the supergravity multiplet the spectrum contains $h^{1,1}_-$ chiral multiplets corresponding
to the K\"ahler moduli, together with $(1+h^{1,2})$ chiral multiplets related to the dilaton and
complex structure deformations. There are no additional vector multiplets since we are taking $h_+^{1,1}=0$ 

Let us first discuss the K\"ahler moduli denoted $T^a$. The K\"ahler form $J$ and the NS-NS 2-form $B$
are odd under the orientifold action. They can thus be expanded as
\beq
J= t^a \omega_a \, , \qquad B=b^a \omega_a \, ,
\label{jbexp}
\eeq
where the so-called saxions $t^a$ and axions $b^a$ are 4-dimensional scalars. These fields combine into the complex K\"ahler moduli 
defined by
\beq
T^a= t^a+ i b^a \, .
\label{Tdef}
\eeq
The $T^a$ are scalar components of chiral multiplets. The K\"ahler potential, which determines in particular the metric
of the K\"ahler moduli space, turns out to be
\beq
K_K = - \log (8 \cv ) \, , 
\label{kkdef}
\eeq
where $\cv$ is the Calabi-Yau volume in the 10d string frame, given by
\beq
\cv = \frac16 \int_\cam J \wedge J \wedge J = \frac16 \kappa_{abc} t^a t^b t^c \, .
\label{volm}
\eeq
The $\kappa_{abc}$ are triple intersection numbers characteristic of $\cam$.

We next turn to the moduli arising from deformations of $\Omega$. Special geometry of the Calabi-Yau moduli space
allows to make the expansion
\beq
\Omega = X^K \alpha_K - \cf_K \beta^K \, .
 \label{omexp}
 \eeq
Here $(X^K, \cf_K)$ are periods of $\Omega$ and furthermore $\cf_K=\partial \cf/\partial X^K$, with $\cf$ the prepotential
function. The complex structure K\"ahler potential is defined as
\beq
K_{CS} = - \log\left(i\int_\cam \Omega \wedge \bar\Omega \right) \, . 
\label{kcsdef}
\eeq
The orientifold action $\R \Omega = \bar\Omega$ still needs to be imposed. It requires in particular $\text{Im} X^K=0$.
Taking into account the freedom to scale $\Omega$ this condition implies that there are $h^{1,2}$ real free parameters
in $\Omega$. However, as explained in \cite{Grimm:2004ua}, it is more convenient to keep the scaling freedom and
introduce a compensator field $C$ so that $C\Omega$ is scale invariant and depends on $1+ h^{1,2}$ real parameters.
The axionic partners come from the RR 3-form $C_3$ which is even under the orientifold action so it can be written as
$C_3 = \xi^K \alpha_K$.

The complex structure moduli are encoded in the complexified 3-form
\beq
\Omega_c = C_3 + 2\sqrt2 i \text{Re}(C\Omega) \, .
\label{omcdef}
\eeq
Concretely, the $1+h^{1,2}$ complex moduli denoted $N^K$ are derived from
\beq
N^K = - i \int_\cam \Omega_c \wedge \beta^K \, .
\label{nkdef}
\eeq
It remains to specify the compensator $C$. Analysis of the effective action obtained by dimensional reduction reveals that
\beq
C = e^{-\phi_4} e^{K_{CS}/2} \, ,
\label{cfield}
\eeq
where $\phi_4$ is the 4-dimensional dilaton given by $e^{\phi_4} = e^\phi/\sqrt{\cv}$. Finally, the K\"ahler potential
of the $N^K$ moduli is found to be 
\beq
K_Q = - 2 \log\left(2\int_\cam \text{Re}(C\Omega) \wedge {}^* \text{Re}(C\Omega) \right) 
= - 2\log\left(\tfrac14 {\mathcal B}_{KL} n^K n^L \right)\, ,
\label{kqdef}
\eeq
where $n^K= \preal N^k$ and ${\mathcal B}_{KL} =\int_\cam \alpha_K \wedge {}^*\alpha_L$. It follows that
$e^{-K_Q}$ is homogeneous of degree four in the $n^K$. Furthermore,
it can be shown that $K_Q=4\phi_4$.

At this stage the moduli are massless. Their vevs can be fixed by turning on fluxes to generate a potential. 
Under the orientifold action, the RR forms  $F_0$ and $F_4$ are even whereas the NS-NS 3-form $H_3$ as well as the 
RR forms  $F_2$ and $F_6$ are odd . Thus, their fluxes enjoy the expansions
\beq
\bar F_0 = - m \, , \quad \bar F_2 = q^a \omega_a \, , \quad \bar F_4 = e_a \tilde\omega^a \, , \quad 
\bar F_6 = e_0 d\text{vol}_6 \, , \quad \bar H_3 = h_L \beta^L \, .
\label{allfluxes}
\eeq
The fluxes induce a superpotential that can be written as \cite{Grimm:2004ua}
\begin{equation}
\begin{split}
W&=W_Q + W_K \, , \\
W_Q &= \int_\cam \Omega_c \wedge \overline{H}_3  \, , \\
W_K &= \int_\cam e^{B + i J} \wedge \left(\bar{F}_0  + \bar{F}_2  + \bar{F}_4  + \bar{F}_6 \right)  \, .
\end{split}
\label{wtot}
\end{equation}
Inserting previous definitions leads to
\begin{equation}
\begin{split}
W_Q & = i h_L N^L \,  ,\\
W_K & = e_0 + i e_a T^a - \frac12 \kappa_{abc} q^a T^b T^c + \frac{i}{6} m \kappa_{abc} T^a T^b T^c \, ,
\end{split}
\label{wqk}
\end{equation}
In units of $1/\ell_s$ the various flux parameters are quantized.

The scalar potential takes the standard form of
$\cn=1$ supergravity in four dimensions,  namely
\beq
\label{fpot}
V= e^K\left\{K^{I\bar J} D_I W \overline{(D_J W)} - 3 |W|^2\right\} \, , 
\eeq
where $K$ is the full K\"ahler potential $K=K_K + K_Q$.
As usual, $K^{I\bar J}$  is the inverse of $K_{I\bar J} = \partial_I\partial_J K$, $D_IW = \partial_I W + K_I W$, 
and $I$ runs over all moduli. 

The fluxes also contribute to RR tadpoles. In general tadpole cancellation implies the condition in $H_3(\mathcal{M, \mathbb{Z}})$
\begin{equation}
\sum_\a \left([\Pi_\a] + [\mathcal{R} \Pi_\a]\right) - m [\Pi_{H}] - 4 [\Pi_{\rm O6}]\, =\, 0 \, ,
\label{RRtadpole}
\end{equation}
where $\Pi_\alpha$ and $\Pi_{\mathrm{O6}}$ refer respectively to
the $3$-cycles wrapped by spacetime filling D6-branes and O6-planes, $[\Pi_H]$ is the Poincar\'e dual of the NS 
flux class $[\bar H_3]$ and  $m \in \IZ$ is the RR 0-form flux introduced in eq.~\eqref{allfluxes}. In the absence of $H_3$ and $F_0$ fluxes, we must introduce spacetime filling D$6$-branes in order to cancel the tadpole and their corresponding open string moduli must be taken into account. 
As explained in \cite{Carta:2016ynn, Herraez:2018vae}, the open string moduli will redefine the holomorphic variables in the K\"ahler potential, and also contribute to the scalar potential in the presence of extra open string fluxes. However, later we will justify that
open string moduli can be ignored in our analysis.

To end this section we exemplify the orientifold construction in the simple setup where $\cam$ is the orbifold
$T^6/\IZ_2 \times \IZ_2^\prime$, whose geometry is summarized in appendix \ref{ap:z2z2}.
We focus on the untwisted moduli. The real part of the K\"ahler moduli are the $t^i$ introduced in \eqref{jt6}.
Their K\"ahler potential is  
\beq
K_K = - \log (8 t^1 t^2 t^3) = -\log\left((T^1+\bar T^1) (T^2+\bar T^2) (T^3+\bar T^3)\right) \, .
\label{kkz2z2}
\eeq
The choice of $\Omega$ in \eqref{omt6} leads to $C=e^{-\phi_4}/\sqrt{2 \tau_1 \tau_2 \tau_3}$, with
$e^{\phi_4}=e^\phi/\sqrt{t^1 t^2 t^3}$. For the complex structure moduli we obtain
\beq
n^0 = \frac{e^{-\phi_4}}{\sqrt{\tau_1 \tau_2 \tau_3}}\, , 
\qquad n^i = e^{-\phi_4} \sqrt{\frac{\tau_j \tau_k}{\tau_i}} \, , \ i\not=j\not=k  \, .
\label{nkz2z2}
\eeq
From \eqref{kqdef} we find the K\"ahler potential 
\beq
K_Q= -\log\left(n^0 n^1 n^2 n^3\right) = 4 \phi_4\, .
\label{kqz2z2}
\eeq
The superpotential is the sum of $W_Q$ and $W_K$ given in \eqref{wqk}, with $\kappa_{123}=1$.

\section{Towers of  tensionless branes in type IIA orientifolds}
\label{sec:IIAtowers}

Before beginning the systematic study of towers of tensionless objects, let us stress a basic issue concerning scales.
The spirit of all Quantum Gravity Conjectures is to make statements about  low energy EFTs when gravity is not decoupled, that is, when the ratio between the cutoff scale of the EFT and the Planck scale is non-vanishing. This implies that whenever we argue about a dimensionful quantity in the context of Swampland Conjectures, the physically sensible approach is to compare it with the Planck scale $M_P$.
With this in mind, the statement that a state becomes massless in an EFT of gravity actually means that the ratio between its mass and $M_P$ goes to zero. This clarification is important because the string scale, $M_s=1/\ell_s$, actually depends on the moduli when expressed in terms of $M_P$ and this is crucial in order to obtain meaningful results.

The relation between $M_s$ and $M_P$, obtained writing the 4d action in Einstein frame after dimensional reduction, reads
\begin{equation}
\label{stringmass}
M_s^2=\dfrac{g_s^2 M_P^2}{4\pi (\cv/2)} =\dfrac{e^{K_Q/2} M_P^2}{2\pi} \, .
\end{equation}
Here it is understood that the internal volume $\cv$, defined in \eqref{volm}, as well as $g_s=e^{\phi}$, are evaluated 
at the moduli vevs. The factor of 2 in $\cv/2$ is due to the orientifold action. In the second equality we have used $K_Q=4\phi_4$.
For future use we also record the Kaluza-Klein mass scale can be estimated by
\begin{equation}
\label{KKmass}
M_{KK}\sim \dfrac{M_s}{\cv^{1/6}} \sim \dfrac{g_s M_P}{\cv^{3/2}} \, ,
\end{equation}
but we will give a more accurate expression for the toroidal orientifold (see Appendix \ref{ap:z2z2}). 
The units in the K\"ahler potential and the superpotential are restored by inserting suitable factors of $M_P$.
Notice that $M_P$ is constant and always finite, so that gravity is not decoupled as we move through moduli space, as required in order for Swampland Conjectures to be non-trivial.

Let us now briefly describe our strategy to identify the infinite towers of 
domain walls that become exponentially tensionless (with respect to $M_P$) as we approach an infinite distance point. 
The towers that we have found consist of bound states formed by  $\text{D}p$-branes 
and/or NS5-branes 
wrapped along cycles in the internal Calabi-Yau threefold. First we identify a basis 
defined by wrapping one single $\text{D}p$-brane or NS5-brane 
on every possible homology class.
For example, for domain walls there is a basis comprising (D8,  $\tD 6_a$, $\tD 4^a$, D2) 
wrapping respectively the whole Calabi-Yau manifold, each of the 4-cycles labeled by $a=1 \dots h_+^{2,2}$, each 2-cycle
labeled by $a=1 \dots h_-^{1,1}$, and a point. 
We compute the tensions of the basis objects by making use of the DBI action 
(or the corresponding modification for the NS5-branes) 
and study which subset of the basis  becomes tensionless at the different infinite distance points. This guarantees that all the bound states built from combinations of this subset of objects also become tensionless. After that, we explain how to construct the infinite towers and check that at every infinite distance point we can build at least one within the subset of states that becomes tensionless.

\subsection{Tensionless domain walls}
\label{sec:tensionlesswalls}

In this section we explore towers of domain walls that become tensionless as we approach infinite distances in moduli space. We will study the spectrum of domain walls in the $\mathcal{N}=1$ theory after the orientifold projection in the absence of fluxes, so that we can move freely through moduli space. Before we begin, several comments are in order. First of all, even though we do not perform in detail , one expects 
that our results for domain walls can be straightforwardly generalized to the $\mathcal{N}=2$ unorientifolded case, given the fact that the structure of the moduli space is inherited from the parent Calabi-Yau, specially in the K\"ahler sector. 
Second, without fluxes the RR tadpole cancelation condition \eqref{RRtadpole} requires to introduce D6-branes and the accompanying
open string moduli. Now, 
if all fluxes are turned off, both the closed and the open string moduli represent flat directions in moduli space. Thus, we can always move along the open string moduli space to adjust the values of the open string moduli to the reference ones, in which the closed string holomorphic variables take the usual form and the moduli space keeps its factorized structure. This is precisely the reason why we are allowed to ignore the open string moduli from now on and focus only on the closed string moduli space. 

The strategy is to consider domain walls formed by wrapping $\text{D}p$-branes along 
$(p-2)$-cycles, and NS$5$-branes along 
$3$-cycles of the internal manifold. First, we will compute the tensions of all the domain walls that arise from wrapping one brane along a 
given cycle, which constitute what we have called the basis of domain walls. We will also make contact with the usual BPS bound in terms of the superpotential that is generated on the other side of the wall. We will then show how the tensions of some of these objects go to zero as we move towards infinite distance along any direction in closed string moduli space. Once this has been done, we will introduce the candidates for the infinite towers of domain walls, whose tensions are proportional to the ones previously introduced, implying the same asymptotic behavior for the whole tower. In order to construct these infinite towers, we will make use of monodromies as generators of infinite orbits of states as explained in \cite{irene, Grimm2, Corvilain}, but in this case applied to the orbits of domain walls in type IIA. Finally, we will comment on the exponential dependence of the tensions with the proper field distance, as required by the Swampland Distance Conjecture.

The tensions of 4d domain walls obtained by wrapping a $\text{D}p$-brane on a $(p-2)$-cycle can be obtained from the DBI action, 
which for (unmagnetized) $\text{D}p$-branes  in the 10d string frame is given by \cite{BOOK}
\begin{equation}
S_{{\rm DBI}} = - \mu_p \int_{W_{p+1}} \hspace*{-5mm} d^{p+1}\xi \ e^{-\phi} \sqrt{-\det(P[g_{mn}+B_{mn}])} \, ,
\label{sdbiax}
\end{equation}
where $\mu_p=2\pi/ \ell_s^{p+1}$ is the 10d tension of the brane in the string frame, $W_{p+1}$ is the  $(p+1)$-dimensional worldvolume and $P[g_{mn}+B_{mn}]$ is the pullback  on the worldvolume of the tensor obtained by adding the background metric and the $B$-field. 
In the following we will neglect the background of the $B$-field along the internal dimensions (i.e. the $b^i$ axions) since
its contribution will not be relevant when we approach infinite distances.
We take $W_{p+1}$ to be the product of the internal cycle $\gamma_{p-2}$ and the domain wall worldvolume.
Integrating over the internal cycles gives
\begin{equation}
\label{sdbiDbrane}
S_{{\rm DBI}}=-\dfrac{2\pi \mathcal{V}_{p-2}}{g_s \ell_s^3} \int d^3 \xi \sqrt{-g^{(3)}} \, ,
\end{equation}
where $\cv_{p-2}$ is the volume of $\gamma_{p-2}$.
In terms of the Planck scale the tension is then
\begin{equation}
T_{\tD p}(\gamma_{p-2})= M_P^3 \dfrac{g_s^2 \cv_{p-2}}{\sqrt{2\pi} \cv^{3/2}} \, .
\label{tensionDp1}
\end{equation}
We can further use eqs. \eqref{kkdef}, \eqref{kqdef} and the definition of the 4d dilaton $\phi_4$, to relate the 10d dilaton  
to the K\"ahler potential and the internal volume through
\begin{equation}
\label{kdilaton}
e^K=\dfrac{e^{4\phi}}{8(\mathcal{V})^3}\, .
\end{equation}
Substituting in \eqref{tensionDp1} the tension can be finally expressed as
\begin{equation}
T_{\tD p}(\gamma_{p-2})=\dfrac{M_P^3}{\sqrt{4\pi}}\, 4 e^{K/2}\mathcal{V}_{p-2}.
\label{tensionDp}
\end{equation}

Since D2 and D8-branes wrap respectively a point and the whole manifold we have that
$\cv_0=1$ and $\cv_6=\cv$. We would like to take all $\gamma_{p-2}$ to be supersymmetric
but in a general Calabi-Yau they are not known explicitly. However, 
we can still calculate their volumes by exploiting the fact that the K\"ahler form and the  holomorphic 3-form are calibrations and that the volumes of the supersymmetric cycles are given by integrals of
these calibrated forms along any cycle in the same homology class. In particular, the volume of even cycles
can be deduced
by integrating suitable powers of the K\"ahler form along a cycle in the same homology
classes, which we take to be the Poincar\'e duals of the harmonic even forms. 
Then, the volumes of the  supersymmetric even cycles follow from 
\begin{equation}
\cv_2^a = \int_{\gamma_2^a}\!\!\! J = \int_\cam \!\!\! J\wedge \tilde \omega^a \, , 
\quad \cv_{4, a} = \frac12 \int_{\gamma_4^a} \!\!\! J \wedge J = \int_\cam\!\!\! J\wedge J\wedge \omega_a \, ,
\quad
\cv_6 = \frac{1}{3!} \int_\mathcal{M} \!\!\! J \wedge J \wedge J \, .
\label{vols24}
\end{equation}
So far we have neglected the $B$-field background. Including it amounts to replacing
$J \rightarrow J_c$ and taking the absolute value at the end.

Let us now clarify why we have made particular emphasis in the objects forming the basis of 4d domain walls,  that is, the objects which are constructed by wrapping only one kind of $\text{D}p$-brane along one supersymmetric cycle once. The key point is that these are, in general, the only ones for which the tension of the final BPS domain wall can be obtained from the  DBI action, since for arbitrary combinations they will usually form bound states, not superpositions. Nevertheless, the important point is that for these general BPS bound states the tension is always bounded from above by the addition of the DBI tensions of each of the components, guaranteeing that all the BPS bound states that are formed by an arbitrary combination of the subset of basis branes that are tensionless, will also be tensionless.  We will now consider these general BPS domain walls that are bound states of arbitrary combinations of basis domain walls. These bound states can usually be understood in microscopical terms, but since we are interested in their tensions we can resort to the BPS formula to understand the fact that they must actually form bound states. The BPS formula for the tension of a domain wall is given by \cite{Gukov:1999ya, Taylor:1999ii} (see also e.g. \cite{Ceresole}) 
\begin{equation}
\label{bps}
T=2 \Delta  \left|e^{K/2} W\right|,
\end{equation}
which relates the tension with  the difference between the modulus of the $\mathcal{N}=2$ central charge, given by the covariantly holomorphic superpotential, $e^{K/2}W$, at both sides of the domain wall. In particular, since we are studying the case without fluxes, the superpotential on one side of the wall will always be zero and the tension of the corresponding BPS domain wall can be computed from the induced superpotential on the other side. It is clear from here that a 4d domain wall consisting on an arbitrary superposition of D$p$-branes wrapping $(p-2)$-cycles will not, in general, saturate the BPS bound, except when their  superpotentials have the same phase, that is $\left| \sum_ i W_i \right| \leq \sum_i \left| W_i \right|$ and the inequality is only saturated  if the central charges are aligned, i.e. ${\rm Arg}(W_i)=\theta$ for all $i$. Hence, we can only expect to be able to reproduce the tension of BPS domain walls by adding their corresponding DBI tensions when their superpotentials are aligned, which is the case if we choose only one element of the basis of domain walls  and wrap it along the same cycle several times. These are precisely the BPS states that are unstable against decay to their BPS constituents, whereas the rest of BPS states, whose tension is strictly lower than the sum of the DBI tensions of their constituents form bound states, hence stable against decay.

To support our arguments we will recover the modulus of the superpotential given in eq. \eqref{wqk}  from the DBI computation \eqref{tensionDp} for the cases in which, as explained above, they must match. To this end, notice that 
the number of $\tD p$-branes wrapping a given $(p-2)$-cycle can be related to the corresponding flux at the other side of the wall and
in fact, each $\tD p$-brane wrapping one of the supersymmetric cycles of the Calabi-Yau once, does shift the corresponding flux by two units\footnote{The factor of 2  appears because we are considering D$p$-branes wrapping the $(p-2)$-cycles of the parent Calabi-Yau. After the orientifold projection there are new cycles which correspond to ``half-cycles'' in the parent Calabi-Yau, in such a way that after taking charge quantization into account branes wrapping these ``half-cycles'' carry integer charges and  the ones wrapping cycles on the parent Calabi-Yau carry even ones}. Taking this factor of two into account, eq. \eqref{tensionDp} yields the
domain walls tensions
\begin{equation}
\label{matchbps}
\left(T_{2}, \ T_{4}, \ T_{6} , \ T_{8}\right) = 
\frac{M_P^3}{\sqrt{4\pi}} 2 e^{K/2} 
\left( e_0 , \  e_ a \cv_{2}^a , \  q^a \cv_{4, a} , \ m \cv_{6} \right) \, .
\end{equation}
These tensions effectively match the BPS formula \eqref{bps} with the superpotential given by $W_K$ in eq.~\eqref{wtot}, 
and the volumes calculated as in \eqref{vols24} with the replacement $J\rightarrow J_c$ to include the $B$-field background .
Besides, the mass dimensions of the superpotential are reinserted throught the factor $M_P^3/\sqrt{4\pi}$, which arises in dimensional
reduction \cite{Conlon:2005ki}.
We have then computed the tensions of 4d domain walls coming from wrapping  $\tD p$-branes and checked that when they can be understood as a superposition of branes, the BPS formula matches the tensions from the DBI action.
On the other hand, when the domain walls come from bound states of $\tD p$-branes we can use the BPS formula \eqref{bps} to 
compute their tensions. 

We now turn to domain walls constructed from NS5-branes wrapping 3-cycles. 
We will first compute their tension from their action in the probe aproximation and show how it coincides with the BPS formula 
when only one cycle is wrapped. The rest of the arguments concerning bound states extends to the NS5 case in a straightforward way.
Thus, we can again limit ourselves to studying which subsets of the basis of NS5 domain walls become tensionless as we move 
towards infinite distance along directions in moduli space, to ensure that all the bound states formed with them will also be tensionless. 
The 10d tension of an NS5 brane differs from that of a D5-brane by a factor of $e^{-\phi}$. Thus, from \eqref{tensionDp}, with $p=5$,
we read the tension
\begin{equation}
\label{TNS52a}
T_{\text{NS}5}(\gamma_3)=\dfrac{M_P^3}{\sqrt{4\pi}} \, 4 \, e^{K/2} \, e^{-\phi} \mathcal{V}_3 .
\end{equation} 
The volume of the supersymmetric 3-cycles is computed integrating $\text{Re} \left(e^{-\mathcal U} \Omega\right)$ around a representative
in the cohomology class dual to $\beta^K$. The normalization factor is such that 
$\frac{i}8 e^{-2\mathcal U} \int_\cam \Omega \wedge \bar\Omega = \frac16 \int_\cam J^3=\cv$ \cite{Grimm:2004ua}. 
It follows that $\left| e^{-\mathcal U}\right| =2\sqrt2 e^\phi C$, where $C$ is the compensator field introduced in \eqref{cfield}.
We then have\footnote{We only consider the 3-cycles Poincar\'e dual to $\beta^K$ (and not to $\alpha_K$) since the dual of the B-field gives rise to 3-forms when expanded in terms of the $\alpha^K$ due to its even parity under the orientifold  action. These are the 3-forms that couple to the NS5 domain walls, implying that they arise from NS5's wrapped along the 3-cycles dual to $\beta^K$, which are calibrated with respect to Re$\left(2\sqrt{2} e^\phi C \Omega \right)$.} 
\begin{equation}
\label{vol32a}
\cv_3 = 2\sqrt2 e^\phi \int_{\cam} \rm{Re}(C\Omega) \wedge \beta^K \, . 
\end{equation}
Axions are included making the replacement $\preal (C \Omega) \rightarrow \Omega_c$ and taking the modulus at the end.
To compare with the tension obtained from the BPS formula, we can again reason that the number of NS5-branes constituting the domain wall is counted by one half of the corresponding flux at the other side of the wall.
Given the definition of the complex moduli in \eqref{nkdef}, we conclude that \eqref{TNS52a}
reduces to the BPS tension \eqref{bps} with the superpotential given in \eqref{wqk}. 

\begin{table}[t]\begin{center}
\renewcommand{\arraystretch}{2.00}
\begin{tabular}{|c|c|c|}
\hline
Brane & Cycle & Tension (in units of $ M_P^3/\sqrt{4\pi}$) \\
\hline \hline
D$2$& -  &$2\,  e^{K/2}\,  e_0$  \\
\hline
D$4$& P.D. $[ \omega_a ]$ &$2 \, e^{K/2} \, \left| e_ a \, T^a \right|$   \\
\hline
D$6$& P.D. $[ \tilde{\omega}^a ]$  & $2 \, e^{K/2} \, \left|\frac12  \sum_{b,c}\,  \kappa_{abc}\, q^a \, T^b \,  T^c\right| $ \\
\hline
D$8$& $\mathcal{M}$ &  $2 \, e^{K/2} \, \left| \frac16 \sum_{a, b,c}\,  m\,  \kappa_{abc}\, T^a \, T^b \,  T^c\right| $ \\
\hline
NS5&  P.D. $[ \beta^K]$& $2 \, e^{K/2} \, \left| h_K \, N^K \right|$  \\
\hline
\end{tabular}

\caption{Tensions of the different domain walls obtained by wrapping one kind of D$p$ or NS5-brane around a given supersymmetric cycle on the Calabi-Yau manifold. The number of branes wrapped on the same cycle is counted by the corresponding flux at the other side of the wall divided by 2. Summation over repeated indices should only be understood when explicitly indicated.}
  \label{tab:tensionsCY}\end{center}\end{table}

Let us summarize what we have done so far. We have calculated the tensions of the basis of domain walls and extended it to the case in which one object of that basis wraps its corresponding supersymmetric cycle several times, relating this number to the flux at the other side of the wall. Moreover, we have argued that if a subset of this basis becomes tensionless in some infinite distance point, all the BPS bound states formed from that subset will also become tensionless. These tensions are collected in Table \ref{tab:tensionsCY}. 
It is actually interesting to consider the typical energy scales of these objects, which can be obtained by naive dimensional analysis by just taking the cube root of the tensions, in order to compare them with the other relevant energy scales in the problem, namely the string mass and the KK mass given in eqs.~\eqref{stringmass} and \eqref{KKmass}.
We postpone this discussion to section \ref{ss:towersn2}, in which we present these scales in the toroidal orbifold $T^6/\IZ_2\times \IZ_2^\prime$, including those of towers of particles and strings that appear only in the $\mathcal{N}=2$ setup.

\subsubsection{Perturbative region of the moduli space}

Before looking into the tensions in more detail, it is important to discuss which regions 
of the moduli space are reliable to explore, in the sense that our approximations are valid and the effective field theory is under control. In particular,  one important restriction is to stay within the regime of validity of string perturbation theory, that is
\begin{equation}
\label{dilaton}
e^\phi\, =\,  2^{-3/2}\, e^{K_Q/4} \, e^{-K_K/2} \, \lesssim \, 1,
\end{equation}
where we have used eqs. \eqref{kdilaton} and \eqref{kkdef} to express the 10d dilaton in terms of the K\"ahler potentials in type IIA.
From this expression it can be seen that if we keep the K\"ahler moduli fixed and send one or several complex structure moduli to infinity, we are guaranteed to stay within the perturbative region. However, if we consider the case in which the internal volume goes to infinity by making one or several K\"ahler moduli diverge, the requirement that we stay within the perturbative region implies an important constraint, namely $e^{-K_K}\, \rightarrow\, \infty$ must be acompanied by $e^{K_Q} \, =\, A\,  e^{\, q \, K_K}$, with $A$ constant and $q\geq2$  in order to mantain $e^\phi\, \lesssim \, 1$. Let us remark that, even though the points at infinite distance in which the complex structure moduli are not divergent are out of the perturbative region, we will still consider them since, as we will see in the toroidal case, this regime can be nicely matched to M-theory. 

\subsubsection{Tensionless domain walls at different infinite distance points}

Let us now analyze the behavior of the tensions of the different elements in the basis of domain walls at various infinite distance points in moduli space. These are characterized by the subset of the moduli that go to infinity (we will show in section \ref{sec:distances} that they are actually at infinite distance) and it is essential to distinguish the cases in which only one or several moduli are taken to infinity. The reason being that whereas in the first case path dependence is trivial since we are dealing with a one dimensional problem, in the second it becomes a critical issue and different paths towards the infinite distance point may, in general, yield different results\footnote{If the moduli space is multidimensional, we fix all the moduli that do not become divergent to a finite value without loss of generality, since they will not affect the divergent behavior of the tensions. We then refer to a one or multidimensional problem referring only to the subspace spanned by the divergent moduli, and path dependence within that subspace}. In the second case, a full analysis would require to study all possible paths towards the infinite distance points and an identification of the geodesics of the moduli space of a general Calabi-Yau threefold, which is beyond the scope of this work. We will then treat the one-divergent-modulus case, which can be studied in full generality and restrict ourselves to a particular subset of paths for the other cases. In this section, this subset will include paths in which all the divergent K\"ahler moduli are proportional to each other, and similarly for all the divergent complex structure moduli.  We leave the consideration of more general paths for section \ref{sec:distances}. However, it is important to remark that even if the aforementioned paths do not include the geodesic, we should still be able to identify infinite towers of tensionless states as we approach the singularity along them, since it would be senseless to be able to find a path along which we can approach the singularity and avoid the existence of the infinite tower if it exists along a geodesic. This then seems like a necessary (though maybe not sufficient) condition for the existence of a tower along the geodesic.

In the following we present a list of different infinite distance points and examine them in some detail. Without loss of generality,
when a subset of $m$ K\"ahler, or complex structure, moduli goes large it will be taken to be $\{t^i\}$, or $\{n^j\}$, with $i=1,2,\ldots ,m$ and $j=0, 1,2,\ldots, m-1$. 
The moduli which are not explicitly taken to infinity are understood to be kept fixed. The list reads:

\begin{itemize}

{\item[{\bf (CS.I)}] 
{\bf One complex structure modulus going to infinity: $n^0 \to \infty$.}\\
In this situation, every domain wall coming from a D$p$-brane on a $(p-2)$-cycle becomes tensionless as we go to infinity. For the domain walls coming from the NS5 branes, if they wrap a cycle in the homology class of the Poincar\'e dual of $\beta^K$,  with $K\neq0$ they are also tensionless. The ones wrapping the 3-cycle that diverges are tensionless only if $e^{-K_Q}$ 
goes to infinity faster than $(n^0)^2$. }

{\item[{\bf (CS.II)}] 
{\bf Several complex structure moduli going to infinity along a path 
$n^0 \propto n^1... \propto n^{m-1} \to \infty$, $1 < m  <h_+^{1,2}$.}\\
All domain walls coming from  D$p$-branes on  $(p-2)$-cycles becomes tensionless as the singularity is approached. For the domain walls 
built from the NS5-branes wrapping a cycle  dual to $\beta^K$, the tension is proportional to
$e^{K_Q/2} n^K$.Taking into account that $e^{-K_Q}=\left(\tfrac14 {\mathcal B}_{KL} n^K n^L\right)^2$ is homogeneous
of degree four in the $n^K$,
there are two possibilities:  
\begin{itemize}
{\item[{\bf a)}] If all the terms in $e^{-K_Q}$ are homogeneous of degree two or less in the variables $\{ n^0, \dots, n^{m-1}\}$, only the domain walls which wrap $3$-cycles whose volume does not diverge become tensionless at the infinite distance point.}

{\item[{\bf b)}] If any of the terms in $e^{-K_Q}$ is homogeneous of degree three or more in the variables  that go to infinity, all the domain walls from NS5-branes become tensionless.}
\end{itemize}
} 

{\item[{\bf (CS.III)}] 
{\bf All complex structure moduli going to infinity: $n^{0} \propto n^{1} ...\propto  n^{h_+^{1,2}} \rightarrow \infty$.}\\
As in case (CS.II), all domain walls coming from D$p$-branes become tensionless in this limit. Besides, also
those formed by NS5-branes wrapping $3$-cycles are tensionless, since $e^{-K_Q}$ is a homogeneous function of degree 4 in the $n^K$.
}

{\item[{\bf (K.I)}]
{\bf One K\"ahler modulus going to infinity: $t^1 \rightarrow \infty$.}\\
In this case, it can be seen that all the domain walls obtained by wrapping D$2$-branes and NS$5$-branes become tensionless.
Additionally:
\begin{itemize}
{\item[{\bf a)}]
 If $\kappa_{111}=0$, the domain walls coming from de D$4$'s that wrap $2$-cycles whose volume is not controled by $t^1$ and the ones constructed from D$6$-branes wrapping $4$-cycles that do not contain this $2$-cycle (i.e. the ones wrapping $4$-cycles dual to  $\omega_a$, such that $\kappa_{ab1}=0=\kappa_{a11}$) also become tensionless. } 
 {\item[{\bf b)}]
 If $\kappa_{111}\neq0$, all the domain walls associated to D$4$'s become tensionless and also the ones from D$6$'s that wrap $4$-cycles that contain the divergent $2$-cycle only once or do not include it  (i.e. if we label the cycle by its Poincar\'e dual $2$-form $\omega_a$, the ones that satisfy $\kappa_{a11}=0$).} 
 \end{itemize}  
  The domain walls constructed by D$4$ and D$6$-branes wrapping the rest of the cycles and the ones corresponding to the D$8$ do not become tensionless. In particular, all bound states of D$2$'s, NS5's and the aforementioned D$4$'s and D$6$'s become tensionless.}

{\item[{\bf(K.II)}]
{\bf Several K\"ahler moduli going to infinity: $t^1 \propto t^2 ... \propto t^{m} \rightarrow \infty$,
$1<m<h_{-}^{1,1}$.}\\
As in case (K.I), all the domain walls obtained by wrapping D$2$-branes and NS$5$-branes become tensionless. In addition:
 \begin{itemize}
  {\item[{\bf a)}] If all the $\kappa_{ijk}=0$ for $i,\ j,\ k\, =\, 1,...,m$, the domain walls that consist on  D$4$'s that do not wrap any of the $2$-cycles that diverge become tensionless. So do the ones obtained from D$6$'s in $4$-cycles which do not include any of the infinite volume $2$-cycles (i.e. the ones wrapping $4$-cycles dual to  $\omega_a$, such that $\kappa_{abi}=0=\kappa_{aij}$ for all $i,\ j\, =\, 1,...,n$) become tensionless. }
  
   {\item[{\bf b)}]If any of the $\kappa_{ijk}\neq 0$,  the domain walls that become tensionless are  the ones constructed from D$4$'s and from D$6$'s wrapping a $4$-cycle that does not contain only divergent $2$-cycles (i.e. if we label the 4-cycle by its Poincar\'e dual $2$-form $\omega_a$, the ones that satisfy $\kappa_{aij}=0$ for every $i,\ j\, =\, 1,...,n$). }
   \end{itemize}
 As before, the rest of the D4's, D6's and the D8 do not become tensionless.
}

{\item[{\bf (K.III)}] 
{\bf All the K\"ahler moduli going to infinity: $t^1\propto t^2 ... \propto t^{h_{-}^{1,1} }\rightarrow \infty$.}\\
As in the previous cases, the domain walls that consist on D2's or NS5's become tensionless. Furthermore, 
every domain wall from a D4 on a $2$-cycle becomes tensionless, too. None of the domain walls from  D6's  and D8's are tensionless in this case.
}
\end{itemize}

Having shown that there is always some set of basis domain walls that become tensionless as we approach a singular point, we are in a position to propose candidates for the infinite towers of tensionless branes,  formed by bound states of this subset of tensionless basis branes. We address this problem in section \ref{sec:towers}. At this stage 
let us remark that in general there are many more 4d domain walls apart from the ones that we have discussed 
and these could turn on other kinds of fluxes (e.g. metric and non-geometric fluxes). The 10d picture might not be clear in some cases but, from the 4d point of view, as long as the $|W|$ associated to them does not cancel the whole factor of $e^{K/2}$ in 
eq.~\eqref{bps} they will become tensionless at some infinite distance point. 
We will not consider exotic domain walls to construct towers that become tensionless, but we will return to them and comment on possible implications in section \ref{ss:discussion}.

\subsection{Infinite distances and monodromies}
\label{sec:distances}

In this section we show that the points in moduli space at which any of the real parts
of the complex structure or K\"ahler moduli
tends to infinity, are actually at infinite proper distance. 
Additionally, we will relate this behavior to monodromy matrices and generators, in the spirit of \cite{irene, Grimm2, Corvilain}.
These concepts will play a central role in the construction of the towers in next section.

The proper distance between two points $P$ and $Q$ in moduli space, joined by a curve $\gamma$,  is defined as
\begin{equation}
\label{propddef}
d_\gamma (P,Q)\, =\, \int_\gamma \, \sqrt{2 \, K_{I \bar{J}}\, \dot{z}^I \, \dot{\bar{z}}^J}\, ds \, ,
\end{equation}
where $\dot{z}^I=\partial z^I / \partial s$ and $K_{I\bar{J}}$ is the  K\"ahler metric. The two pieces in the
full K\"ahler potential, $K=K_K+K_Q$,  are given in eqs.~\eqref{kkdef} and \eqref{kqdef}.
It is easy to see that $K_K$ diverges if one or more of the K\"ahler moduli $t^a\rightarrow\infty$\footnote{This is actually straightforward only when we assume that all $\kappa_{abc}\geq 0$, in order to avoid subtle cancelations that could spoil the divergence. However, as explained in \cite{Grimm2, Corvilain}, this can be proven in general by means  of the growth theorem of \cite{GrowthTh}.}. 
This is also the case for $K_Q$ when all $\{ n^K \} \rightarrow\infty$, and we will assume  it also holds when some subset of them are sent to infinity. With this in mind, the goal is to prove that the proper distance along any path,  from any point $P$ at which every modulus takes finite values,  to a point $Q$ characterized by one or more moduli going to infinity, is bounded from below by the value of $K$ at the point at infinity. 
Since the latter diverges, so will do the proper distance.

To analyze the proper distance we basically adapt the
arguments in \cite{Corvilain} to include the complex structure sector in the $z^I$. 
The integrand fulfills
\begin{equation}
\label{boundint}
\left( 2 K_{I\bar{J}}\dot{z}^I  \dot{\bar{z}}^J \right)^{\frac{1}{2}}  \geq 
\dfrac{1}{\sqrt{\cc}} \left( 2 K_I K^{I\bar{J}} K_{\bar{J}}  \right)^{\!\frac{1}{2}} \! \left( 2 K_{I\bar{J}} \dot{z}^I  \dot{\bar{z}}^J \right)^{\frac{1}{2}}  \geq 
 \dfrac{1}{\sqrt{\cc}} \left| K_I \dot{z} ^I + K_{\bar{J}} \dot{\bar{z}}^J  \right|  =   \dfrac{1}{\sqrt{\cc}}  \left| \dot{K} \right|\, ,
\end{equation}
where in the first step we used that
\begin{equation}
2 K_I K^{I\bar{J}} K_{\bar{J}} \, \leq  \, \cc,
\end{equation}
with $\cc$ a finite constant. This condition can be  straightforwardly met for any $\cc >14$, by virtue of the no-scale condition 
$K_I K^{I\bar{J}} K_{\bar{J}} =7$. The inclusion of $\alpha^\prime$ corrections generically breaks this no-scale condition for the K\"ahler sector, but it can be seen that the deviation goes to zero as the volume increases \cite{Corvilain, Escobar:2018rna}, so that a finite constant 
$\cc$ can always be found. The second inequality relies on the Cauchy-Schwarz inequality \mbox{$|u| \cdot |v| \geq |u \cdot v|$}, with vectors  $u=(\dot{z}^I, \dot{\bar{z}}^J)$, $v=(K^{I\bar{L}}K_{\bar{L}}, K^{\bar{J}L}K_L )$, and inner product given by block diagonal matrix with the K\"ahler metric in the diagonal blocks. Substituting the bound \eqref{boundint} in the definition \eqref{propddef} gives
\begin{equation}
\label{dbound}
 d_\gamma (P,Q) \, \geq \, \dfrac{1}{\sqrt{\cc}}  \int_\gamma \left| \dot{K} \right| ds \geq  \dfrac{1}{\sqrt{\cc}} \left| \int_\gamma dK \right|\, = \, \dfrac{1}{\sqrt{\cc}} \left| K(Q)-K(P) \right| \, .
\end{equation}
Thus, the proper distance from $P$ where all moduli are finite, to $Q$ where the K\"ahler potential diverges because at least one moduli does,
is actually bounded from below by infinity and must be infinite along any path. 

The infinite distance to points where one or more moduli tend to infinity can actually be understood in terms of monodromy
transformations of the period vector around singularities.
In our setup the period vector in the large volume limit takes the form \cite{Herraez:2018vae} (see Appendix \ref{ap:details} for more details on the period vector and its precise relation to the K\"ahler potential)
\begin{equation}
\label{periodvector}
\Pi^{\, t}(T^a, N^K) \, =\, (1, i T^a, -\oh \CK_{abc} T^bT^c, -\frac{i}{6} \CK_{abc} T^aT^bT^c, i  N^K) \, ,
\end{equation}
It is convenient to denote the moduli $\{T^a, N^K\}$ generically by $Z^\ci$. Under shifts of the axions $\text{Im}\, Z^\ci$, the period
vector transforms as $\Pi \to R_\ci \Pi$. The monodromy transformations are explicitly given by   
\begin{equation}
\label{monodromymatrices}
R_a= \left(\begin{array}{ccccc} 1 & 0 & 0 & 0 & 0 \\ \vec{\d}_a  & \d_b^{\,c} & 0 & 0 & 0 \\  \frac{1}{2}\CK_{aab} & \CK_{abc} & \d_b^{\,c}  & 0 & 0  \\ \frac{1}{3!} \CK_{aaa} & \frac{1}{2}\CK_{aac} &  \vec{\d}_a^{\,t} & 1 & 0  \\ 0 & 0 & 0 & 0 & \delta_M^{\, L}  \end{array} \right)\, , \quad 
R_K =\left(\begin{array}{ccccc}1 & 0 & 0 & 0 & 0\\ 0 & \d_b^{\,c}  & 0 & 0 & 0\\ 0 & 0 & \d_b^{\,c} &0 & 0\\0 & 0 & 0 &1 & 0\\ \vec{\d}_K & 0 & 0 &0 & \delta_M^{\, L} \end{array}\right)\,.
\end{equation}
In turn they can be written as $R_\ci = e^{P_\ci}$, in terms of monodromy
generators \cite{Herraez:2018vae}\footnote{We do not include $\alpha^\prime$ corrections in this work. They can be 
incorporated, without changing the conclusions,
by substituting our period vector and generators by those in  \cite{Grimm2, Corvilain, Escobar:2018rna}.}
\begin{equation}
\label{monodromygenerators}
P_a =\left(\begin{array}{ccccc}0 & 0 & 0 & 0 & 0 \\ \vec{\d}_a  & 0 & 0 & 0& 0 \\  0 & \CK_{abc} & 0  & 0 & 0 \\ 0 & 0 &  \vec{\d}_a^{\,t}&0& 0\\0 & 0 & 0 &0& 0\end{array}\right)\, ,\qquad \qquad
P_K =\left(\begin{array}{ccccc}0 & 0 & 0 & 0 & 0\\ 0 & 0 & 0 & 0 & 0\\ 0 & 0 & 0 &0 & 0\\0 & 0 & 0 &0 & 0\\ \vec{\d}_K & 0 & 0 &0 & 0\end{array}\right)\, .
\end{equation}
It can be checked that the $P_\ci$ are nilpotent and fulfill $\left[ P_\ci, \, P_\cj \right]=0$. 

For a more general shift of a a set of axions $\{\text{Im}\, Z^i \}$, the monodromy transformation is just
\begin{equation}
R_i=\exp \left( \sum_ i  k^i  P_ i \right),
\end{equation}
where the $k^i$ are positive integers. This means that the monodromy generator
associated to a subset of the moduli is obtained by taking the appropriate linear combination of the corresponding generators.
Note that as long as the $k^i$ are positive we can take all of them to be equal to 1 without loss of generality. We then define the generator of simultaneous shifts of $n$ axions by
\begin{equation}
\label{compmonodromygen}
P_{(n)}=\sum_ i^n  P_ i,
\end{equation}
in analogy to \cite{Grimm2, Corvilain}.

We now describe how the period vector \eqref{periodvector} has an expansion consistent with the nilpotent orbit theorem of \cite{nilpotent}.  
Consider a singularity in moduli space described by $Z^i\rightarrow \infty$, where $\{Z^i\} \subseteq \{ T^a, \, N^K\}$ is the subset of the moduli which diverge, and call $\{\zeta^\flat\}$ the rest of the moduli. In general, about this singularity 
the period vector has the local expansion
\begin{equation}
\label{nilpotentorbit}
\Pi (Z^i, \zeta^\flat )=\mathrm{exp}\left[ \sum_i i Z^i P_ i\right] \left({\textbf{a}_0} (\zeta^\flat)+ 
\sum_{j}{\textbf{a}_j}(\zeta^\flat) \, e^{-2\pi Z^j} + \ldots \right)\, ,
\end{equation}
where $\ldots$ stands for higher order terms in $e^{-2\pi Z^j}$. The singular behaviour is captured by the exponential in front,
acting on  the vector $\textbf{a}_0$. This vector depends only on the non-divergent moduli and can be deduced from the expansion.
In this language,
the relation between infinite distances and nilpotent orbits is encoded in the fact that the point being at infinite distance implies that for some $P_i$ around that point we must have
\begin{equation}
\label{infdistmonodromy}
 P_i \, {\bf a_0} \, \neq \, 0 \, .
\end{equation}
In \cite{irene} it was conjectured that this implication goes both ways, that is, if the condition is fullfilled for some $P_i$, the singular point is at infinite distance\footnote{This result was actually proven in 
\cite{irene} for the case in which only one modulus diverges, and conjectured to be true for the rest of the cases.}. 

We can actually check the conjecture of \cite{irene} in some cases, since we have shown that all the points where
some moduli diverge are at infinite distance.  
Expanding the period vector around one of these points allows us to obtain the corresponding vector $\textbf{a}_0$. 
We can then see that it is not annihilated by the monodromy generator $P_{(n)}$ about that singular point, implying the existence of some $P_i$ fulfilling \eqref{infdistmonodromy}. For instance, at the point where all moduli diverge, namely
$\preal Z^\ci \rightarrow \infty, \ \forall \, \ci$, the expansion of \eqref{periodvector} yields
\begin{equation}
\textbf{a}_0^t=\left( 1, 0,0,0,0 \right) , \qquad P_{(n)}=\sum_{a=1}^{h_{-}^{1,1}} P_a+ \sum_{K=1}^{h_+^{1,2}} P_K \, .
\end{equation}
Clearly $\textbf{a}_0$ is not anhilitated by any of the monodromy generators $P_a$ and $P_K$ given in \eqref{monodromygenerators}.
It is straightforward to determine the different $\textbf{a}_0$'s and $P_{(n)}$'s associated to other infinite distance points
where only a subset of the moduli goes to infinity. Details are presented in Appendix \ref{ap:details}.

\subsection{Towers of branes}
\label{sec:towers}

In section \ref{sec:tensionlesswalls} 
we have identified a basis of 4d domain walls 
and characterized how their tensions behave as we approach infinite distance points, emphasizing that this is a highly path dependent question when more than one modulus diverges.  We have identified that along all the paths towards infinite distance points that we have studied, there is a subset of this basis that becomes tensionless. The next step is to identify an infinite tower of 
domain walls formed by bound states of the aforementioned subset of tensionless ones, so that at the infinite distance point the whole tower becomes tensionless. 

If we consider the tower of BPS domain walls that arises from wrapping the same brane  $n$ times along the same cycle, it could be argued that the tension of each of these states is given by $n$ times the tension of the corresponding element of the basis (see Table \ref{tab:tensionsCY})  and it might then be unstable against decay to its constituents, implying that one cannot ensure that the tower is populated by physical states. Hence, if we want to consider infinite towers composed by branes wrapping cycles several times, we must make sure that they consist of BPS states that are bound states (and not just superpositions) of branes, so that they are stable and therefore populated by physical states. In any case, let us anticipate that since we have seen that at all the infinite distance points the  D$2$  domain walls  always becomes tensionless  and so does (at least) one kind of D$4$, we can always form bound states with $n$ of these  D$2$-branes  and one  D$4$, resulting in an infinite tower, labeled by $n$, that is definitely stable at the infinite distance point.

\begin{figure}[tb]
    \begin{center}
        \includegraphics[width=400pt]{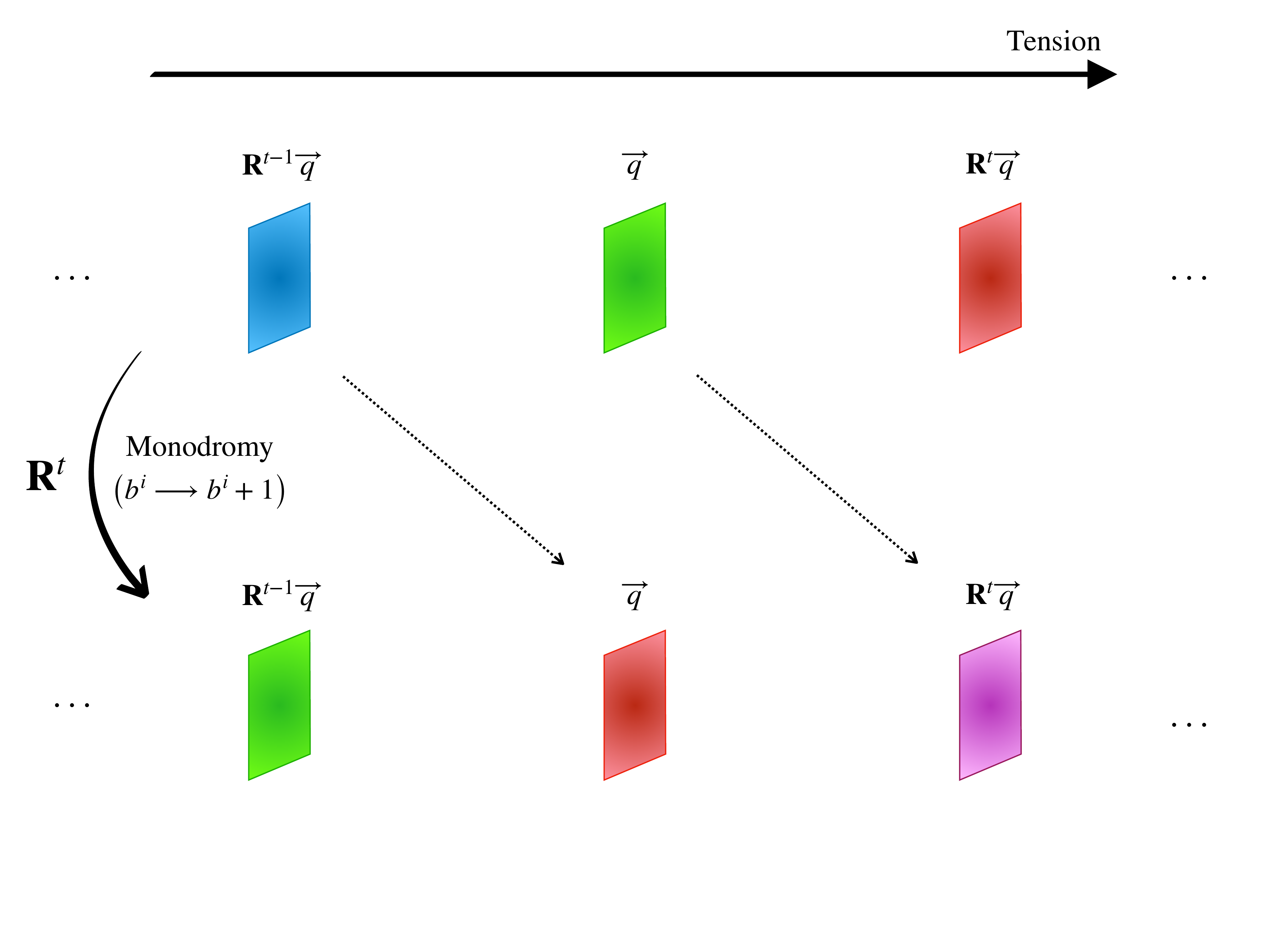}
     
        \caption{\footnotesize The monodromy maps some states into others with different mass and charge but the whole tower is mapped to itself.} 
          \label{im:monodromy}
          \end{center}
\end{figure}

The generation of these towers of BPS states can be studied using the language of monodromies introduced in the previous section, as first proposed in \cite{irene} for the tower of BPS particles that come from D$3$-branes wrapping certain 
$3$-cycles, and further generalized in \cite{Grimm2, Corvilain}. The logic behind the tower of states that is generated by a monodromy can be summarized as follows (see Fig. \ref{im:monodromy}). Suppose we have a BPS state (a bound state of domain walls from  wrapped branes) characterized by a central charge, proportional the modulus of the superpotential that is generated at the other side of the wall. For a fixed point in moduli space, the charge of this object is fixed. Now suppose we perform a monodromy transformation, that is, we shift some axions by their period (or, equivalently, shift the corresponding fluxes accordingly). As can be seen, the central charge and the tension of the state after the shift have changed (equivalently, one can keep the axions fixed and shift the fluxes in the complementary way, corresponding to a change in the set of wrapped branes that form the bound state giving rise to the domain wall). However, since monodromies come from gauge redundancies in the higher dimensional theory, we know that the physics of the theory after the monodromy transformation cannot be different, that is, it should not map a state in the theory to one which is not  a part of the spectrum, because the whole set of states of the theory must be unchanged under this transformation.  That is, the monodromy transformation will, in general, map one bound state formed by a particular set of branes on cycles to another bound state with a different set of branes (i.e. a BPS domain wall with different central charge than the original one). Hence, following this logic, the only way to reconcile this non-trivial mapping is that the whole set of domain walls that are connected by the monodromy transformation is mapped to itself, that is, the monodromy can only map states belonging to that set to other states within the same set, requiring the existence of a whole tower if one of its states exists in the theory, as shown in Fig. \ref{im:monodromy}. Moreover, the existence of one state on one infinite order monodromy orbit automatically implies that the whole orbit must exist for consistency of the theory. Let us note that the validity of this argument depends on the assumption that the states do not become unstable when they undergo the monodromy, otherwise we could not argue that the whole monodromy tower is populated by physical states. This aspect was analyzed in some detail in \cite{irene} for the towers of particles in terms of crossing walls of marginal stability when the monodromy transformation is performed. We will not analyze this aspect in general, but we will comment on some specific examples of towers that are stable at the infinite distance points. Let us mention, whatsoever, that we expect these monodromy orbits to  capture the infinite towers of states. 

In analogy with \cite{irene}, we can then use an infinite order monodromy to show the existence of an  infinite tower of BPS states. These will actually become massless as we approach the singularity if they are formed by bound states of the subset of basis domain walls that become massless at that point . We can express eq. \eqref{bps} in the following form \cite{Gukov:1999ya,Taylor:1999ii} 
\begin{equation}
\label{tensionW}
T=2 \, e^{K/2}W \, = \, 2 \, e^{K/2} \, \vec{\Pi}^t \cdot \vec{q},
\end{equation}
where the charge vector has the form (see Appendix \ref{ap:details} for a precise definition of the charge vector and details on the superpotential)  
\begin{equation}
\label{chargevector}
\vec{q}\,^t=\left( e_0, \, e_a , \, q^a, \,  -m, \, h_K \right).
\end{equation}
The monodromy transformations can then act on the period vector and induce an action in the charge vector in the following way
\begin{equation}
T\, = \, 2 \, e^{K/2} \, \vec{\Pi}^t \cdot \vec{q} \,  \xrightarrow {R_i } \, T^\prime \, = \, 2 \, e^{K/2} \, \left[ R_ i\, \vec{\Pi} \right]^t \cdot \vec{q} \, = \,  \, 2 \, e^{K/2} \,  \vec{\Pi} ^t \cdot \left[( R_i)^t \, \vec{q} \, \right] \, .
\end{equation} 
Thus, if we perform the monodromy defined by $R_i$ to a state of  charge $(R_i)^{t-1} \cdot \vec{q}$ it has the same tension as a state of charge $\vec{q}$ before the monodromy. This defines the action of the monodromy on the charge vector and by repeating this process one can study the whole monodromy orbit. It can be seen that for the cases of interest to us, the monodromy orbit being infinite is equivalent to
\begin{equation}
\label{infmonodromy}
 P_i^{\, t} \, \vec{q}\, \neq\, 0, 
 \end{equation}
for some $i$, since this means that the charge vector is not mapped to itself and thus an infinite orbit is generated. With this in mind, once  we have the subset of the basis of domain walls that are tensionless at the singular point (see section \ref{sec:tensionlesswalls}) the construction of the infinite tower which becomes massless at that point boils down to showing that, taking an element of the aforementioned subset, there exist a monodromy transformation which generates an infinite orbit consisting of bound states of the elements of that subset only. Along the way, we will also be able to understand the subset of the basis of domain walls which become tensionless in terms of the conditions on the generators of monodromies explained in \cite{irene, Corvilain}.

 For a general domain wall, characterized by a set of charges (i.e. the fluxes at the other side of the wall), the action of the generators 
$(P_a)^{t}$ and $(P_K)^{t}$ on the charge vector is easily found to be  
\begin{equation}
\label{actmon}
(P_a)^{ t} \, \vec{q}  = \left( e_a, \CK_{abc} q^c, -m \, \vec{\d}_a , 0, 0 \right)^{t} \,  , 
\quad
(P_K)^{t} \, \vec{q}  = \left( h_K, 0, 0, 0, 0 \right)^{t} \, .
\end{equation}
We are now ready to elaborate on the construction of the towers. Below we will discuss the cases with K\"ahler moduli
growing to infinity. The examples with complex structure moduli are presented in Appendix \ref{ap:moretowers}.

\begin{itemize}
{\item[\textbf{(K.I})]{\bf One K\"ahler modulus going to infinity: $t^1 \rightarrow \infty$.}\\
The subset of domain walls that were found to be tensionless in this case are the following:
\begin{itemize}
{\item[{\bf a)}] If $\CK_{111}=0$, the domain walls with $e_1=0$, $q_a=0$ for every $a$ such that $\kappa_{1ab}\neq0$, and $m=0$. From eqs. \eqref{actmon} it can be seen that the action of $(P_{a=1})^{ t} \, \vec{q}=0$ but for the rest of the generators $(P_{a\neq 1})^{ t} \, \vec{q}\neq 0\neq (P_K)^{t} \, \vec{q}$ and, moreover, the set of charges that were required to be zero in order to stay within the tensionless subset remain zero throughout the whole orbit.}
{\item[{\bf b)}] If $\CK_{111}\neq 0$, the tensionless domain walls fulfill  $q_a=0$ for every $a$ such that $\kappa_{11a}\neq0$, and $m=0$. Again, using eqs. \eqref{actmon} it can be observed that the action of all  the generators $(P_{a})^{ t} \, \vec{q}\neq 0\neq (P_K)^{t} \, \vec{q}$. In addition, the action of the monodromy on a tensionless brane always is such that the charges that needed to be zero for the brane to be tensionless remain zero along the whole monodromy orbit.}
\end{itemize}
Hence we have identified the infinite towers of tensionless domain walls by checking that there is some monodromy that acts non-trivially on the charge vector but never connects a tensionless wall with another one with non-zero tension. These towers actually include the bound states of D4's and D2's that would be analogous to the bound states of D2's and D0's found in \cite{Corvilain} for the case of the particles. Finally, note that whereas in case (a) the monodromy that generates the tower must be different from the one about the infinite distance point, in case (b) the tower can also be generated by the monodromy about the singular point\footnote{In the language of \cite{irene, Grimm2, Corvilain}, this can be understood from the fact that case (a) implies a singularity of the so called type II or III, whereas case (b) signals a type IV.}. 
}
\end{itemize}

It is worth mentioning at this point that any linear combination of the monodromies that individually generate a tower will also generate a valid infinite tower that becomes tensionless at the singular point. Let us now continue with examples in which several moduli diverge. As mentioned before, in these cases the problem becomes highly path dependent. Our approach is to consider first the linear paths studied in section \ref{sec:tensionlesswalls} and build the towers from the elements in the basis of domain walls that become tensionless as we approach the infinite distance point along these paths. Afterwards, we will briefly discuss the generalization to the growth sectors considered in \cite{Grimm2, Corvilain}.

\begin{itemize}
{\item[\textbf{(K.II})]{\bf Several K\"ahler moduli going to infinity: $t^1 \propto t^2 ...\propto t^{m} \rightarrow \infty, \ 1<m<h_{-}^{1,1}$.}\\
We distinguish two different cases:
\begin{itemize}
{\item[{\bf a)}] If $\CK_{ijk}=0$ for all $i, j, k=1,2...m$, the domain walls that become tensionless are the ones that fullfill the conditions $e_i=0$, $q_a=0$ for every $a$ such that $\kappa_{abi}\neq0$, and $m=0$. Using eqs. \eqref{actmon} we conclude that for tensionless domain walls $(P_{a=i})^{ t} \, \vec{q}=0$ but $(P_{a\neq i})^{ t} \, \vec{q}\neq 0\neq (P_K)^{t} \, \vec{q}$ and, as before, the charges that needed to be zero to stay within the tensionless subset remain zero throughout the whole orbit, as expected.}
{\item[{\bf b)}] If $\CK_{ijk}\neq 0$ for some $i, j, k=1,2...n$, the tensionless domain walls obey  $q_a=0$ for every $a$ such that $\kappa_{ija}\neq0$, and $m=0$. Using eqs. \eqref{actmon} once more, one realizes that the action of all  the generators $(P_{a})^{ t} \, \vec{q}\neq 0\neq (P_K)^{t} \, \vec{q}$. In addition, the action of the monodromy on a tensionless brane always gives another brane that fulfills the tensionless conditions for the fluxes, ensuring that the whole orbit remains tensionless.}
\end{itemize}
We have then identified an infinite tower of domain walls that become tensionless as we approach the infinite distance point along the aforementioned paths. As explained previously, even when these family of paths does not contain the geodesic, it is crucial that we still find an infinite tower as we approach the infinite distance point, since it would make no sense to find a path for which this does not happen if we expect to find the tower when traversing the geodesic. 
If we want to generalize this to include all the paths within a given growth sector given by (see \cite{Grimm2, Corvilain}), 
\begin{equation}
\left\{T^i=t^i +i b^i \ : \  \dfrac{t^1}{t^2}>\lambda, \, ... \, , \dfrac{t^{n-1}}{t^n}>\lambda, \, t^n>\lambda  \right\}
\end{equation}
 for some positive constant $\lambda$, where we have introduced a particular ordering for the $T^i$'s, the conditions for the tower to remain tensionless become more restrictive and imply
\begin{equation}
\begin{split}
&(P_{(i)})^{ t} \, \vec{q}= 0, \qquad \mathrm{if} \ \CK_{jkl}=0 \ \mathrm{for \ all } \ j,k,l =1,2...,m  \\
&(P_{(i)}^2)^{ t} \, \vec{q}= 0, \qquad \mathrm{if} \ \CK_{jkl}\neq 0 \ \mathrm{for \ some } \ j,k,l=1,2...,m
\end{split}
\end{equation}
where the equality must be fulfilled for all $i=1,2...,n$ with the matrices defined in \eqref{compmonodromygen}.
}
{\item[{\bf (K.III)}] {\bf All the K\"ahler moduli going to infinity: $t^1\propto t^2 ... \propto t^{h_{-}^{1,1} }\rightarrow \infty$}.\\
When approaching  the Large Volume Point along these trajectories all the discussion completely mimics the one in (K.IIb). The tensionless domain walls are those with $q^a=0$ for all $a$ and $m=0$. By means of eqs. \eqref{actmon} it can be seen that all the monodromies act non-trivially on the charge vectors and if we begin with a tensionless brane the whole orbit is tensionless. Hence we can always generate an infinite tower of bound states of D2 and D4-branes at the Large Volume Point by means of the monodromy generators about that point.
If we again want to generalize this to include every path within a given growth sector given by
\begin{equation}
\left\{T^i=t^i +i b^i \ : \  \dfrac{t^1}{t^2}>\lambda, \, ... \, , \dfrac{t^{n-1}}{t^n}>\lambda, \, t^{h_-^{1,1}}>\lambda  \right\}
\end{equation}
 with $\lambda$ some positive constant, the conditions for the tensionlessness of the tower become 
\begin{equation}
(P_{(i)}^2)^{ t} \, \vec{q}= 0, \qquad \mathrm{for \ all } \ i =1,2...,h_-^{1,1}.
\end{equation}
}
\end{itemize}

To recap, we have identified at least one infinite tower that becomes massless as we approach various infinite distance points. In order to construct the tower, we have shown that we can always find a monodromy that acts non-trivially on a charge vector of a tensionless domain wall, generating an  infinite number of domain walls  whose charge vectors fulfill the tensionless conditions. That is, if we begin with a bound state of any of the tensionless domain walls  characterized in section \ref{sec:tensionlesswalls}, there is always a monodromy that relates it with an infinite number of different bound states of tensionless domain walls. The main difference among the different towers resides on whether they are generated by the monodromies around the singular point or around any other non-singular point that can intersect the original singular point at another singularity.

\subsubsection{The exponential behavior}

To close this section, we comment on the exponential behavior predicted by the SDC. The exponential dependence on the proper field distance of the masses and tensions of the objects which form the infinite towers is hard to prove in general, due to the fact that it would require  a calculation of the geodesics that go through the singular point in the first place and then,  a computation of  the distance along these geodesics for the moduli space of a completely general Calabi-Yau orientifold. If the singular point is characterized by only one moduli going to infinity, the exponential behavior was proven in \cite{irene} by means of the nilpotent orbit theorem and this applies to our case as well. Additionally, in \cite{irene, Grimm2} it was argued that this also happens for more general cases, even though it was not proven in full generality. Here we will show how the exponential mass behavior arises along certain trajectories in the cases in which all the moduli in one sector (either all the K\"ahler, all the complex structure or both) diverge. 

The idea is the following. In eq.~\eqref{dbound} we found a lower limit for the geodesic distance to the singular point, so that the distance will be greater or equal to that for any path. Additionally, we can find an upper bound for the geodesic distance by taking any particular path, since the length along the geodesic will always be less or equal than the one computed along that path (with the inequality being saturated if we happen to find the geodesic). Hence, finding a path along  which the asymptotic behavior of the distance coincides with the one in 
eq.~\eqref{dbound}, it is ensured that the geodesic distance will have the same asymptotic behavior, since it is 
constrained by the same bounds both from above and below. We prove this for straight line trajectories towards the infinite volume point (i.e. all K\"ahler moduli going to infinity, the rest fixed), since it is straightforward to repeat the derivation for the complex structure case. To begin we  consider a path parameterized by $\lambda$, given by $T^1/w^1=T^2/w^2 =\dots=T^{h^{1,1}_-}/w^{h^{1,1}_-}=\lambda$, 
with $w^a$ a vector of positive constants. Axions will be fixed to zero without loss of generality. From the fact that $e^{-K_K}$ is a homogeneous function of the $t^a$'s (of degree three), we can conclude that $K_{a\bar{b}}$ is a homogeneous function of degree minus two of the $t^a$'s. Along this path, this implies
\begin{equation}
K_{a\bar{b}}\left( t^a (\lambda)\right)=\dfrac{1}{\lambda^2} K_{a\bar{b}}\left( t^a  = 1 \right),
\end{equation}
where   $K_{a\bar{b}}\left( t^a  = 1 \right)$ is a positive definite matrix of constants. Moreover, since 
$\frac{\partial T^a}{\partial \lambda}=w^a$ the distance takes the form
\begin{equation}
d_\gamma (P,Q)\, =\,  \int_P^Q \, \sqrt{\dfrac{2}{\lambda^2} K_{a \bar{b} } \left( t^a  = 1 \right)    w^a w^{\bar{b}}}\  d\lambda \, =\,  \alpha \, \int_P^Q \, \dfrac{1}{\lambda}\, d\lambda
\end{equation}
where we have defined the constant $\alpha^2= 2 \sum_{a,\bar{b}} K_{a\bar{b}} \left( t^i  = 1 \right)    w^a w^{\bar{b}} $, which is positive since $K_{a\bar{b}}$ is positive-definite. Finally, the distance takes the form
\begin{equation}
d_\gamma (P,Q)\,=\, \alpha \log \left| \lambda \right|^Q_P\, = \, \dfrac{\alpha}{3} \left| K_K(Q)-K_K(P)\right|,
\end{equation}
where in the last step we have used that, along this trajectory $e^{K_K}\propto \lambda^{-3}$. 
Note that this can be repeated for the complex structure sector with the only change that $e^{-K_Q}$ is a homogeneous function of the $n^I$'s of degree  four, yielding the same conclusion and also for all the moduli going to infinity at the same time, since $e^{-K}$ is homogeneous of degree seven. Note that the expressions of the tensions that we have calculated allow us to conclude that the prefactors always decrease exponentially with the proper distance and for the given paths this is also the case for the other factors, resulting in the exponential behavior predicted by the SDC.

Finally, we note that in the cases in which  the dependence of the function $e^K$ on a particular set of moduli  can be factorized and the factor constitutes a homogeneous function (of any degree) of the aforementioned subset of moduli, we could automatically reproduce the above argument to show the exponential behavior with the proper distance at that infinite distance point. In fact, the toroidal orientifold that we explore in detail in the next section provides the typical example of this situation, in which we can factorize the homogeneous function of degree seven $e^K$ into seven functions, each of them depending only on one modulus and homogeneous of degree one in that modulus. Hence it is guaranteed that when we send any combination of the moduli towards infinity the growth of the distance will be asymptotically logarithmic in the moduli, as can be explicitly computed. In this case, it can also be shown that these straight lines are actually the geodesics towards the infinite distance point.

\subsection{Tensionless branes in toroidal orientifolds}
\label{sec:toroidal}

In this section we describe the tensionless towers in the particular example of the $T^6/\IZ_2 \times \IZ_2^\prime$
orientifold introduced at the end of  section \ref{ss:2a}. 
The tensions of 4d domain walls for this toroidal orientifold are summarized in Table~\ref{tab:domainwallsstorus}.
We now discuss the subsets of them that become tensionless at different infinite distance points.

\begin{table}[tb]\begin{center}
\renewcommand{\arraystretch}{2.60}
\begin{tabular}{|c|c|c|}
\hline
Brane & Cycle & Tension (in units of $ M_P^3/\sqrt{4\pi}$) \\
\hline \hline
D$2$& -  &$  \dfrac{1}{(8 \, n^0 n^1 n^2 n^3 t^1 t^2 t^3)^{1/2}} $  \\
\hline
D$4$& $i$-th 2-torus & $\ds \dfrac{\left|T^i \right|}{\left(8\, n^0 n^1 n^2 n^3 t^1 t^2 t^3\right)^{1/2}}$  \\
\hline
D$6$&$i$-th and $j$-th 2-tori & $\ds \dfrac{ \left|  T^i  T^j \right| }{\left(8\, n^0 n^1 n^2 n^3 t^1 t^2 t^3\right)^{1/2}}$  \\
\hline
D$8$& All tori &  $\ds \dfrac{\left| T^1 T^2 T^3\right|}{\left(8\, n^0 n^1 n^2 n^3 t^1 t^2 t^3\right)^{1/2} }$ \\
\hline
NS5 & $y^1=y^2=y^3=0$  & $\ds \dfrac{ \left|N^0 \right|}{\left(8\, n^0 n^1 n^2 n^3 t^1 t^2 t^3\right)^{1/2} }$  \\
\hline
NS5 &  $y^i=x^j=x^k=0$ & $\ds \dfrac{ \left| N^i \right|}{\left(8\, n^0 n^1 n^2 n^3 t^1 t^2 t^3\right)^{1/2} }$  \\
\hline
\end{tabular}
\caption{Tensions of the basis of domain walls obtained from D$p$ or NS5-branes around the different $(p-2)$-cycles or 3-cyles of the $T^6/\IZ_2\times \IZ_2^\prime$,  respectively. Note that $i\neq j \neq k \neq i$}
  \label{tab:domainwallsstorus}\end{center}\end{table}  
  
\begin{itemize}
{\item[{\bf (TK.I)}]{\bf One K\"ahler modulus going to infinity: $t^1 \rightarrow \infty$ }.\\
All domain walls
obtained by wrapping D$2$-branes and NS$5$-branes become tensionless. The same happens with the two domain walls from  D$4$'s which wrap the 2nd or the 3rd 2-tori and with the D$6$ that wraps the $4$-cycle formed by the 2nd and 3rd tori. The other D$4$, the other two D$6$'s and the D$8$ do not become tensionless. Notice that this matches case (K.Ia) in the general analysis.
}
{\item[{\bf(TK.II)}]{\bf Two K\"ahler moduli going to infinity: $t^{1}\propto t^{2} \rightarrow \infty, \  (\ i\neq j \neq k \neq i)  $}.\\
The domain walls
obtained from D$2$-branes and NS$5$-branes become tensionless. In addition, the one constructed by wrapping a D$4$ along the 3rd 2-torus also becomes tensionless. The other two  from D$4$'s wrapping the 1st and 2nd 2-tori, the three  D$6$'s and the D$8$ have a non-vanishing tension. This matches case (K.IIa) for the general Calabi-Yau.  
}

{\item[{\bf (TK.III)}] {\bf All K\"ahler moduli going to infinity: $ t^1 \propto  t^2\propto t^3\rightarrow \infty$}\\
As in the previous cases, the domain walls that consist on D$2$'s or NS5's become tensionless and so do the three domain walls formed by wrapping a D$4$ along any of the three 2-tori. The three D$6$'s and the D$8$ do not. This matches case (K.III) for the general Calabi-Yau orientifold.
}

{\item[{\bf (TCS.I)}] {\bf One complex structure moduli going to infinity $n^{i} \rightarrow \infty$}.\\
All the domain  walls coming from a $\tD p$-brane on a $(p-2)$-cycle becomes tensionless as we go to infinity. From the NS5-branes, the ones that do not wrap the 3-cycle whose tension if proportional to  $|N^i|$  also become tensionless but the other one does not. This matches case (CS.Ia) for the general Calabi-Yau.} 

{\item[{\bf (TCS.II)}] {\bf Two complex structure moduli going to infinity $ n^{i}\propto n^{j} \rightarrow \infty$}.\\
All the domain  walls constructed from a D$p$-brane on a $(p-2)$-cycle are tensionless at the infinite distance point. Additionally, the NS5-branes wrapping a $3$-cycle with a tension not proportional to  $|N^i|$ or $|N^j|$  become tensionless but the other two do not. This situation also matches case (CS.IIa) for the general Calabi-Yau.} 

{\item[{\bf (TCS.III)}] {\bf Three or four complex structure moduli going to infinity $n^{i}\propto n^{j} \propto n^{k}  \rightarrow \infty \ {\rm or} \  n^0 \propto  n^1 \propto  n^2 \propto  n^3\rightarrow \infty$}.\\
All domain walls in table \ref{tab:domainwallsstorus} become tensionless as we approach the infinite distance point. 
Notice that this matches cases (CS.IIb) and (CS.III) for the general Calabi-Yau.
}
\end{itemize}

\subsection{Charges and the Weak Gravity Conjecture}
\label{ss:charges2a}

In this section we relate our earlier results to the WGC for extended objects. We will show how the states that conform the towers of domain walls that become tensionless at the infinite distance point also fulfill the WGC. To be precise, we use the form of the electric WGC given in \cite{WGC16}, which for domain walls in 4 dimensions translates into
\begin{equation}\label{WGC}
\left[\dfrac{\alpha^2}{2}-\dfrac{3}{2}\right]T^2\leq e^2 Q^2 M_P^{2}\, .
\end{equation}
Here $\alpha$ is the dilatonic coupling to the field strength, $e$ is the gauge coupling and $Q^2$ is the modulus of the charge vector in a framework with conventional normalizations\footnote{Concretely, for domain walls coupled to 3-forms 
the 4d kinetic term reads $\frac{1}{2e^2}\int e^{-\alpha \varphi} \widehat \cf_4\wedge {}^*\widehat \cf_4$, where 
$\cf_4=d \ca_3$. The coupling of the domain wall with worldvolume $W_3$ is $\int_{W_3} \ca_3$.}. 
As argued in \cite{imuv}, we can assume that this particular form of the WGC for domain walls, which actually arises from a naive generalization of the general formula in \cite{WGC16},  is well defined as long as the dilaton coupling $\alpha$ is large enough to ensure that the LHS is positive (i.e. $\alpha^2>3$, which holds in our case where $\alpha^2=7$ as we will see).

The relation between the towers predicted by the SDC and the WGC is an interesting problem in itself.
Both conjectures aim at making more precise and quantitative the hypothesis that there are no global symmetries in Quantum Gravity.
It is then reasonable to think that their towers may be related. 
From the SDC perspective, the obstruction to the presence of global symmetries can be understood from the fact that at the infinite distance points, where we would recover the global shift symmetries, the infinite tower of states becomes tensionless, invalidating the EFT. 
This can also be nicely connected with the WGC because the fact that towers fulfill it, ensures that at weak coupling points, which usually lie at infinite distance, the states in the tower become tensionless as required by eq.~\eqref{WGC} when $e\to 0$. 
In the following we will calculate the electric charges of the different elements of the basis of domain walls 
in the particular example of the $T^6/\IZ_2 \times \IZ_2^\prime$
orientifold,  and check that the WGC bound is saturated.

We wish to determine the electric charge of a domain wall built by a
$\tD p$-brane wound around a $(p-2)$-cycle. The coupling of the domain wall to a 3-form follows from the Chern-Simons
(CS) action \cite{BOOK}
\begin{equation}
S^{(p)}_{\rm{CS}} = \mu_p \int_{W_{p+1}} \hspace*{-5mm} P[\textstyle{\sum_q} C_q \wedge e^{-B}] \, .
\label{scsax}
\end{equation}
To simplify the analysis we will neglect axions from the $B$-field. 
The $\tD p$-brane worldvolume is taken to be the product of the
domain wall worldvolume $W_3$ and the internal cycle $\gamma_{p-2}$. Besides,  the RR potentials
are expanded as $C_{p+1} = c_3 \wedge \omega_{p-2}$,
where the $\omega_{p-2}$ are harmonic forms of $\cam$ .
Integrating over $\gamma_{p-2}$ we see that the CS action gives rise to a coupling $\int_{W_3} \ca_3$, with
$\ca_3 = 2\pi c_3/\ell_s^3$. The next step is to look at the 4d kinetic terms for $\cf_4 = d \ca_3$, which descend from the 
10d action by dimensional reduction. Luckily, this calculation has been done in \cite{Herraez:2018vae} as we now review.

In the notation of \cite{Herraez:2018vae} the RR $(p+1)$-forms are expanded as
\begin{equation}
C_3 = c_{3}^0\, ,  \qquad C_5 = c_{3}^{a} \wedge \omega_a\, , \qquad
C_7 = \tilde{d}_{3a} \wedge \tilde \omega^a\, , \qquad  C_9= \tilde{d}_{3} \wedge \omega_6 \, . 
\label{all3}
\end{equation}
Thus, for D2, D4, D6 and D8-branes  the relevant 4-forms are 
$F_4^0$, $F_4^a$, $\tilde F_{4a}$ and $\tilde F_4$, given by the exterior derivatives
of the 3-forms in \eqref{all3}\footnote{We are setting the axions to their background values. In this case the 4-forms are exact, i.e. the field strengths of the corresponding 3-forms. Otherwise we would need to first rotate to the so-called A-basis \cite{Herraez:2018vae}.}. 
To go to 4-forms with the normalization of \cite{WGC16} we take $\cf_4 = 2\pi F_4/\ell_s^3$. 
The resulting 4d kinetic terms for a general Calabi-Yau orientifold turn out to be
\begin{equation}
S_{\text{kin}}
=  \frac{\pi}{2 M_P^4}\int \frac{e^{-K}}{8} \left[ \cf_4^0 \wedge\!{}^*\cf_4^0  
+ 4 g_{ab} \cf_4^a \wedge\!{}^*\cf_4^b 
+ \frac{1}{4\cv^2} g^{ab} \tilde \cf_{4a} \wedge\!{}^*\tilde \cf_{4b}
+ \frac{1}{\cv^2} \tilde \cf_{4} \wedge\!{}^*\tilde \cf_{4}
\right] \, ,
\label{kinF1}
\end{equation}
where $g_{ab}$ is the metric in K\"ahler moduli space. To arrive at this result we have used \eqref{stringmass}
to trade $M_s$ for $M_P$, after going to Einstein frame with the transformation 
$g_4 = \big(\frac{e^{2\phi}}{\cv} \frac{\cv_0}{e^{2\phi_0}}\big) g_{4E}$, where subscript $0$ stands for vev.

Let us now consider the $T^6/\IZ_2 \times \IZ_2^\prime$ model, which is particularly simple because the metric $g_{ab}$ is diagonal.
We readily find
\begin{equation}
S_{\text{kin}} =  \frac{\pi}{2 M_P^4}\int \frac{e^{-K}}{8} 
\left[ 
\cf_{4}^0 \wedge\!{}^* \cf_{4}^0 \op + \!
\sum_{i=1}^3  \!\! \left[ \frac{1}{(t^i)^2} \cf_{4}^i \wedge\!{}^* \cf_{4}^i 
 + \frac{(t^i)^2}{\cv^2} \tilde \cf_{4i} \wedge\!{}^*\tilde \cf_{4i} \right]
\! + \frac{1}{\cv^2} \tilde \cf_{4} \wedge\!{}^*\tilde \cf_{4} 
 \right] \, ,
\label{kinF2}
\end{equation} 
with $\cv=t^1 t^2 t^3$ and $e^{-K} = 8 n^0 n^1 n^2 n^3 t^1 t^2 t^3$.
Additionally, with this K\"ahler potential, the conventionally normalized saxions are
$\tilde t^i = \log t^i $ and $\tilde n^I = \log n^I$. Thus, all kinetic terms are of  the form 
$e^{\tilde n^0 + \tilde n^1+ \tilde n^2 + \tilde n^3 \pm  \tilde t^1 \pm  \tilde t^1 \pm \tilde t^3} \cf_{4} \wedge\!{}^* \cf_{4}$.
This shows that they are all of type
$e^{-\alpha \varphi} \cf_{4} \wedge\!{}^* \cf_{4}$, with $\alpha^2=7$. 
The charges of the different domain walls can be read off from the above kinetic terms. For instance,
for the domain wall from the D2-brane, $e^2=\frac{8}{\pi} M_P^4 e^K$. On the other hand, the squared
tension from \eqref{tensionDp} is $T^2=\frac{4}{\pi} M_P^6\op  e^K$. Hence, the WGC bound is saturated.
It can be verified that this is also true for domain walls from D4, D6, and D8-branes.

\subsection{Towers in $\mathcal{N}=2$}
\label{ss:towersn2} 

In previous sections we have considered the $\mathcal{N}=1$ theories which arise from compactifying type IIA on  Calabi-Yau orientifolds. We restricted ourselves to orientifolds with $h^{1,1}_+=0$, which implies that 4d particles coming from D$p$-branes wrapping $p$-cycles are projected out, since the gauge fields to which they couple are also projected out. If these condition were relaxed,  4d particles would arise from D$p$-branes along these new cycles and they would, in principle, form towers at the infinite distance points. In the case of 4d strings, the ones arising from D$4$-branes wrapping 3-cycles dual to the $\alpha_K$ 3-forms are also projected out by the orientifold, whereas those from D4's on the 3-cycles dual to the $\beta^K$'s and from NS5's on even 4-cycles are not. 
In this section we relax the orientifold projection and consider type IIA compactification on a Calabi-Yau manifold leading
to ${\mathcal N}=2$ supersymmetry in 4d. In this way, we can study not only the towers of particles and strings that could be present in the orientifold (e.g. if we allowed $h^{1,1}_+\neq 0$, for the case of particles) but also the ones that were projected out in that case. 
Without the orientifold projection, none of the 4d particles or strings that come from D$p$-branes or NS5's are  eliminated.
All these can then form towers of particles and strings that become exponentially massless or tensionless as we travel to points at infinite distance in moduli space.
We first review the known infinite towers of particles, explored in detail in \cite{Corvilain}, which 
are dual to the infinite towers of particles obtained by wrapping D3-branes along 3-cyles in type IIB compactifications
\cite{irene}. We then discuss strings. 
Additionally,  we particularise the general results to the $T^6/\IZ_2 \times \IZ_2^\prime$ toroidal model
in order to develop a more intuitive understanding of the energy scales involved and compare those of particles and strings with the ones associated to domain walls.

\subsubsection{Towers of particles}

The basis of particles from which the whole infinite towers can later be constructed consists of single D0, D2, D4 and D6-branes wrapped on the corresponding even cycles of the internal space. To calculate the masses of the 4d particles we make use of the 
DBI action \eqref{sdbiax}. 
Now we take $W_{p+1}$ to be the product of an internal cycle $\gamma_p$ and the particle worldline. Then, integrating over the
internal cycle leads to
\begin{equation}
\label{sdbipart}
S_{{\rm DBI}}=-\dfrac{2\pi \cv_p}{g_s \ell_s} \, \int d\xi \sqrt{-g^{(1)}} \, ,
\end{equation}
where $\cv_p$ is the volume of $\gamma_p$.

From the action we deduce that the mass of the 4d particles is given in general by
\begin{equation}
\label{mparticles}
M_{p}(\gamma_p)= \dfrac{2 \pi \cv_p}{g_s} M_s = \sqrt{\pi} M_P \dfrac{\mathcal{V}_p}{\sqrt{\mathcal{V}}} \, ,
\end{equation}
where in the second step we have used $M_s^2=\frac{g_s^2 M_P^2}{4\pi (\cv)}$, which  differs from the $\mathcal{N}=1$ case by the factor of two appearing in the compactification volume due to the orientifold action. For $p=0$ and $p=6$ we just have $\cv_0=1$ and $\cv_6=\cv$, since D0 and D6-branes wrap 
respectively a point and the whole manifold. 
For $\gamma_2$ and $\gamma_4$ we would like to take supersymmetric
(holomorphic) cycles, but in a general Calabi-Yau they are not known explicitly. 
However, as mentioned before,
we can still calculate their volumes by exploiting the fact that the K\"ahler form $J$ is a calibration
so that the volumes of the supersymmetric cycles are given by integrating powers of $J$
along any cycle in the same homology class. 
In particular, for the even cycles we consider Poincar\'e duals of the even harmonic forms and compute 
the volumes according to eq.~\eqref{vols24}. Again, the $B$-field is taken into account by replacing
$J \rightarrow J+ i B$ and taking the modulus at the end.
The resulting particle masses are summarized in Table \ref{tab:massparticles}, using $K_K=-\log(8\cv)$.

\begin{table}[h!!]\begin{center}
\renewcommand{\arraystretch}{2.00}
\begin{tabular}{|c|c|c|}
\hline
Brane & Cycle & Mass (in Planck units) \\
\hline \hline
D$0$& -  &$ \sqrt{8\pi} \, e^{K_K/2} $  \\
\hline
D$2$&  $[\tilde{\pi}^a] \in H_2(\mathcal{M,\IZ}) $ &$ \sqrt{8\pi} \, e^{K_K/2} \, \left| T^a \right|$   \\
\hline
D$4$& $[\pi_a ] \in H_4(\mathcal{M,\IZ})$ & $\sqrt{8\pi} \, e^{K_K/2} \, \left|\frac12  \sum_{b,c}\,  \kappa_{abc}\, \, T^b \,  T^c\right| $ \\
\hline
D$6$& $\mathcal{M}$ &  $\sqrt{8\pi} \, e^{K_K/2} \, \left|\frac16  \sum_{a, b,c}\,    \kappa_{abc}\, T^a \, T^b \,  T^c\right| $ \\
\hline
\end{tabular}
\caption{Masses of the different particles obtained by wrapping one kind of $\text{D}p$-brane around a given even cycle on the Calabi-Yau threefold.}
\label{tab:massparticles}\end{center}\end{table}  

We now want to see which particles become massless as we move towards infinite distance along any direction in moduli space. 
If we can then form an infinite tower of particles by bound states of them, the whole tower would become massless and it could be a candidate for the tower predicted by the SDC. 
To begin with, if we send one $t^a \to \infty$ there is one  2-cycle whose volume goes to infinity implying that the whole Calabi-Yau volume diverges, too. 
Thus, the particles coming from branes not wrapping that 2-cycle will become massless. Moreover, if all the K\"ahler moduli are taken to be proportional to each other and sent to infinity all the particles associated to D0 and D2-branes will become massless. In fact, the particles coming from
D0-branes always become massless as we go to infinite distance in K\"ahler moduli space whereas those formed by D6-branes wrapping the whole Calabi-Yau never do. For a more systematic analysis of the subset of elements of the basis that become massless at different infinite distance points we refer to the end of section \ref{sec:tensionlesswalls}, where this is performed for domain walls but can straightforwardly be adapted to particles. However, let us remark that there is no particle coming from a D$p$-brane that becomes massless at any large complex structure limit, since their masses (in Planck units) do not depend on the complex structure moduli, as opposed to the tensions of domain walls. 

In order to get more intuition, we can consider the toroidal orbifold introduced in section \ref{ss:2a} but without 
imposing the orientifold projection.
In this case the masses of the basis of 4d particles turn out to be:
\begin{equation}
\label{mparticlestorus}
\left(M_{0}, \ M_{2_i}, \ M_{4_i} , \ M_6 \right) = 
\sqrt{\pi} M_P \left( \dfrac{1}{\sqrt{ t^1 t^2 t^3}}, \ \sqrt{\dfrac{t^i}{ t^j t^k}}, \  
 \sqrt{\dfrac{t^j t^k}{ t^i}}, \ \sqrt{t^1 t^2 t^3}\right) \, ,
\end{equation}
with $i\neq j \neq k \neq i$ . The subindex in the masses refers to the type of $\text{D}p$-brane from which the particle arises.
Besides, the $i$ in $M_{2_i}$ and $M_{4_i}$ indicates that the 2-cycle wrapped by the $\text{D}2$-brane is the $i$th 2-torus, 
and that the 4-cycle wrapped by the D4-brane  is the product of the $j$th and $k$th 
2-tori. Note that in the toroidal setup these cycles are all holomorphic, thus supersymmetric.
The results show that when the volume of one of  the 2-tori goes to infinity (i.e. $t^i \rightarrow \infty$), the particles
that become massless are those coming from the D0, the D2's which do not wrap that 2-torus and the D4 that wraps the other two 2-tori. 
The same game can be played if we make any pair of $t^i$ go to infinity, and also if we make all of them diverge. We revisit this in more detail in section \ref{sec:toroidal} but the main point is that in all the aforementioned cases there are particles that become massless.

Once we know that there is (at least) one particle of this kind that becomes massless as we move towards infinite distance along any direction in K\"ahler moduli space, two more things are needed in order for the SDC to be fulfilled. The first one is to build  the infinite tower of particles whose mass is proportional to that in eq. \eqref{mparticles}, so that the whole tower becomes massless if one of the particles does. The second is to show that the mass of those particles goes to zero exponentially in the proper field distance. 

Regarding the infinite towers of particles, they can be generated by the induced action
of the monodromy transformations on the charge vector of the particles. This was performed in \cite{Corvilain} and we will not repeat it here, but it can be straightforwardly adapted from the corresponding discussion for domain walls in section \ref{sec:tensionlesswalls}

Finally, let us comment on the exponential behavior of the masses as we approach infinite proper distances.
Consider for instance moving towards infinity in moduli space along the direction of $T^i$. Since 
$K_{T^i \bar{T}^i}\, \sim \, 1/(t^i)^{2}$, the proper distance between two points $P$ and $Q$ is given by 
\begin{equation}
d(P,Q)=\int^Q_P \, \sqrt{2 K_{T^i \bar{T}^i}}\, dt^i \, \sim \, \left.\mathrm{log}(t^i)\right|^Q_P 
\end{equation}
Thus, the power dependence in $t^i$ that arises in \eqref{mparticles} and \eqref{mparticlestorus} translates 
into an exponential dependence in $d(P,Q)$. This can be generalized beyond the toroidal orbifold as explained in
section \ref{sec:distances}.

\subsubsection{Towers of strings and domain walls}

Let us now go back to the general Calabi-Yau case. It is also natural to consider towers of tensionless strings, which 
can result from D4-branes or NS5-branes wrapping 3-cycles or
4-cycles in the internal Calabi-Yau manifold. For the D4-branes, the tension can be obtained from the DBI action \eqref{sdbiax},
taking $W_5$ to be the product of an internal cycle $\gamma_3$ and the string worldvolume.
Integrating over $\gamma_3$ we read the tension
\begin{equation}
\label{TsD4}
T_{\tD 4}(\gamma_3)\, = \dfrac{2\pi \cv_3}{g_s \ell_s^2} = \dfrac{M_P^2}{2} \,  \dfrac{e^{\phi_4}}{\mathcal{V}^{1/2}}\,\mathcal{V}_3 \, ,
\end{equation} 
where we used the definition of the 4d dilaton and expressed the string mass in Planck units. As before, the volume of the supersymmetric 3-cycle is computed integrating 
the normalized calibrating form $\text{Re} \left(e^{-\mathcal U} \Omega\right)$ around a 3-cycle
dual to the 3-forms $\alpha_K$ or $\beta^K$. Hence, $\cv_3$ is given by\footnote{Notice this is the same as in \eqref{vol32a}, but here we express it in more appropriate variables for the $\mathcal{N}=2$ case}
\begin{equation}
\cv_3=\dfrac{e^{K_{CS}/2} \, e^{-K_K/2}}{2^{3/2}}\int_{\gamma_3}  \rm{Re}(\Omega).
\end{equation}
where the integral selects the real part of one the periods of the holomorphic 3-form, $\left( X^K, \mathcal{F}_K \right)$, defined in \eqref{omexp}. 
These periods depend on the $h_{2,1}$ complex structure moduli, which can be identified with the special coordinates $X^K/X^0$, whereas the $\cf_K$ can be obtained as derivatives with respect to $X^K$ of a prepotential. 
Replacing $\preal(\Omega)$ by $|\Omega|$, in order to account for different phases for the calibrations,
yields
\begin{equation}
\left(\cv_3^{A^K}, \, \cv_3^{B_K}\right) =\dfrac{M_P^2}{2}e^{\phi_4} e^{K_{CS}/2} \left( \, |X^K| ,\,   |\mathcal{F}_K|  \,  \right)\, ,
\end{equation}
and the full tension displayed in Table~\ref{tab:tensionsstringsCY}.

The tension of the string obtained from the NS5-brane on a 4-cycle $\gamma_4$
can be derived from the DBI action including an extra factor of $e^{-\phi}$. Upon integrating over $\gamma_4$ we find
the tension
\begin{equation}
\label{TsNS5}
T_{\text{NS}5}(\gamma_4) = \dfrac{2\pi \cv_4}{g_s^2 \ell_s^2}= 8 M_P^2\,  e^{K_K}\,\mathcal{V}_4 .
\end{equation}
The 4-cycle is taken to be the Poincar\'e dual of the harmonic 2-form $\omega_a$. Computing the volume according to
\eqref{vols24} yields the tension shown in Table~\ref{tab:tensionsstringsCY}. 

\begin{table}[t]\begin{center}
\renewcommand{\arraystretch}{2.00}
\begin{tabular}{|c|c|c|}
\hline
Brane & Cycle & Tension (in units of $M_P^2$) \\
\hline \hline
D$4$&  P.D. $[ \beta^K]$  &$  \dfrac{1}{2}e^{K_{CS}/2}\, e^{\phi_4} \,\left| X^K \right|$  \\
\hline
D$4$&  P.D. $[ \alpha_K]$  &$  \dfrac{1}{2}e^{K_{CS}/2}\, e^{\phi_4} \,\left| \cf^K \right|$  \\
\hline
NS5 & P.D. $[\omega_a]$ & $ 8 \,e^{K_K} \, \left|\frac12  \sum_{b,c}\,  \kappa_{abc}\, \, T^b \,  T^c\right|$ \\
\hline
\end{tabular}
\caption{Tensions of 4d strings formed by D4 and NS5-branes wrapping the indicated internal cycles.}
\label{tab:tensionsstringsCY} 
\end{center}\end{table}  

It is interesting to notice that D4-branes give 4d strings whose tensions are controlled by the 4-dimensional dilaton and
the complex structure moduli, whereas strings from NS5-branes have tensions depending exclusively on the K\"ahler moduli. 
Performing a case by case analysis shows 
that there is always some subset of the basis of NS5's that becomes tensionless when approaching an infinite distance point in K\"ahler moduli space. Note that this still holdes for the orientifold case with $\mathcal{N}=1$, since these objects are not projected out by the orientifold action. For the infinite distance points within the CS moduli space we present concrete results on the toroidal orientifold. In order to build the towers of strings, the same logic as for the towers of particles and domain walls would apply, considering the towers formed by bound states of the subset of strings that become tensionless at the infinite distance point. 

In order to get some intuition, we focus again on the toroidal orbifold and restrict ourselves to the rectangular lattices for the three 2-tori, since these are the ones that are compatible with the orientifold projection. This will allow to straightforwardly translate our results to the $\mathcal{N}=1$ case. Instead of expressing the results in terms of the complex structure moduli that arise from identification with the special coordinates from the periods of the holomorphic 3-form, we write them  in terms of the moduli defined in \eqref{nkz2z2} so that these results can be straightforwardly applied to the orientifold setup. Let us stress, however, that in order to include more general lattices for the 2-tori one can just substitute into the general formulas of Table~\ref{tab:tensionsstringsCY}. The tensions of the different strings under consideration in the $T^6/\IZ_2 \times \IZ_2^\prime$ take the form
\begin{equation}
\label{tstringstorus}
\left(T_{\tD 4}^{A^K}, \ T_{\tD 4}^{B_K}, \ T_{\text{NS}5}^{i} \right) = 
 M_P^2 \left( \sqrt{\dfrac{n^K}{2 n^I n^J n^L}}, \ \dfrac{1}{ 4 n^K}, \  
 \ \dfrac{2}{t^i}\, \right) \, ,
\end{equation}
with all $I,\ J,\ K,\ L$ different. If we take the orientifold projection, the last two entries correspond to the 4d strings that survive, whereas the first one is projected out.

Regarding domain walls, the towers that we have constructed for the orientifold case are all present in the parent Calabi-Yau with $\mathcal{N}=2$. In fact, since we were restricting ourselves to Calabi-Yau threefolds with $h^{1,1}_+=0$, the unorientifolded case has exactly the same towers of domain walls. If we allowed $h^{1,1}_+\neq 0$ we would only obtain more domain walls arising from the D$p$-branes wrapping the extra even cycles which are Poincar\'e dual to the $h^{1,1}_+$ 2-forms and 4-forms. The general result is, in fact, that from all the towers of extended objects that appear in the parent $\mathcal{N}=2$ case only the subset of them that couple to forms that are not projected out by the orientifold survive after the projection to $\mathcal{N}=1$.

\subsubsection{Energy scales in a IIA toroidal orbifold example}
\label{sss:scales2a} 

The energy scales associated to the strings and domain walls can be obtained (applying naive dimensional analysis) by taking the square or cube root of the tensions, respectively.  
It instructive to compare these energy scales with those of the particles and study them for the objects that become masless/tensionless at different infinite distance points. For that purpose, we consider a isotropic version of the toroidal orbifold, that is, we set
$T^1=T^2=T^3\equiv T=t+ib$, $N^1=N^2 =N^3 \equiv U$, where $U$ are the complex structure deformations of the toroidal orbifold and half of their degrees of freedom are killed by the orientifold projection, i.e. the rectangular lattice is selected. Finally, we set $n^0=s$. As emphasized at the beginning  of this section, we write everything in Planck units and include the string, winding and KK scales in the comparison. First, we consider trajectories towards the infinite distance points  which remain inside the perturbative region of the moduli space, that is, paths that fulfill $e^\phi<1$. From the definition of the 4d dilaton $\phi_4$ and eq. \eqref{nkz2z2} we obtain, for this isotropic toroidal orientifold 
 \begin{equation}
 e^\phi=\dfrac{t^{3/2}}{s}.
\end{equation}
\begin{figure}[tb]

     \begin{center}
     	\subfigure[]{	
        \includegraphics[height=280pt]{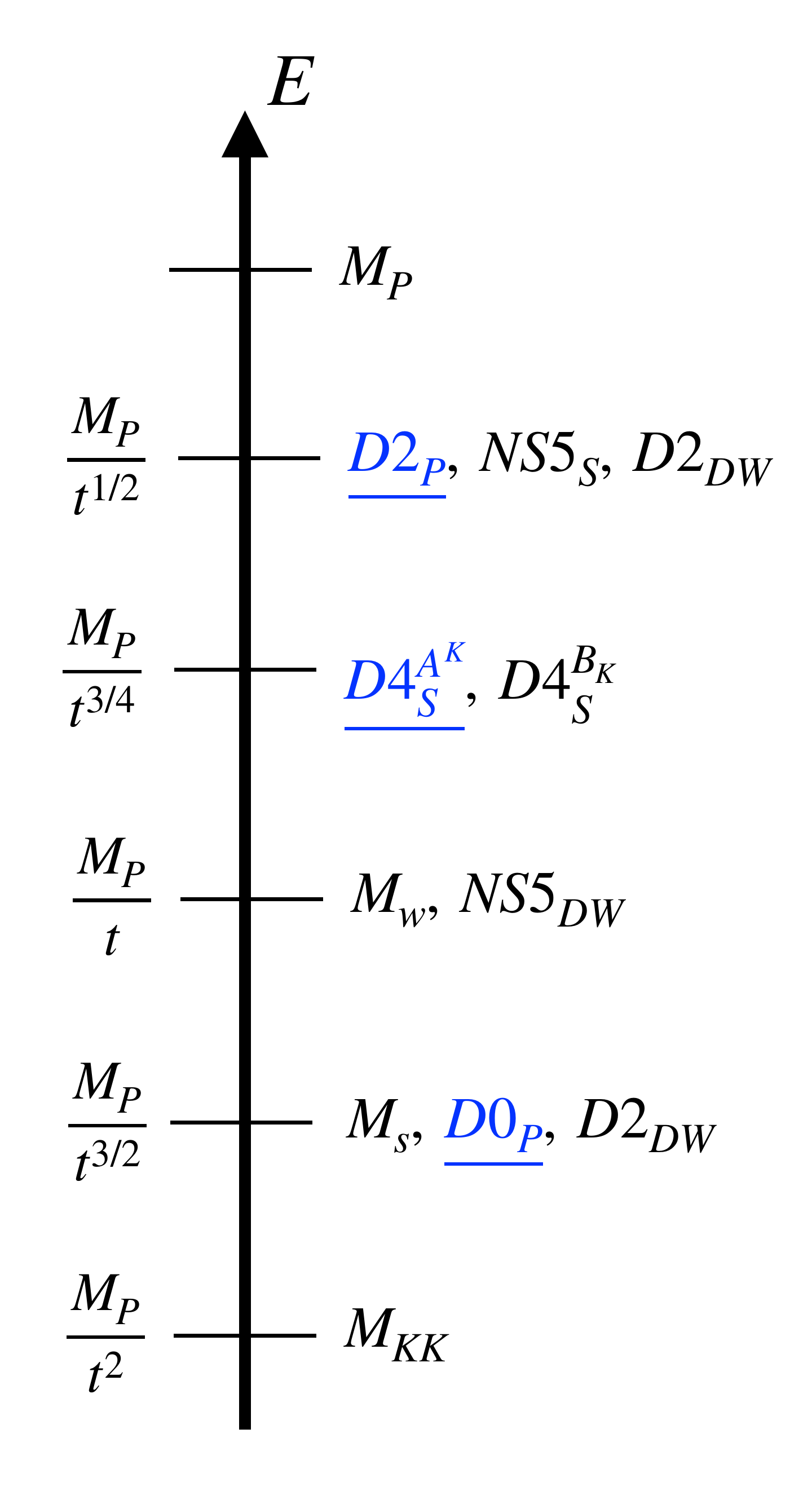}
        \label{fig:scalestu32}
        }
        \subfigure[]{
        \includegraphics[height=280pt]{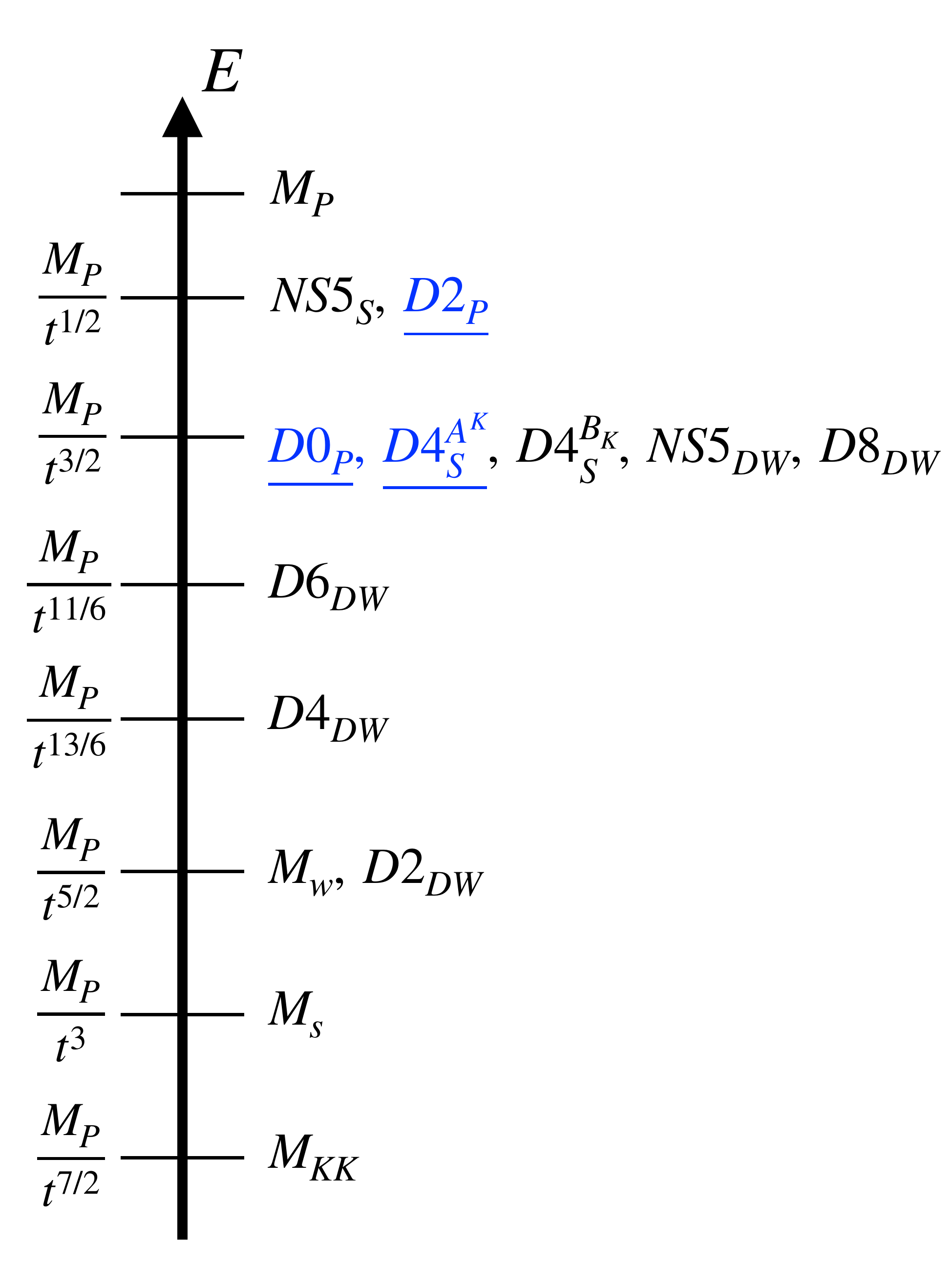}
        \label{fig:scalestu3}
        }
        \caption{Energy scales associated to the 4d particles, strings and domain walls that become massless/tensionless and remain within the perturbative regime as we approach  the infinite distance points given by {\bf (a) } $ s\propto u \propto t^{3/2} \rightarrow \infty$  and {\bf (b) } $ s\propto u \propto t^{3} \rightarrow \infty$. The string, KK and winding scales are also included. The subindices $P$, $S$ and $DW$ indicate whether the object is a particle, a string or a domain wall and the ones in blue and underlined are projected out by the orientifold action.}
        \label{fig:scalestu}
     \end{center}
\end{figure}

\begin{figure}[tb]
     \begin{center}
      \subfigure[]{
        \includegraphics[height=230pt]{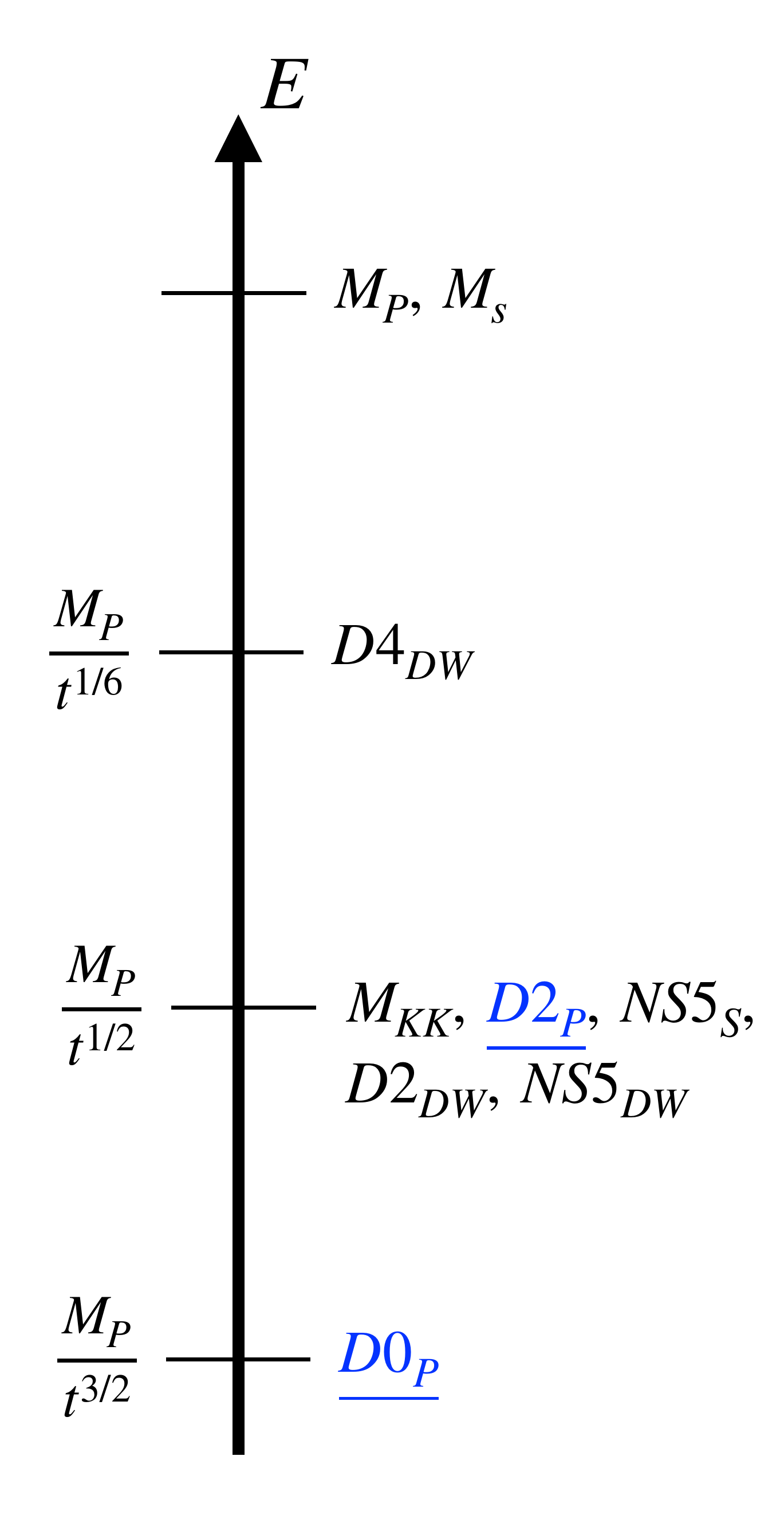}
        \label{fig:scalest}
        }
		\subfigure[]{
        \includegraphics[height=230pt]{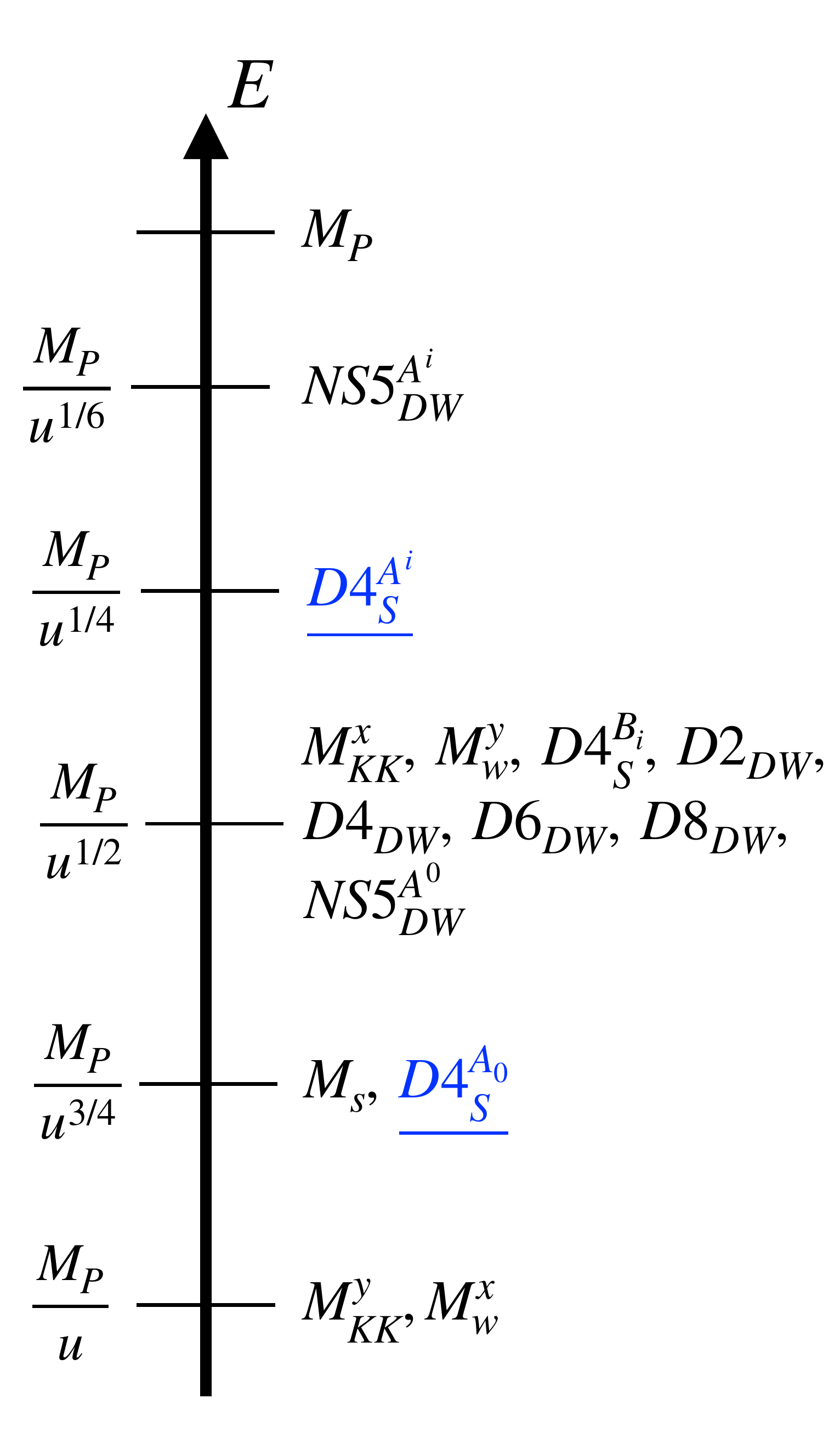}
        \label{fig:scalesu}
        }        
     	\subfigure[]{	
        \includegraphics[height=230pt]{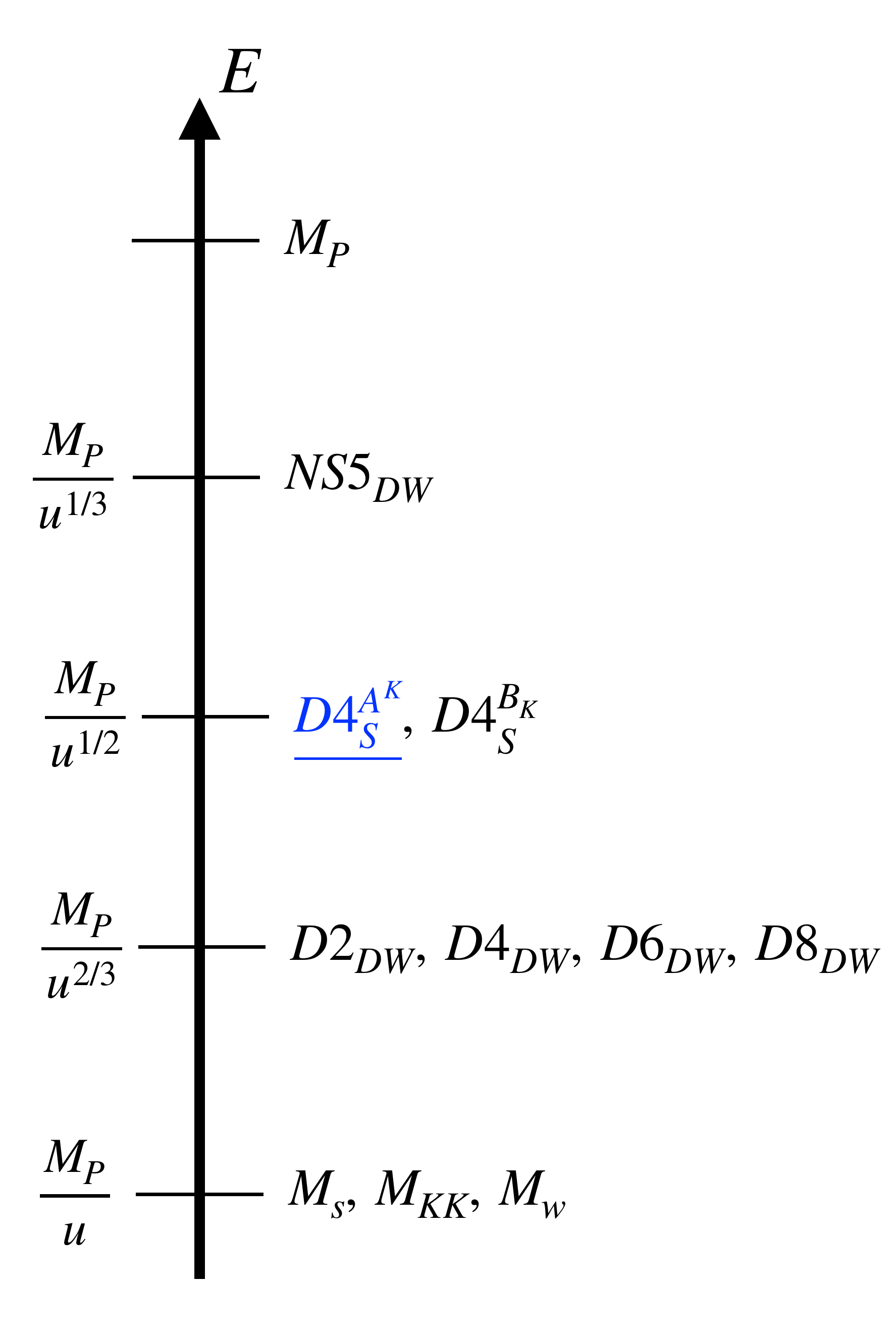}
        \label{fig:scalessu}
        }
       
         \caption{Energy scales associated to the 4d particles, strings and domain walls that become massless/tensionless as we approach  the infinite distance points given by {\bf (a) } $ t \rightarrow \infty, \ s, \ u$ fixed, {\bf (b) } $ u  \rightarrow \infty,\  s \ \mathrm{ and }\ t $ fixed  and {\bf (c) } $ s \propto u \rightarrow \infty, \ t $ fixed. The string, KK and winding scales are also included. The subindices $P$, $S$ and $DW$ indicate whether the object is a particle, a string or a domain wall and the ones in blue and underlined are projected out by the orientifold action.}
        \label{fig:scalesuort}
     \end{center}
\end{figure}

We begin by considering the infinite volume point, characterized by $t\to \infty$. Paths towards this point can only stay within the perturbative region  if $s \sim t^r$ with $r\geq 3/2$\footnote{$r$ is related with the $q$ used in the discussion below eq. \eqref{dilaton} as $r=\frac{3}{4}q$}. The energy scales associated to the objects that become massless/tensionless as we approach this point along two different trajectories, characterized by two different values of $r$, are shown in Fig. \ref{fig:scalestu}. Additionally, the infinite volume point can be approached through a path which is outside the perturbative region and along which  $u$ is kept fixed. The corresponding energy scales, together with the ones associated to the infinite distance point at which $s\propto u\to \infty$ with $t$ fixed are shown in Figure \ref{fig:scalesuort}. The reason why we include the path towards $t\to \infty$ which is outside the perturbative region is that it can be nicely interpreted from the M-theory point of view. The states with different D0 charges can be understood as corresponding to different KK modes along the extra circle of M-theory. In fact, the tower of BPS particles that are formed by bound states of D0-D2 branes in type IIA could be interpreted as the KK tower of the particles coming from M2 branes wrapping the same two cycles as the D2 branes (see e.g. \cite{Corvilain}) and this can be extrapolated to other domain walls with different D0 charge. 

\begin{figure}[tb]
 
    \begin{center}
        \includegraphics[width=450pt]{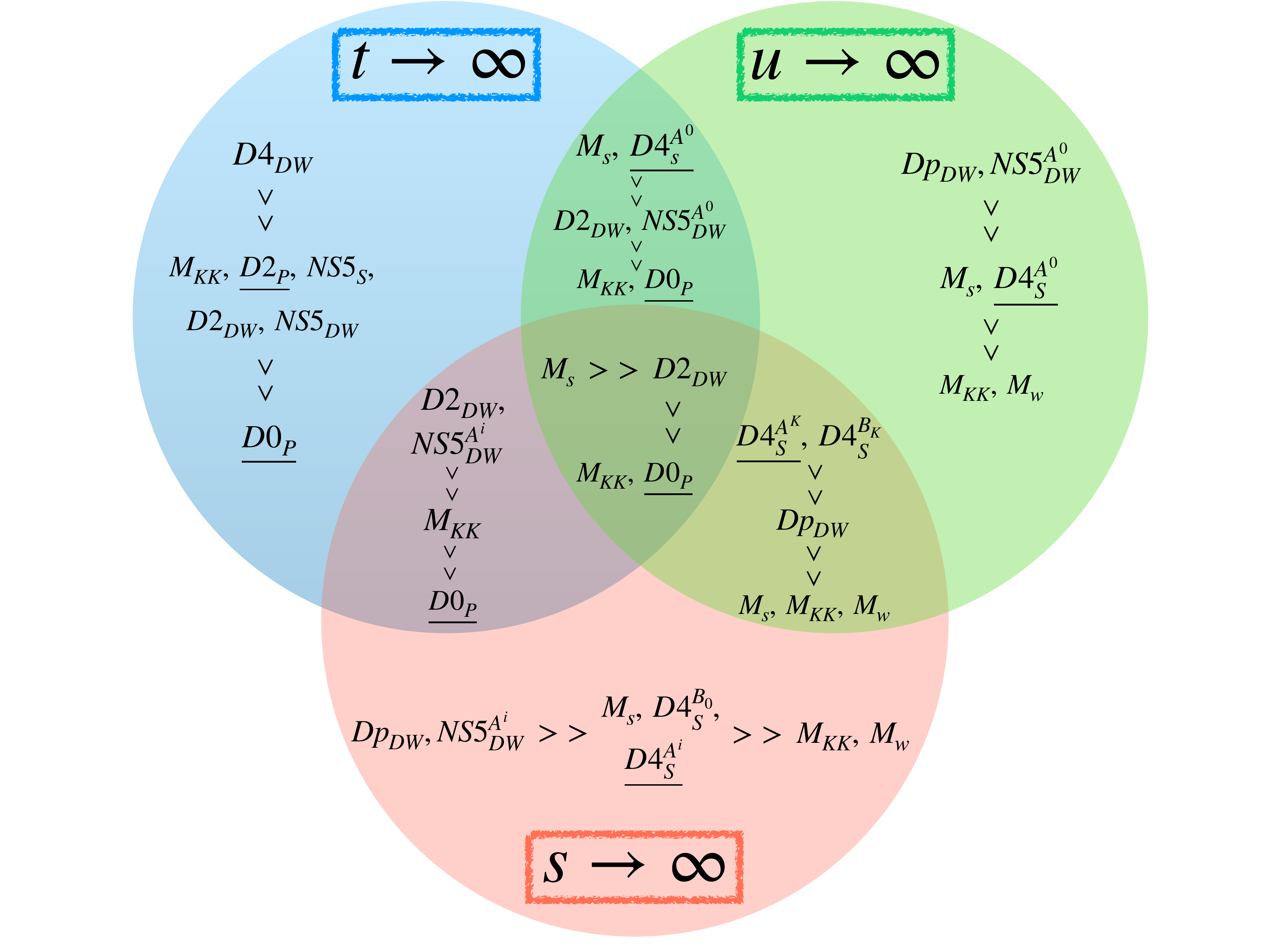}
        
     \caption{Spectra of towers of  lightest particles and branes for different infinite limits in moduli space, for the $T^6/\IZ_2 \times \IZ_2^\prime$. The subindices $P$, $S$ and $DW$ indicate whether the object is a particle, a string or a domain wall and the underlined ones are projected out by the orientifold action.} 
      \label{fig:circulos}
          \end{center}
        
\end{figure}

Figures \ref{fig:scalestu} and \ref{fig:scalesuort} show that it is a common feature to find the energy scales of strings and/or domain walls below those of particles, so that these new towers of extended objects could have an impact on the cutoff scale of the EFT even before the infinite tower of particles appears. Moreover, in the first two cases and also in the last one, the string and KK scales are below all the other mass scales (in the third the particle from the D0-brane is below the KK scale).
In Fig. \ref{fig:circulos} we show the lightest spectra of particles and branes for different infinite limits in (universal) K\"ahler, complex structure and complex dilaton
in the type IIA case. One can observe that the posibilities are varied but in general towers of particles come along with tensionless branes. 

Finally, it is remarkable that in the orientifold case the towers of particles and several towers of strings are projected out by the orientifold action. This is always the case when the 1-forms or 2-forms that couple to the particle or string in question are projected out by the orientifold, as in the toroidal example. It is, moreover, consistent with the relation between the towers coming from the WGC and those from the SDC and the idea that these towers prevent the appearance of a global symmetry. This is due to the fact that when the corresponding $q$-form fields are projected out there is no gauge symmetry giving rise to a global one at the infinite distance point, hence not requiring the presence of the corresponding tower.

\section{Towers of tensionless branes in type IIB orientifolds}
\label{s:iib}

In this section we extend the preceding study of type IIA tensionless branes to the IIB context.
The basics of type IIB Calabi-Yau orientifolds are reviewed in Appendix \ref{ap:2b}.
In the following we will first examine the towers of tensionless domain walls,
formed by $\text{D}5$-branes, or NS5-branes, wrapping 3-cycles in the internal Calabi-Yau manifold.
Afterwards we will relax the orientifold projection in order to discuss towers of particles and strings.

The tensions of domain walls described by D5-branes wound on 3-cycles can be computed from dimensional reduction 
of the DBI action given in eq.~\eqref{sdbiax}. 
After integrating over the internal 3-cycle $\gamma_3$ we find
\beq
T_{\text{D}5}(\gamma_3) = \frac{2 \pi M_s^3}{g_s} V_{3}(\gamma_3) \, ,
\label{td52b}
\eeq
where $V_{3}(\gamma_3)$ is the volume of $\gamma_3$ in string units. To derive the behavior 
compared to the fixed Planck mass $M_P$ we simply substitute $M_s$ using the relation \eqref{msp2}.

The volume of a supersymmetric internal 3-cycle can be computed via
\beq
V_3(\gamma_3) = \left|\int_{\gamma_3}\hspace*{-3mm} N\op \Omega \right| \, ,
\label{vol32b}
\eeq
with $N$ a normalization such that $\frac{i}8 N^2 \int_\cam \Omega \wedge \bar\Omega = \frac16 \int_\cam J^3=V_6$.
It follows that $N^2=8\, e^{K_{CS}} g_s ^{3/2} \cv$. Inserting in \eqref{td52b} yields
\beq
T_{\text{D}5}(\gamma_3) = \frac{2 M_P^3}{ \sqrt{\pi}}\, e^{K/2} \left|\int_{\gamma_3}\hspace*{-2mm} \Omega \right| \, ,
\label{t32b2}
\eeq
where we used \eqref{kpot2b} for the full K\"ahler potential.
For NS5-branes the dilaton dependence $e^{-\phi}$ in the DBI action is replaced by $e^{-2\phi}$. Therefore,
$T_{\text{NS}5}(\gamma_3) =T_{\text{D}5}(\gamma_3) /g_s$.
Taking the 3-cycles in the symplectic basis
$\{A^\lambda, B_\lambda\}$, dual to  the 3-forms $\{\beta^\lambda\, , \alpha_\lambda\}$, and evaluating
the integrals of $\Omega$ using the expansion \eqref{om2bs} 
gives the tensions shown in Table \ref{table_dws}. Note that they are consistent with the results of Table \ref{tab:tensionsCY} for the type IIA mirror.

\begin{table}[h!]\begin{center}
\renewcommand{\arraystretch}{1.25}
\begin{tabular}{|M{2cm}||c|c|c|c|}
\hline
 \small{Cycle} &  $A^0$ & $A^i $  & $B_i$ & $B_0$ \\
 \hline
\small{Tension (in units of $2 M_P^3/\sqrt{\pi}$)} & $e^{K/2}$  & $ e^{K/2}\, \left|U^i\right|$ &  
$e^{K/2} \left|\frac12 d_{ijk} U^j U^k\right|$  & $ e^{K/2} \left|\frac16 d_{ijk} U^i U^j U^k\right| $  
\\[2mm]
\hline
\end{tabular}
\caption{DBI tensions of domain walls formed by D5-branes wrapping 3-cycles.}
 \label{table_dws}
\end{center}
\end{table}

The tension of supersymmetric domain walls can also be computed using the BPS formula \eqref{bps}.
For instance, consider the 3-cycle $B_0$ and the corresponding flux $e_0$ given by  
$e_0=\int_{B_0} \bar F_3$.
The value of $e_0$ jumps across a domain wall obtained by wrapping a D5-brane along a 3-cycle $A^0$ 
such that $[A^0]\cdot [B_0]=1$. To be more precise, $\int_{A^0} \alpha_0 =1$. 
Inserting the appropriate factor to restore the mass dimensions in $W$, the tension of this domain wall is then
\beq
T_{\text{D}5}(A^0) = \frac{M_P^3}{\sqrt{4\pi}}  4\,  e^{K/2} \, , 
\label{tD5A0}
\eeq
where we have considered that the flux jumps by two units. This factor of 2 was explained around eq.~\eqref{matchbps}.
For a NS5-brane wrapping the same cycle the value of $h_0$ jumps and the domain wall tension becomes
\beq
T_{\text{NS}5}(A^0) = \frac{M_P^3}{\sqrt{4\pi}} 4\, e^{K/2} \text{Re}\, S= \frac1{g_s} T_{\text{D}5}(A^0) \, .
\label{tNS5A0}
\eeq
Here we have set the axion $\text{Im} S$ to zero. 
Notice that we recover the tensions derived in \eqref{t32b2} in the case $\gamma_3 =A^0$, with $\int_{A^0} \Omega=1$.

Let us now analyze the behavior of the tensions in the limit of large complex structure. To this end recall that
$e^{K_{CS}/2}=\left(\frac43 d_{ijk} u^i u^j u^k\right)^{-\frac12}$. The tensions in Table~\ref{table_dws} 
then show that the domain walls from D5-branes wrapping the $A$-cycles
become tensionless in the limit of all $u^i$ going to infinity. 
To examine limits of sets of $u^i$'s going to infinity, it is necessary to specify the
intersection numbers $d_{ijk}$. Below we will consider a simple
example. 

Infinite towers of tensionless domain walls can be identified by inspecting the monodromy action on the
associated charge vectors, which can be taken to be $\vec{q} = (e_0, e_i, -q^i,-m)$ in the
basis $(A^0, A^i, B_i, B_0)$. When all the $u^i$ grow to infinite the tensionless branes have
$q^i=m=0$, while other charges are different from zero. It is easy to  check that the $P_i^t$, with
monodromy generators $P_i$ defined in \eqref{Pi2b}, do not anhilate $\vec{q}$ and 
do not connect tensioless to tensionfull branes. The conclusion is that there is an infinite tower of
tensionless domain walls formed by bound states of D5-branes wrapping $A^0$ $n$ times, together
with D5-branes wrapping the $A^i$ cycles.  
They are T-dual to the infinite IIA towers 
consisting of bound states of $n$ D2-branes and one D4-brane wrapping a 2-cycle. 
 
To illustrate the previous results we consider a simple model with
Hodge numbers $h^{2,1}_-=3$ and $h^{1,1}_+=3$.
Besides, the only non-zero intersection  numbers are taken to be $d_{123}=1$ and $\kappa_{123}=1$. 
In practice such geometry is realized
by the untwisted sector of the orbifold $T^6/\IZ_2 \times \IZ_2^\prime$, with $T^6=T_1^2\times T_2^2 \times T_3^2$. 
The full K\"ahler potential reads
\begin{equation}
\label{K2bex}
K= - \log s -\log(v^1 v^2 v^3) - \log(8 u^1 u^2 u^3) \, ,
\end{equation}
where $s=\preal S=e^{-\phi}$, $v_i =\preal T_i$ and $u^i =\preal U^i$. 

The tensions are easily found substituting
in the general formulas given in Table~\ref{table_dws}.
The IIA and IIB domain walls from D-branes are related by T-duality. For example, a D5-brane wrapping the IIB 
3-cycle denoted $A^3$, i.e. $y^1=y^2=x^3=0$, is T-dual to a D4-brane wrapping $T_3^2$. The IIB tensions are obtained by 
replacing $T^i_{\text{IIA}}\to U^i_{\text{IIB}}$,
$N^0 \to S$ and $N^i \to T^i_{\text{IIB}}$ in Table \ref{tab:domainwallsstorus}. 
The IIA and IIB domain walls from NS5-branes wrapping $A^0$,
i.e. $y^1=y^2=y^3=0$, are also T-dual to each other. To recover T-duality for other domain walls from NS5-branes requires introducing geometric and non-geometric fluxes \cite{acfi}.

\subsection{Charges}
\label{ss:charges2b}

The charges of IIB domain walls can be obtained proceeding along the lines
explained in section \ref{ss:charges2a}. After deriving general results we will
compute the charges in the toroidal model.
Our main purpose is to show that in that setup the electric WGC bound in \eqref{WGC} is saturated.

The domain walls are formed by D5-branes wrapping 3-cycles. Their gauge potentials are obtained from
the general expansion of the RR 6-form 
\begin{equation}
C_6= c_{3}^\lambda \wedge \alpha_\lambda +  \tilde c_{3\lambda} \wedge \beta^\lambda\, , 
\label{all3b}
\end{equation}
where $(\alpha_\lambda, \beta^\lambda)$ is the basis of 3-forms in the internal manifold.  
The couplings of the 4d potentials $c_{3}^\lambda$ and $\tilde c_{3\lambda}$ to the domain walls follow from the Chern-Simons action
for $C_6$, after integrating over the internal cycles.
To deduce the charges we need to determine the kinetic terms involving field strengths normalized according to the conventions in 
\cite{WGC16}. For instance, for D5-branes wrapping the $A^\lambda$ cycles, the suitable
field strengths are $\cf_4^\lambda = 2\pi \op dc_3^\lambda/\ell_s^3$. 

The relevant kinetic terms are derived by dimensional reduction of the appropriate term in the 10d action.
In the democratic formulation \cite{Bergshoeff:2001pv} we start from
\begin{equation}
S_{\text{kin}}^{(10)} = \frac{2\pi}{\ell_s^8} \int  \tfrac14 F_7 \wedge\!{}^*F_7  \, .
\label{skin10b}
\end{equation}
Dimensional reduction will involve integrals such as $\int_\cam \alpha_\kappa \wedge {}^*\alpha_\lambda$,
which depend only on the complex structure moduli $U^i$. They are given in \eqref{abint} in the case that
the axions ${\text{Im}}\, U^i$ are set to zero, which will be assumed in the following.
The 4d action will also pick up a dependence on the dilaton and the K\"ahler moduli, arising from the change
to 4d Einstein frame. Since $V_6= e^{3\phi/2} \cv$, 
in IIB the transformation is $g_4 = \big(\frac{e^{\phi/2}}{\cv} \frac{\cv_0}{e^{\phi_0/2}}\big) g_{4E}$,
with subscript $0$ denoting vev. 
Notice that $e^{-\phi} \cv^2 = e^{K_{CS}-K}$, where $K$ is the full K\"ahler potential
defined in \eqref{kpot2b}. 
Putting all pieces together and using \eqref{msp2}, we finally arrive at the 4d kinetic terms
\begin{equation}
S_{\text{kin}}
=  \frac{\pi}{2 M_P^4}\int \frac{e^{-K}}{8} \left[ \cf_4^0 \wedge\!{}^*\cf_4^0  
+ 4 g_{ij} \cf_4^i \wedge\!{}^*\cf_4^j 
+ \frac{1}{4\cad^2} g^{ij} \tilde \cf_{4i} \wedge\!{}^*\tilde \cf_{4j}
+ \frac{1}{\cad^2} \tilde \cf_{40} \wedge\!{}^*\tilde \cf_{4 0}
\right] \, .
\label{kinF1b}
\end{equation}
Here $g_{ij}=\partial_i\partial_{\bar\jmath} K_{CS}$ is the metric in complex structure moduli space, $g^{ij}$ is its inverse 
and $\cad =\frac16 d_{ijk} u^i u^j u^k=e^{-K_{CS}}/8$.
Notice the complete analogy with the IIA results in eq.~\eqref{kinF1}. In fact, this is in agreement with the type IIB results of \cite{Bielleman:2015ina}

It is straightforward to specialize to the toroidal model introduced above, in which
$\cad = u^1 u ^2 u^3$ and $e^{-K}= 8 s\op v^1 v^2 v^3 u^1 u^2 u^3$. The metric $g_{ij}$ is diagonal
with $g_{ii}=1/4(u^i)^2$. The kinetic terms reduce to
\begin{equation}
S_{\text{kin}} =  \frac{\pi}{2 M_P^4}\int \frac{e^{-K}}{8} 
\left[ 
\cf_{4}^0 \wedge\!{}^* \cf_{4}^0 \op + \!
\sum_{i=1}^3  \!\! \left[ \frac{1}{(u^i)^2} \cf_{4}^i \wedge\!{}^* \cf_{4}^i 
 + \frac{(u^i)^2}{\cad^2} \tilde \cf_{4i} \wedge\!{}^*\tilde \cf_{4i} \right]
\! + \frac{1}{\cad^2} \tilde \cf_{4 0} \wedge\!{}^*\tilde \cf_{4 0} 
 \right] \, .
\label{kinF2b}
\end{equation} 
Since the conventionally normalized saxions turn out to be  $\tilde s = \log s $,
$\tilde t^i = \log t^i $ and $\tilde u^i= \log u^i$, all kinetic pieces involve exponential dilatonic couplings 
$e^{-\alpha \varphi}$,  with $\alpha^2=7$. 
The charges of the different domain walls can be easily read off and compared to the tensions
in Table \ref{table_dws}.  The WGC bound is saturated in all cases. For example,
for the domain wall from the D5-brane wrapping the cycle $A^1$, $e^2=\frac{8}{\pi} M_P^4 e^K (u^1)^2$, while the squared
tension is $T^2=\frac{4}{\pi} M_P^6\op  e^K (u^1)^2$.

\subsection{Tensionless extended objects in $\cn=2$ compactifications}
\label{ss:}

Infinite towers of states becoming exponentially massless at infinite distance in moduli space have been identified in 
type IIB compactifications on a Calabi-Yau manifold $\cam$ \cite{irene}. The states are particles  described by D3-branes wrapping 
certain 3-cycles in $\cam$. The masses of these states turn out to depend only on the complex structure moduli, a result that
follows immediately from the DBI action as we review shortly. In the $\cn=2$ framework, these particles  couple to 4d gauge
vectors arising from the reduction of the RR 4-form, whose expansion includes terms $V^K \wedge \alpha_K$, $K=0,\ldots,h^{1,2}$.
Notice that imposing the orientifold projection to go to $\cn=1$ would eliminate the 4d vectors $V^K$ if $h_+^{1,2}=0$ and if not,
the volume of the corresponding 3-cycles could not be computed integrating the holomorphic 3-form which is odd under the
orientifold involution. In the $\cn=2$ compactification it is also natural to look at 4d strings formed by wrapping branes. A 
clear example is the string associated to the D1-brane which is absent in the orientifold because the RR 2-form is odd. 
Domain walls coupling to 4d 3-forms can be considered as well. Below we will examine how the
mass/tension of the various objects tied to branes behave in limits of infinite distance. The findings will be illustrated in a 
toroidal example in which the different energy scales can be compared.  

For the essentials of type IIB compactifications on a Calabi-Yau manifold we refer to \cite{Grimm:2004uq}.
The needed features can actually be borrowed, with a few adjustments, from the short review in Appendix \ref{ap:2b}.
We will again work in the large complex structure limit characterized by the prepotential in \eqref{prep2b} and the
corresponding holomorphic 3-form in \eqref{om2bs}, but now with index $i$ replaced by $I$ running from 1 to $h^{1,2}$.
There are $h^{1,2}$ complex structure moduli denoted $U^I$. These modifications further apply to the period vector
and the monodromy generators defined in \eqref{pivec} and \eqref{Pi2b}.
On the other hand, in the basis of 2-forms and 4-forms we change the index $\alpha$ to $A$ running from 1 to $h^{1,1}$.
Besides, now $\cv=\kappa_{ABC} t^A t^B t^C$. 

Tensions of extended objects can be computed from the DBI action, cf. \eqref{sdbiax}.
It is useful to work out the generic case of a \mbox{$\text{D}p$-brane} wrapping a \mbox{$k$-cycle} with $k=p+1-n$.
Particles correspond to $n=1$, while strings and domain walls to $n=2$ and $n=3$ respectively. 
The tension is determined integrating over the internal cycle. We find
\beq
T_{\text{D}p}(\gamma_k) = \frac{2 \pi M_s^n}{g_s} V_{k}(\gamma_k) \, ,
\label{tpn}
\eeq
where $V_{k}(\gamma_k)$ is the volume of $\gamma_k$ in string units. Expressions in
terms of the fixed Planck mass $M_P$ are obtained substituting 
$M_s = \frac{g_s^{1/4}\op M_P}{\sqrt{4\pi} \cv^{1/2}}$, which differs slightly from \eqref{msp2}
because now there is no orientifold projection. 

In \cite{irene} it was discussed how particles obtained by wrapping D3-branes around certain 3-cycles
are natural candidates to populate the infinite tower of light states postulated by the SDC.
It is instructive to reproduce this proposal. The masses obtained from the DBI action are simply given by
\eqref{tpn} with $p=k=3$. The volume of the internal 3-cycle is again given by \eqref{vol32b}, with
$N=2\sqrt2 \, e^{K_{CS}/2} g_s ^{3/4} \cv^{1/2}$.
Substituting in \eqref{tpn} gives
\beq
M_{\text{D}3}(\gamma_3) = 2\sqrt{2\pi} M_P\, e^{K_{CS}/2} \left|\int_{\gamma_3}\hspace*{-3mm} \op \Omega \right| \, .
\label{tp1}
\eeq
These masses depend only on the complex structure moduli through $\Omega$ and the K\"ahler potential given by
$K_{CS}=-\log\left(\frac43 d_{IJK} u^I u^J u^K\right)$, in the limit of large complex structure.
Taking the 3-cycles to be the
duals to  the basis of 3-forms leads to the masses shown in Table \ref{table_particles}.

\begin{table}[h!]\begin{center}
\renewcommand{\arraystretch}{1.25}
\begin{tabular}{|M{1.5cm}||c|c|c|c|}
\hline
 \small{Cycle} &  $A^0$ & $A^I $  & $B_I$ & $B_0$ \\
 \hline
\small{Mass (in units of $2\sqrt{2\pi} M_P$)} & $e^{K_{CS}/2}$  & $e^{K_{CS}/2}\, \left|U^I \right|$ &  
$e^{K_{CS}/2}\,\left|\frac12 d_{IJK} U^J U^K\right|$  & $e^{K_{CS}/2}\, \left|\frac16 d_{IJK} U^I U^J U^K\right|$  
\\[2mm]
\hline
\end{tabular}
\caption{DBI masses of particles formed by D3-branes wrapping 3-cycles. $I$ runs from 1 to $h^{1,2}$.}
 \label{table_particles}
\end{center}
\end{table}  

The behavior of the masses in the limit of large complex structure follows taking into account the form of
$K_{CS}$. In particular,  from
the masses in Table \ref{table_particles} we see that the particles from D3-branes wrapping the $A$-cycles
become massless when all $u^I$ go to infinity. 
Similar results were found in \cite{irene} where the isotropic case $u^I=u$, $\forall I$,
was studied. Limits of sets of $u^K$'s going to infinity depend on the specific
intersection numbers $d_{IJK}$. 
The analysis of the monodromy action for particles is analogous to that done for domain walls in the orientifold case
because the monodromy generators take the same form. It allows to  
identify an infinite tower of massless particles formed by bound states of D3-branes wrapping $A^0$ $n$ times and
D3-branes wrapping the $A^I$ cycles.
These are none but the T-dual of the infinite
towers of massless particles found in IIA, consisting of bound states of $n$ D0-branes and
D2-branes wrapping 2-cycles.
 
In the $\cn=2$ case we can also look at strings. The tensions can be computed from the general formula \eqref{tpn}. 
Specializing to $n=2$ for strings gives
\beq
T_{\text{D}p}(\gamma_k) = \frac{M_P^2}{2 g_s^{1/2}\cv} V_{k}(\gamma_k) \, ,
\label{tp2}
\eeq
where $k=p-1$. The resulting tensions are shown in Table \ref{table_strings}.
The volumes of \mbox{$k$-cycles} for $k=2,4,6$ are derived from \eqref{vols24}, taking
$J=e^{\phi/2} t^A \omega_A$. For NS5-branes wrapping $\gamma_4$ there is an additional factor $1/g_s$. 
Clearly the tensions do not depend at all on the complex structure moduli.
The strings from D7-branes on $\cam$ as well as from D3-branes wrapping even 2-cycles survive the orientifold
projection. The former can become tensionless when $g_s \to 0$ and the latter at infinite distance 
directions in K\"ahler moduli space.

\begin{table}[htb!]\begin{center}
\renewcommand{\arraystretch}{1.5}
\begin{tabular}{|c|c|c|}
\hline
Brane & Cycle & Tension (in units of $M_P^2/2$)\\
\hline \hline
D1& -  &$\dfrac{1}{g_s^{1/2} \cv}$  \\[2mm]
\hline
D3 & P. D. $[\tilde{\omega}^A] $ & $\dfrac{t^A}{ \cv}$    \\[2mm]
\hline
D5 & P. D. $[\omega_A]$ & $\dfrac{g_s^{1/2}}{ \cv}  \, \left( \frac12 \kappa_{ABC} 
\, t^B \,  t^C\right) $ \\[2mm]
\hline
NS5 &  P.D. $[\omega_A]$&  $\dfrac{1} {g_s^{1/2}\, \cv}  \, \left( \frac12 \kappa_{ABC} 
\, t^B \,  t^C\right) $ \\[2mm]
\hline
D7 & $\mathcal{M}$ &  $g_s$  \\[2mm]
\hline
\end{tabular}
\caption{DBI tensions of strings formed by branes wrapping  even cycles. $A$ runs from 1 to $h^{1,1}$.}
 \label{table_strings}
\end{center}
\end{table}  

We finally consider domain walls. The tensions can actually be read off from the results \eqref{t32b2} in the orientifold case.
Wrapping a D5-brane on a 3-cycle $\gamma_3$ gives a domain wall with tension
\beq
T_{\text{D}5}(\gamma_3) = \frac{M_P^3 \, g_s^{1/2}}{ \sqrt{2\pi}\, \cv }\, e^{K_{CS}/2} 
\left|\int_{\gamma_3}\hspace*{-2mm} \Omega \right| \, .
\label{t32b2n2}
\eeq
We also obtain $T_{\text{NS}5}(\gamma_3) =T_{\text{D}5}(\gamma_3) /g_s$, after adjusting the dependence on the dilaton.
Taking the 3-cycles in the symplectic basis yields the tensions displayed in Table \ref{table_dws2}. 

\begin{table}[h!]\begin{center}
\renewcommand{\arraystretch}{1.25}
\footnotesize{
\begin{tabular}{|M{2cm}||c|c|c|c|}
\hline
 \small{Cycle} &  $A^0$ & $A^I $  & $B_I$ & $B_0$ \\
 \hline
\small{Tension (in units of $M_P^3/\sqrt{2\pi}$)} &  $\frac{g_s^{1/2}}{\op \cv } e^{\frac{K_{CS}}2}$  & $\frac{g_s^{1/2}}{\op \cv } e^{\frac{K_{CS}}2}\, \left|U^I\right|$ &  
$\frac{g_s^{1/2}}{\op \cv } e^{\frac{K_{CS}}2} \left|\frac12 d_{IJK} U^J U^K\right|$  & 
$\frac{g_s^{1/2}}{\op \cv } e^{\frac{K_{CS}}2} \left|\frac16 d_{IJK} U^I U^J U^K\right| $  
\\[2mm]
\hline
\end{tabular}
}
\caption{DBI tensions of domain walls formed by D5-branes wrapping 3-cycles.}
 \label{table_dws2}
\end{center}
\end{table}  

Analogous to the $\cn=1$ results, the tensions of domain walls associated to D5-branes wrapping the $A$-cycles
clearly become tensionless when all $u^I$ go to infinity. 
The study of the monodromy action is also similar. We again conclude that 
there is an infinite tower of tensionless domain walls composed by bound states of D5-branes wrapping $A^0$ $n$ times and
D5-branes wrapping the $A^I$ cycles.

\subsubsection{Energy scales in a IIB toroidal orbifold example}
\label{sss:scales2b}

To illustrate the results we will again use a model that can be understood as
the untwisted sector of the orbifold $T^6/\IZ_2 \times \IZ_2^\prime$, with $T^6$ factorized as
$T_1^2\times T_2^2 \times T_3^2$. The Hodge numbers are $h^{2,1}=3$, $h^{1,1}=3$, and
the only non-zero intersection numbers happen to be $d_{123}=1$ and $\kappa_{123}=1$. 
Then, $K_{CS} = -\log(8 u^1 u^2 u^3)$ and $\cv=t^1 t^2 t^3$. We will denote $s=e^{-\phi}$
and $v_I = \frac12 \kappa_{ABC} t^B t^C$.
The masses and tensions follow from
the general formulas given in Tables \ref{table_particles}, \ref{table_strings} and \ref{table_dws2}.

For particles the results are just the T-dual of the masses in \eqref{mparticlestorus}. For instance, 
the D3-branes wrapping the cycles $A^0$ and $B_0$ are mapped respectively to the D0-brane and the
D6-brane wrapping the whole $T^6$. 
The IIA and IIB domain walls from D-branes are connected by T-duality as explained in the orientifold case. 
T-duality is also manifest for strings from D-branes.
The D1-brane is T-dual to the D4-brane along $A^0$, whereas e.g. the D5-brane on $T_1^2\times T_2^2$
is T-dual to a D4-brane around the  $A^3$ 3-cycle $x^1=x^2=y^3=0$. Similarly,  
the T-duals of D3 and D7-branes covering $T_i^2$ and $T^6$ are D4-branes wrapping the 3-cycles $B_i$ and $B_0$
respectively. From the IIA tensions in \eqref{tstringstorus} we see that the IIB tensions are obtained by
replacing $n^0 \to s$ and $n^i \to v_i$. However, this map fails for NS5-branes, as expected since they are generally mapped to different objects under mirror symmetry.

\begin{figure}[tb]
     \begin{center}
      \subfigure[]{
        \includegraphics[height=230pt]{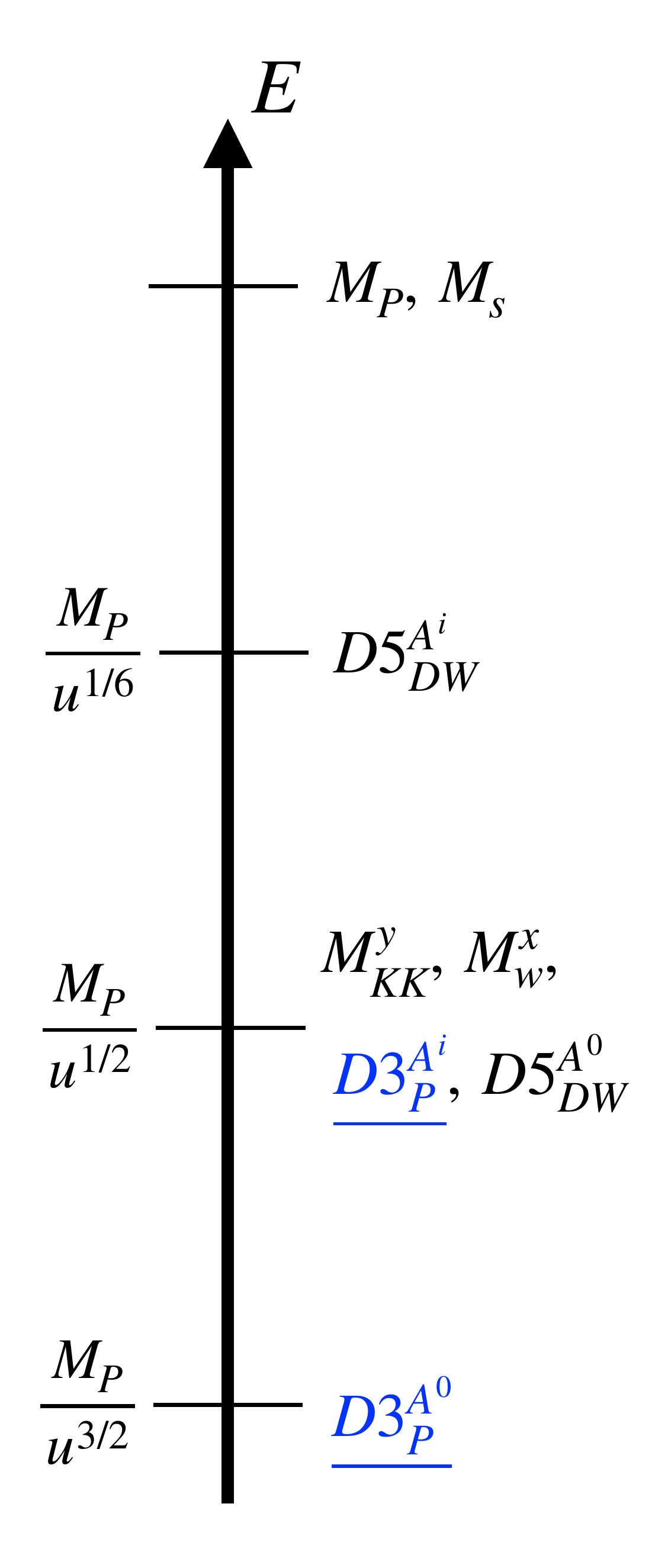}
        \label{fig:scalesIIBu}
        }
     	\subfigure[]{	
        \includegraphics[height=230pt]{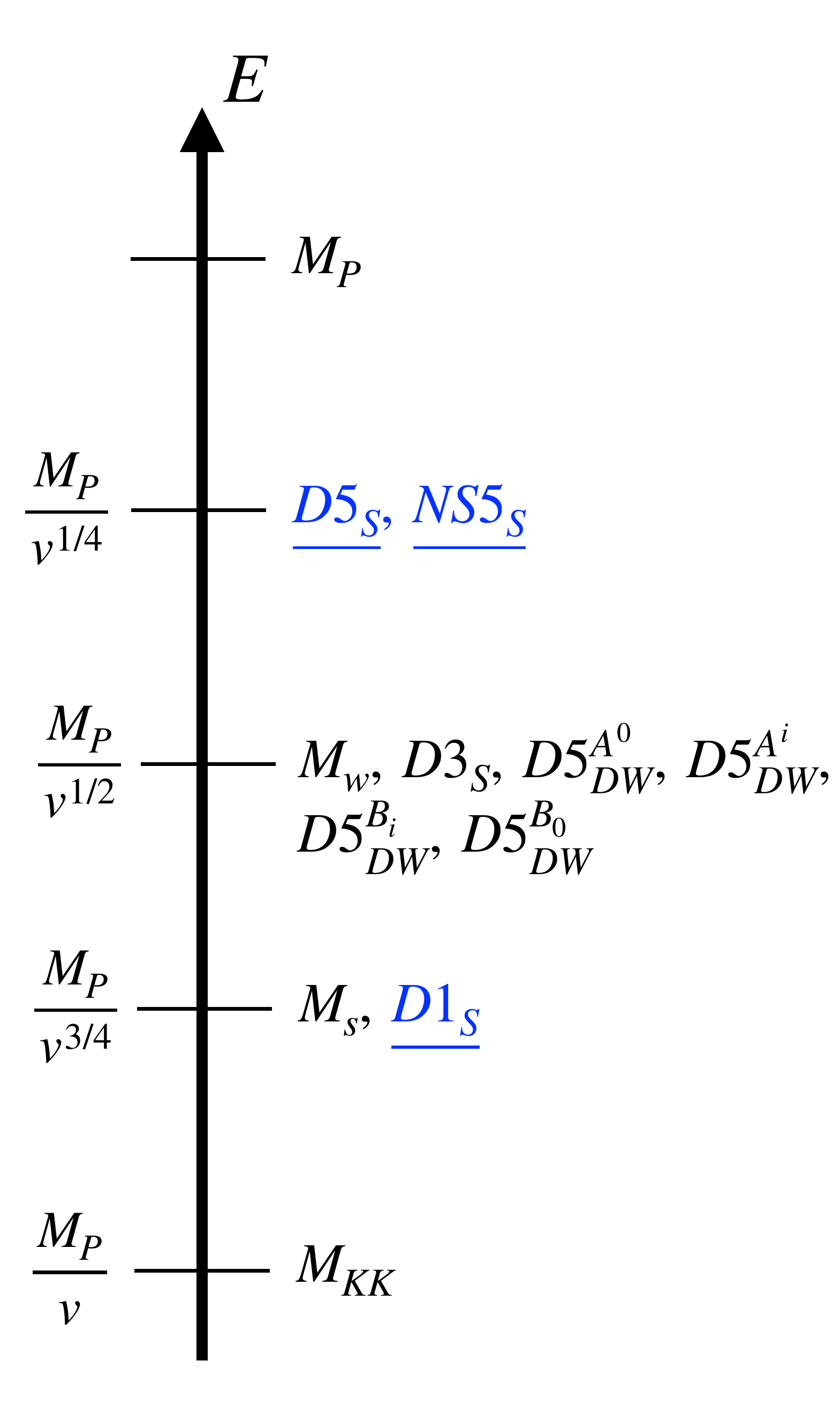}
        \label{fig:scalesIIBv}
        }
       \subfigure[]{	
        \includegraphics[height=230pt]{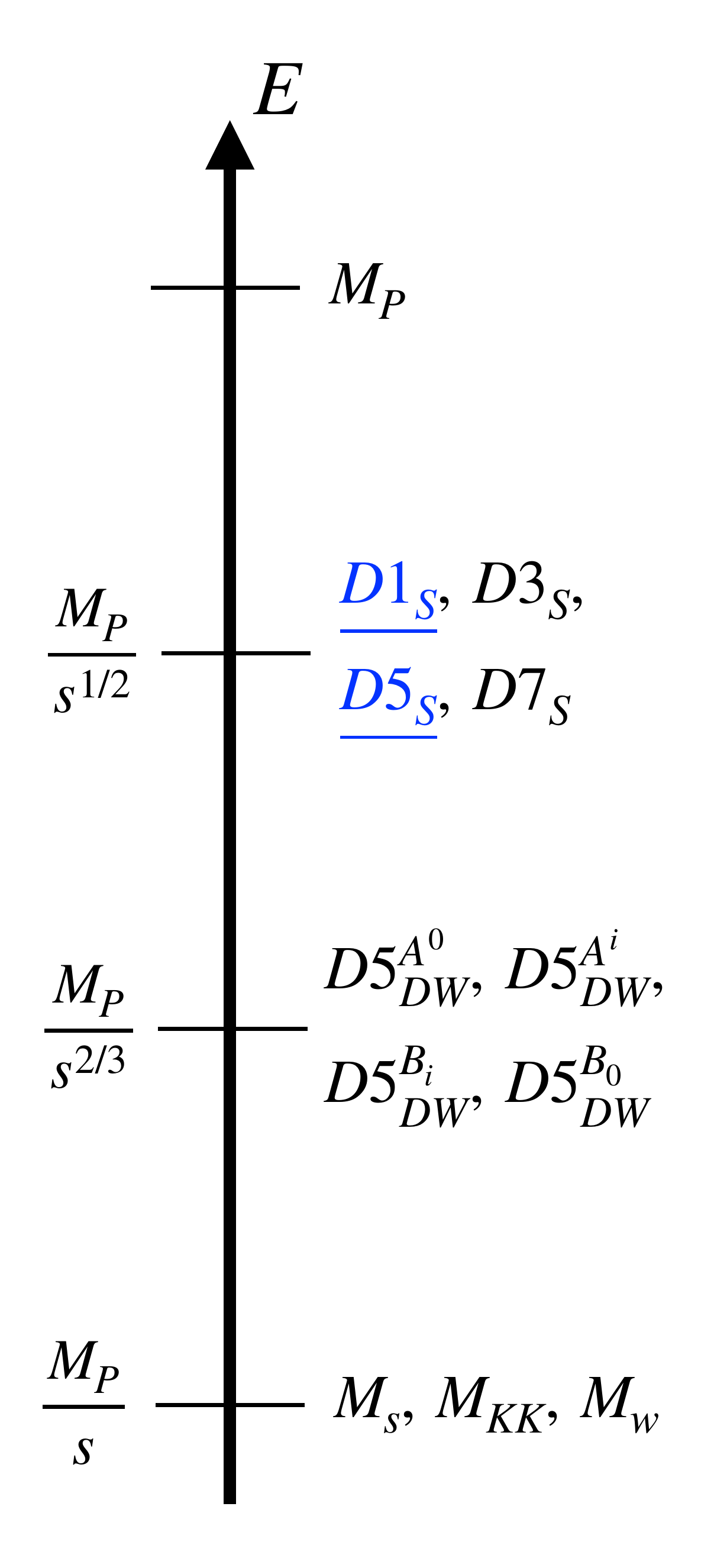}
        \label{fig:scalesIIBsv}
        }
\caption{Energy scales associated to the 4d particles, strings and domain walls that become 
 massless/tensionless at the infinite distance points given by {\bf (a) } $ u \rightarrow \infty$, $s$, $v$ fixed, {\bf (b) } $ v \rightarrow \infty$, $s$, $u$ fixed, and {\bf (c) } $ s \propto v \rightarrow \infty$, $u$ fixed. The string, KK and winding scales are also indicated. These are the T-duals of Fig. \ref{fig:scalesuort}. Objects in blue and underlined are projected out in the orientifold.}
        \label{fig:scalesIIB}
     \end{center}
\end{figure}

Let us finally look at the energy scales of the extended objects that become massless/tensionless at infinite distance
points in moduli space. To simplify we consider the isotropic situation with $v_1=v_2=v_3=v$, and $u^1=u^2=u^3=u$. 
The energy scales of strings and domain walls are estimated by a suitable root of their tensions. In Figure \ref{fig:scalesIIB}
we depict the energy scales in three different limits. Notice that in the limit of large complex structure, $u\to \infty$, at the
same scale there are tensionless domain walls and massless particles, both belonging to infinite towers. Thus, 
it would be important to take into account the implications of towers of tensionless domain walls in the effective field theory. 

\section{Discussion and conclusions}
\label{ss:discussion}

In this paper we have presented a study of the towers of tensionless branes appearing in type II CY compactifications
at  points at infinite distance  in moduli space.
We have first found what elements of a basis of 4d domain walls  become tensionless at different infinite distance points for a general type IIA orientifold. Then, we have explicitely managed to construct the monodromy orbits of domain walls, relating the type of singularity with the fact that the tower is generated by the monodromy around the singular point or by the monodromy around a different point. To this purpose we have shown the  conditions that the elements within the monodromy orbit  have to fulfill in order to remain tensionless, relating this with their construction in terms of the aforementioned subset of elements of the basis that becomes tensionless. We have discussed some aspects of the exponential behavior of the tensions of these towers as we move towards infinite distance. Finally, we have particularised to the type IIA  toroidal orientifold $T^6/\IZ_2 \times \IZ_2^\prime$ in order to get some intuition and discuss the energy scales of the towers of particles, strings and domain walls in this context. 
This analysis has been carried out in some detail for domain walls in $\mathcal{N}=1$ type IIA orientifolds, but we have also analyzed the
towers of particles and tensionless strings appearing in the parent ${\cal N}=2$ compactifications. We have also repeated the 
 discussion for the mirror type IIB orientifold compactifications, and we have checked the  matching  of these towers with the ones found in the mirror type IIA for the  
 toroidal orientifold case.

We have not carried out a full stability analysis of the states within the monodromy orbits,
as e.g. performed in \cite{irene} in terms of walls of marginal stability. This would be important in order to guarantee stability against decay to other elements in the orbit,
ensuring that the infinite tower is populated by  stable states. We have however presented explicit cases where this  stability can be ensured, as is the case for the infinite tower of domain walls formed by D2 and D4 branes with different D2 charges, for which the whole monodromy orbit is stable at the infinite distance point given by $t^1 \rightarrow \infty$. 

We now would like to discuss a number of possible additional consequences  which seem to be implied 
by our general analysis.

The presence of infinite towers of extended objects may have several implications for the emergence proposal  \cite{irene, Grimm2, Corvilain, emergence1, emergence2}. On the one hand, the presence of these towers of strings an domain walls below the Planck scale implies an enormous increase in the total number of degrees of freedom in the EFT. In fact, as shown in Figures \ref{fig:scalestu}-\ref{fig:scalesIIB} the energy scale of these towers of strings and domain walls is not in general greater than that of the towers of particles (it is actually lower in many of the cases under consideration). This would presumably lower the cutoff scale of the theory, argued to be the species bound 
\begin{equation}
\Lambda_{\mathrm{Species}}=\dfrac{M_P}{\sqrt{N}},
\end{equation}
where $N$ is the number of species below this cutoff scale. In fact, one could think that the sole presence of one individual tensionless string or domain wall would already provide an infinite number of degrees of freedom that could account for the vanishing of the cutoff before the infinite towers of particles appear, let alone the presence of an infinite tower of these extended objects. Note that, if true,  this could have a drastic impact in most of the calculations based on emergence in the swampland literature, since they  rely heavily on a counting of the number of species that  so far  has only considered particles as possible species  (the impact of towers of instantons has recently been studied in \cite{infinitons}).
Still, given our lack of understanding about how to integrate out extended objects, one cannot exclude that such counting could somehow still be valid.
In particular, as we discuss below, perhaps the effect is not so drastic at least in some supersymmetric settings. 

Indeed,  one may  argue about a possible interpretation of the scalar potential in terms of emergence
(see also \cite{review} for a related discussion). In \cite{irene} it has been explained how the IR structure  of the kinetic terms of different fields arises from integrating out the towers of particles to which these fields couple. The rough idea is that if the towers of particles couple to these fields, they can modify their two point function at one loop, correcting their kinetic terms. For the case of massless scalars this turned out to explain the asymptotic behavior of the K\"ahler metric and for 1-forms, the running of the gauge coupling. Following this logic, one would expect that the kinetic terms for gauge 3-forms could emerge from integrating out towers of domain walls, which we have identified in this work.  Leaving aside the issue of how to perform the process of integrating out extended objects, the implications would  probably be richer than in the case of the kinetic terms of $p$-forms of lower rank, since in this case we can use the results of \cite{Bielleman:2015ina, Carta:2016ynn, Herraez:2018vae} to relate this kinetic terms with the scalar potential generated by fluxes. In \cite{Bielleman:2015ina,Herraez:2018vae} it was shown that the terms containing 4-forms in the 4d action have  schematically the form
\begin{equation}
S^{\rm 4d}=  - \frac{1}{8\kappa_4^2} \int_{\IR^{1,3}} Z_{AB} F_4^A \wedge *_4 F_4^B + \frac{1}{4\kappa_4^2}\int_{\IR^{1,3}} F_4^A \rho_A\, ,
\label{S4formRR}
\end{equation}
where the indices $A$ and $B$ run over all the fluxes of the compactification and in the basis in which the $F_4^A$ are the field strengths of the corresponding 3-forms the  $Z_{AB}$ depend on all the moduli. The $\rho_A$ take the form of $\vec{q}$ in section \ref{sec:tensionlesswalls} in this basis.
An important observation here is that 
the matrix $Z_{AB}$ contains as sub blocks the K\"ahler metric of both the K\"ahler and the complex structure moduli spaces. 
In the spirit of the emergence proposal, this could point towards a connection between the way in which towers of particles and towers of domain walls are integrated out, since the fact that these metrics can be extracted both from the kinetic term of the moduli and the 4-forms  would imply that the results obtained from integrating out the towers of particles and domain walls should  be compatible. Going back to eq.~\eqref{S4formRR}, since the 4-forms have no propagating degrees of freedom, they can be integrated out to obtain the flux scalar potential, which takes the form 
\begin{equation}
V=\dfrac{1}{8}Z^{AB}\rho_A \rho_B.
\end{equation}
In this context, if the kinetic term of the 3-forms, given by $Z_{AB}$ arose from integrating out the tower of domain walls that become light, the scalar potential would be emergent as well.  This is also of particular interest since it gives a hint on a possible way of approaching the SDC in the presence of potentials via the study of domain walls in flat moduli space. In fact, one could try to tackle the problem of emergence of the potentials from the point of view of bubble nucleation. A possible drastic way in which the EFT could break down due the presence of these infinite towers of domain walls is that their nucleation becomes more and more probable as we approach the infinite distance point and their tension is lowered. It is true that  the energy stored in the interior of the bubble could compensate for this lowering of the tension (as is the case for a the transition from a fluxless vacuum to another supersymmetric AdS vacuum). Still, even if the semiclassical tunneling were forbidden due to energy balance considerations, the quantum formation of these bubbles that nucleate but do not expand should be a right way of thinking about the quantum effects that must be integrated out in order to obtain an effective description of the kinetic term of 3-forms. This kind of approach would lead to a sum over all possible domain walls, that is, over all possible fluxes. This integration could lead possibly to modular covariant functions, in analogy with the non-perturbative examples
in ref.\cite{giu}.

This idea of considering all possible domain walls nicely connects with something that was mentioned at the end of section \ref{sec:tensionlesswalls}, namely ``exotic'' domain walls. As remarked, we did not elaborate on the construction of the corresponding towers.  Still, in spite of the lack of understanding of their microscopic origin in  most  cases, these domain walls could be straightforwardly incorporated to the 4d analysis by using the superpotential that they generate, since it gives us all the information that is needed in order to compute their tensions. Even though these tensions would include higher order polynomials in the moduli, it is likely that some of them become tensionless at certain infinite distance points, too. Take, for instance, the so called non-geometric $\mathcal{P}$ fluxes that appear from the requirement that type IIB in toroidal orientifold be S-duality invariant in the presence of fluxes. One of the terms that appears in the superpotential is \cite{acfi}
\begin{equation}
W_{\mathcal{P}}\, =\, i \sum_{i,j=1}^3 \gamma_{ij} S T^i U^j,
\end{equation}
where $\gamma_{ij}$ is the non-geometric flux that would be sourced by some exotic domain wall. From this superpotential it can be argued that the corresponding domain wall would become tensionless in the same cases as the ones from a D4-wrapping a 2-cycle (whose superpotential is proportional to $T^i$) when the K\"ahler moduli is sent to infinity and the rest kept fixed. Note that
this superpotential is invariant under mirror symmetry, so that we would find the same exotic domain wall in the dual IIA picture. In general, these ``exotic'' domain walls would source all kinds of non-geometric fluxes like those presented in \cite{acfi} and the corresponding contribution to the scalar potential would
possibly  fit into the picture of emergence. 

Another interesting question is whether the towers may affect the effective action of moduli fixing scenarios  in string phenomenology.
In particular, it is important to check whether the presence of these objects invalidates the EFT within the region of moduli space in which the moduli are fixed. We begin by looking at the type IIA flux compactifications in the simple toroidal orientifold introduced at the end of section \ref{ss:2a}. Moduli fixing in this setup was studied in detail in \cite{Camara:2005dc}, where it was found a large class of minima (in the isotropic case) where 
the scaling of the moduli has the form
\begin{equation}
 s \simeq u \simeq t^3,
\end{equation}
in the presence of the fluxes given by the vector $\vec{q}$ in \eqref{chargevector}, and with $s=n^0$, $u=n^i$, in 
our notation. This scaling precisely matches 
Figure \ref{fig:scalestu3}, in which all the moduli lie within the perturbative region and it can be seen that the infinite towers of extended objects are always above the KK and the fundamental string scales, ensuring that as long as these are under control, the towers will not invalidate the EFT. Similar conclusions may be drawn for the vacua in \cite{DeWolfe:2005uu}, since they share the same scaling but with the complex structure moduli projected out. Another interesting scenario that is worth mentioning in this context is the KKLT contruction \cite{KKLT}. In that setup, the complex structure and complex dilaton are fixed at moderate values by the fluxes and the K\"ahler moduli are stabilized at large values by non-perturbative effects. This would match Figure \ref{fig:scalesIIBv} and, as before, this is parametrically safe from the effects of the towers. We will not elaborate more on this but let us remark that the presence of these towers is a general feature of string compactifications. Hence, their energy scales being above the relevant ones for different moduli stabilization scenarios is not in general guaranteed and must be carefully justified. 

Finally, let us mention that the existence of these towers seems consistent with the arguments in \cite{dS3}. There it is argued that the towers of light states that they consider could be formed by particles or by extended objects, in general. In this context, the presence of towers of strings and domain walls motivates the existence of multiple towers of states as an infinite distance point is approached and they might account for some of the entropy needed to saturate the Bousso bound. Notice that, even though the derivation of the de Sitter conjecture in terms of entropy is in agreement with the existence of these towers, their presence could have important implications for cosmology if our universe happened to be in a weakly coupled regime (which is where the arguments of \cite{dS3} apply).

\bigskip

\centerline{\bf \large Acknowledgments}

\bigskip

\noindent We are grateful to  F. Marchesano, E. Palti, S. Theisen,
A. Uranga, I. Valenzuela and M. Wiesner for useful discussions. 
This work has been supported by the ERC Advanced Grant SPLE under contract ERC-2012-ADG-20120216-320421, by the grant FPA2012-32828 from the MINECO,  and the grant SEV-2012-0249 of the 
``Centro de Excelencia Severo Ochoa" Programme.  The work of A.H. is supported by the Spanish FPU Grant No. FPU15/05012.
A.F. thanks the IFT UAM-CSIC, the Max-Planck-Institut f\"ur Gravitationsphysik and the Abdus Salam ICTP for hospitality and support at various stages of this work. 

\bigskip\bigskip

\appendix

\section{The $T^6/\IZ_2 \times \IZ_2^\prime$ orbifold}
\label{ap:z2z2}

We take $T^6$ to be the product $\otimes_{j=1}^3 T_j^2$. Each sub-torus is chosen to have a square lattice
with lattice vectors of sizes $R_x^i$ and $R_y^i$, so that the area and the complex structure of $T_j$ are
respectively $A_i=R_x^i R_y^i$ and $\tau_i=R_y^i/R_x^i$. The $T^6$ metric is diagonal and can be written as
\beq
G = \text{diag}(\frac{A_1}{\tau_1},  A_1\tau_1, \frac{A_2}{\tau_2},  A_2\tau_2, \frac{A_3}{\tau_3},  A_3\tau_3) \, .
\label{t6metric}
\eeq
It is convenient to define complex coordinates $z^i=R_x^i x^i + i R_y^i y^i$. The orbifold and orientifold actions are
\begin{equation}
(z^1,z^2,z^3) \to  
\left\{
\begin{array}{ll}
(-z^1,-z^2,z^3)  \ & \IZ_2 \\
(z^1,-z^2,-z^3) \ & \IZ_2^\prime \\
(\bar z^1,\bar z^2,\bar z^3)  \ & \R
\end{array}
\right. 
\, .
\end{equation}
The orbifold has altogether $h^{1,1}=51$ and $h^{1,2}=3$. In the following we will only consider moduli arising
in the untwisted sector, namely those related to the geometry of $T^6$.

The K\"ahler form is $J=i\sum_{k=1}^3 G_{k\bar k} dz^k \wedge d\bar z^k$. From \eqref{t6metric} we find
$G_{k\bar k}= \frac12$. Thus $J=\sum_{k=1}^3 A_k dx^k \wedge dy^k$. The K\"ahler form is invariant under the
orbifold action and satisfies $\R J = - J$.
Accounting for the orbifold action, $\cam=T^6/\IZ_2 \times \IZ_2^\prime$ has volume
$\cv=\frac14 A_1 A_2 A_3$. We then define $t^k=A_k/2^{\frac23}$. In this way
  \beq
J= t^k \omega_k \,,  \qquad \omega_k= 2^{\frac23} dx^k \wedge dy^k \, .
\label{jt6}
\eeq
The only non-zero triple intersection number is $\kappa_{123}=1$. The basis for dual 4-forms is $\tilde\omega^j$,
with e.g. $\tilde\omega^1= 2^{\frac43} dx^2 \wedge dy^2\wedge dx^3 \wedge dy^3$. 
Notice that the real part of the K\"ahler moduli $T^k$ is precisely $t^k$.  

The holomorphic 3-form is taken to be 
\beq
\Omega= (dx^1 + i \tau^1 dy^1)\wedge (dx^2 + i \tau^2 dy^2)\wedge (dx^3 + i\tau^3 dy^3) \, .
\label{omt6}
\eeq
Clearly $\Omega$ is invariant under the orbifold action and fulfills the condition $\R \Omega = \bar\Omega$.
The normalization of $\Omega$ is conventional. From the definition of the compensator field in \eqref{cfield} we see
that the quantity $C\Omega$, relevant in \eqref{nkdef} and \eqref{kqdef}, is scale invariant.
Therefore, rescaling $\Omega$ will not affect the moduli $N^K$, nor the K\"ahler potential $K_Q$.  

The basis of 3-forms also has to be defined appropriately to have $\int_\cam \alpha_K \wedge \beta^L= \delta_K^L$.
We take for instance
\beq
\alpha_0 = 2\, dx^1 \wedge dx^2\wedge dx^3 \, , \qquad  \beta^0 = 2\, dy^3 \wedge dy^2\wedge dy^1 \, . 
\label{3formst6}
\eeq
The Hodge duals follow easily from standard definitions because the metric is diagonal. For example,
${}^*\alpha_0=\tau_1 \tau_2 \tau_3 \, \beta^0$. 

Finally, the KK and winding scales associated to the 6 internal directions (the three $x^i$ and the three $y^i$), in terms of the 
moduli $s=n^0$, $t_i$ and $u_i=n_i$ take the following form:
\begin{align}
&M_{KK}^{R_x^i}\sim\dfrac{M_s}{R_x^i}=\left( \dfrac{u_j u_k}{s u_i t_i^2} \right)^{1/4}M_s\, , \qquad  M_{w}^{R_x^i}\sim R_x^i M_s =\left( \dfrac{s u_i t_i^2}{u_j u_k} \right)^{1/4}M_s\, ,\\
&M_{KK}^{R_y^i}\sim\dfrac{M_s}{R_y^i}=\left( \dfrac{s u_i }{ u_j u_k t_i^2} \right)^{1/4} M_s\, , \qquad  M_{w}^{R_y^i}\sim R_y^i M_s =\left( \dfrac{ u_j u_k t_i^2}{s u_i} \right)^{1/4}M_s\, .
\end{align}
Note that, upon three T-dualities along the three $x$ axis, the scales of the mirror IIB theory  are obtained by the substitution  $M_{w}^{R_x^i}\leftrightarrow  M_{KK}^{R_x^i}$, $ M_{w}^{R_y^i} \rightarrow  M_{w}^{R_y^i}$ and $ M_{KK}^{R_y^i} \rightarrow M_{KK}^{R_y^i}$ remain the same (after the identification $t_i \leftrightarrow u_i$)

\section{Review of type IIB orientifolds}
\label{ap:2b}

This brief summary of type IIB orientifolds is intended to introduce the basic ingredients. 
For more details we refer to \cite{Grimm:2004uq}, see also \cite{Blumenhagen:2015kja}.
The cohomology of $\cam$ is split into even and odd components according to the action
of the orientifold involution which leaves $J$ invariant and changes the sign of $\Omega$.
To simplify we assume $h_-^{1,1}=h_+^{1,2}=0$. The bases for 3-, 2- and 4-forms
are respectively denoted by $\{\alpha_\lambda, \beta^\lambda\}$, $\lambda=0,\ldots, h_-^{1,2}$, 
$\{\omega_\alpha\}$,  and $\{\tilde\omega^\alpha\}$, $\alpha=1,\ldots, h_+^{1,1}$. They are
chosen so that $\int_\cam \omega_\alpha \wedge \tilde \omega^\beta=\delta_\alpha^\beta$ and
$\int_\cam \alpha_\lambda \wedge \beta^\kappa=\delta_\lambda^\kappa$. The triple intersection
numbers are $\int_\cam \omega_\alpha \wedge \omega_\beta \wedge \omega_\gamma=\kappa_{\alpha\beta\gamma}$.

The moduli comprise $h_+^{1,1}$ K\"ahler moduli $T_\alpha$, $h_-^{1,2}$
complex structure moduli $U^i$, and the universal axio-dilaton $S$. The latter is given by $S=e^{-\phi} - i C_0$, with
$\phi$ the ten-dimensional dilaton and $C_0$ the RR 0-form.  
The K\"ahler moduli are explicitly $T_\alpha =  \frac12 \kappa_{\alpha\beta\gamma}  t^\beta  t^\gamma + i \rho_\alpha$,
where $t^\alpha$ and $\rho_\alpha$ descend from the expansions of the K\"ahler form and RR 4-form according to
$J=e^{\phi/2}  t^\alpha \omega_\alpha$ and $C_4=\rho_\alpha \tilde\omega^\alpha$.

The complex structure moduli enter in the holomorphic 3-form $\Omega$, which has the expansion 
$\Omega = X^\lambda \alpha_\lambda - \cf_\lambda \beta^\lambda$.
The $(X^\lambda, \cf_\lambda)$ are the periods of $\Omega$. This means that $\int_{A^\lambda} \Omega = X^\lambda$
and $\int_{B_\lambda} \Omega = \cf_\lambda$, with $\{A^\lambda, B_\lambda\}$ a basis of 3-cycles dual to 
$\{\beta^\lambda, \alpha_\lambda\}$.
Besides, $\cf_\lambda=\partial \cf/\partial X^\lambda$, where $\cf$ is
the holomorphic prepotential. Supersymmetry requires that $\cf$ is homogeneous of degree two. In the limit of large complex structure
it takes the simple form
\beq
\cf= -\frac16 d_{ijk} \frac{X^i X^j X^k}{X^0} \, , \quad i=1,\ldots, h_-^{1,2} \, ,
\label{prep2b}
\eeq
where $d_{ijk}$ is a completely symmetric constant tensor characteristic of $\cam$.
The complex structure moduli are given by $U^i= -i \frac{X^i}{X^0}$. 
In the limit of large complex structure the holomorphic (3,0) form then reads
\beq
\Omega=\alpha_0 + i U^i \alpha_i - \frac12 d_{ijk} U^i U^j \beta^k + \frac{i}{6} d_{ijk} U^i U^j U^k \beta^0 \, ,
\label{om2bs}
\eeq
where we used the freedom to rescale $\Omega$ to set $X^0=1$. 
The metric in the complex structure moduli space is given by $K_{CS}=- \log(i\int_\cam \Omega \wedge \bar\Omega)$. 
In the limit of large complex structure we find
\beq
K_{CS} = - \log \left(\frac16 d_{ijk}(U^i+\bar U^i) (U^j+\bar U^j) (U^k+\bar U^k)\right) 
=- \log \left(\frac43 d_{ijk}u^i u^j u^k\right) \, ,
\label{kcs2bs}
\eeq 
where $u^i =\text{Re}\, U^i$. Notice that $K_{CS}$ is invariant under shifts of the axions $\text{Im}\, U^i$.

We also need to introduce the period vector $\Pi$ whose components are periods of $\Omega$.
In the limit of large complex structure the transpose of $\Pi$ takes the form
\beq
\Pi^t = \left(1, i U^i, \frac12 d_{ijk} U^j U^k, -\frac{i}{6} d_{ijk} U^i U^j U^k \right) \, ,
\label{pivec}
\eeq
in a basis $(A^0, A^i, B_i, B_0)$ of 3-cycles. Under integer shifts in the axions $\text{Im}\, U^i$, namely
under $U^j \to U^j - i$, the period vector undergoes monodromy $\Pi \to R_j \Pi$. It is straightforward to obtain
the monodromy matrices $R_j$ from this definition and to verify that they are unipotent. Concretely, 
$(R_j - \mathbbm{1})^{n_j}\not=0$, $(R_j - \mathbbm{1})^{n_j+1}=0$, with $1\leq n_j \leq 3$. 

The monodromy generators $P_j=\log R_j$ are explicitly given by
\begin{equation}
\label{Pi2b}
P_i =\left(\begin{array}{cccc}0 & 0 & 0 & 0  \\ \vec{\d}_i  & 0 & 0 & 0 \\  0 & -d_{ijk} & 0  & 0  \\ 0 & 0 &  -\vec{\d}_i^{\,t}&0
\end{array}\right)\, .
\end{equation}
These generators are nilpotent and satisfy $[P_i, P_j]=0$.
Furthermore, it can be shown that the period vector enjoys the expansion
\beq
\Pi = \exp\left( iU^j P_j\right) \mathbf{a}_0 \, ,
\label{piexp}
\eeq
where $ \mathbf{a}_0^t =(1,0,0,0)$. Besides, it can be checked that
\beq
P_j \mathbf{a}_0 \not= 0 \, .
\label{moncond}
\eeq
This is a concrete example of the general expansion around a point in complex moduli space discussed in
\cite{irene}.
The condition \eqref{moncond} is a necessary requirement for the point to be at infinite distance. 
This is the expected result in our case, since the expansion is around the point of large complex structure.

In the computation of charges of domain walls, or particles, there appear integrals of the type
$\int_\cam \alpha_\kappa \wedge {}^*\alpha_\lambda$, and others involving $\beta^\lambda$.  
They can be determined from the period metric, which in turn is derived from the prepotential \cite{Grimm:2004uq}. 
Generic expressions in the large complex structure limit were obtained in \cite{Louis:2002ny}, see also 
\cite{Palti:2008mg}. They simplify when the axions $\text{Im}\, U^i$ are set to zero. In this case
the only non-vanishing results are
\beq
\begin{split}
\int_\cam \hspace*{-2mm} \alpha_0 \wedge {}^*\alpha_0 &= \frac18 e^{-K_{CS}}  \, , \qquad
\int_\cam \hspace*{-2mm} \alpha_i \wedge {}^*\alpha_j = \frac12 e^{-K_{CS}}\, g_{ij} \, , \\[2mm]
\int_\cam \hspace*{-2mm}\beta^0 \wedge {}^*\beta^0 &= \frac1{8\cad^2} e^{-K_{CS}} \, , \
\int_\cam \hspace*{-2mm} \beta^i \wedge {}^*\beta^j = \frac1{32\cad^2} e^{-K_{CS}}  g^{ij} \, .
\end{split}
\label{abint}
\eeq
where $K_{CS}$ is given in \eqref{kcs2bs} and $\cad=\frac16 d_{ijk} u^i u^j u^k$.
Besides, $g^{ij}$ is the inverse of $g_{ij}=\partial_i\partial_{\bar\jmath} K_{CS}$.  

Turning on background fluxes induces a moduli potential which can be
expressed in the $\cn=1$ supergravity form with K\"ahler potential 
\beq
K = -\log\left(\frac{S+\bar S}2\right) - 2\log \cv + K_{CS} \, .
\label{kpot2b}
\eeq
Here $\cv = \frac16 \kappa_{\alpha\beta\gamma} t^\alpha  t^\beta  t^\gamma $ 
and $K_{CS}$ was defined above.
The RR and NS-NS 3-form fluxes, denoted $\bar F_3$ and $\bar H_3$,  generate the superpotential 
\cite{Gukov:1999ya}
\beq
W= \int_\cam (\bar F_3 - i S \bar H_3) \wedge \Omega \, .
\label{w2b}
\eeq
Expanding the fluxes in the basis of 3-forms as
\beq
\bar F_3 = q^\lambda \alpha _\lambda - e_\lambda \beta^\lambda \, , \qquad
\bar H_3 = -\bar h^\lambda \alpha _\lambda  + h_\lambda \beta^\lambda \, ,
\label{flux2b}
\eeq 
then gives
\beq
\begin{split}
W = {}& e_0 + i e_i U^i -\frac12 d_{ijk} q^i U^j U^k + \frac{i}6 q^0 d_{ijk} U^i U^j U^k \\
&+ iS\left[ h_0 + i h_i U^i -\frac12 d_{ijk} \bar h^i U^j U^k + \frac{i}6  \bar h^0 d_{ijk} U^i U^j U^k \right] \, .
\end{split}
\label{w2bs}
 \eeq
The flux coefficients are quantized. 

Dimensional reduction of the 10d action gives the relation between the string and Planck scales 
$M_s^2 = g_s^2\op M_P^2/4\pi (V_6/2)$,

where the internal volume is $V_6=\int_\cam \! J^3/6$, and the factor of 2 is due to the orientifold projection.

In IIB, $V_6= e^{3\phi/2} \cv$, since $J=e^{3\phi/2} t^\alpha \omega_\alpha$. Then
\beq
M_s = \frac{g_s^{1/4}\op M_P}{\sqrt{2\pi} \cv^{1/2}} \, .
\label{msp2}
\eeq
The KK scale is taken to be
\beq
M_{KK} = \frac{M_s}{R_s} = 2\pi \frac{M_s}{V_6^{1/6}} = \frac{\sqrt{2\pi} M_P}{\cv^{2/3}}\, .
\label{mkk}
\eeq
The mass units in the K\"ahler potential and the superpotential are restored by introducing appropriate factors of $M_P$.
Specifically, $K \to M_P^2 K$ and $W \to \frac{M_P^3}{\sqrt{4\pi}} W$ \cite{Conlon:2005ki}.

\section{Details on the periods and charge vectors}
\label{ap:details}

In this appendix we give some details regarding the period vectors, their explicit relation with the K\"ahler potential and also the simplified form that we use in main text. We also give the precise definition of the charge vector $\vec{q}$. Let us begin by recalling that the period vectors for a Calabi-Yau orientifold in the large volume limit take the form 
\begin{equation}
\label{periodvectors}
\begin{split}
&\vec{\Pi}^{\, t}_K(T^a ) \, =\, \left( 1, i T^a, -\oh \CK_{abc} T^aT^b, -\frac{i}{6} \CK_{abc} T^aT^bT^c \right)\, , \\ 
&\vec{\Pi}^{\, t}_Q(N^k, T_\lambda ) \, =\, \left( i N^k,  \, i T_\lambda,\  \pim (C \cf_k),\  \pim (C X^\lambda)\right) \, .
\end{split}
\end{equation}
where $\vec{\Pi}^{\, t}_K(T^a )$ comes from the periods of the different powers of the complexified K\"ahler form and $\vec{\Pi}^{\, t}_Q(N^k, T_\lambda )$ from the ones of the complexified holomorphic 3-form. Note that, regarding this last period vector, the first and third entries are the periods with respect to the 3-forms $\left\{ \beta^k, \alpha_k \right\}$. The second and fourth entries are the periods with respect to  $\left\{ \alpha_\lambda, \beta^\lambda \right\}$, which are taken to be absent in the rest of this work but not in this appendix for completeness. This is the reason why the new moduli $T_\lambda$ appear. These period vectors encode the information of the K\"ahler potential, so that they can be combined with the so called pairing matrices, $\Upsilon$, in order to express \eqref{kkdef} and \eqref{kqdef} as 
\begin{equation}
K_K=-\log \left(i\, \Pi_K^t\, \Upsilon_K \, \bar{\Pi}_K \right)\, , \qquad K_Q=-2 \log \left( \frac{1}{8}\, \Pi_Q^t\, \Upsilon_Q \, \bar{\Pi}_Q \right),
\end{equation}

where the pairing matrices take the form
\begin{equation}
\Upsilon_K=\left(\begin{array}{cccc} 0 & 0 & 0 & -1 \\ 0 & 0 & \delta_b^a & 0 \\  0 & -\delta_b^a  & 0 & 0 \\ +1 & 0 & 0& 0\end{array}\right)\, ,\qquad  \Upsilon_Q=\left(\begin{array}{cccc}0 & 0 & \delta_K^L & 0 \\ 0 & 0 & 0& -\delta_K^L  \\  -\delta_K^L &0  & 0 & 0 \\ 0& \delta_K^L & 0& 0\end{array}\right)\, ,
\end{equation}
which can be obtained from eqs. \eqref{int243}.  This $\Upsilon$ is the same pairing matrix as the $\vartheta$ of \cite{Corvilain} when $\alpha^\prime$ corrections are neglected, but they could be included by replacing these matrices and the monodromy generators by those of \cite{Corvilain, Escobar:2018rna}. With this in mind, it is easy to see that the monodromy transformations \eqref{monodromymatrices} preserve the structure of the K\"ahler potential, making the axionic shift symmetry manifest. Moreover, these pairing matrices can also be used to reexpress the flux superpotential \eqref{wtot} as
\begin{equation}
W=\vec{\tilde{\Pi}}^t \Upsilon \vec{\tilde{q}}, \,
\end{equation}
where $\vec{\tilde{\Pi}}^t=\left( \vec{\Pi}^{\, t}_K, \vec{\Pi}^{\, t}_Q \right)$, $\Upsilon=\mathrm{diag} \left(\Upsilon_K, \, \Upsilon_Q \right)$ and $\vec{\tilde{q}}^t=\left( -m, -q^a, e_a, -e_0, 0, 0, h_k, h_\lambda \right)$ is the flux vector in this the same basis as the period vectors. In order to make contact with section \ref{sec:IIAtowers}, let us first define the charge vector $\vec{q}$ that is used in order to characterize the domain walls in \eqref{tensionW}. In order to define $\vec{q}$, we first note that the last two entries of the combination $\Upsilon \vec{\tilde{q}}$ are zero and hence we can remove them since they do not contribute to the superpotential. Additionally, since in the text we are considering the elements $\left\{ \alpha_\lambda, \beta^\lambda \right\}$ to be absent, this implies that there is no $h_\lambda$ so we can also also remove the third entry from the bottom, yielding the charge vector $\vec{q}$ given in \eqref{chargevector}. Regarding the period vector, since we have just argued that the three last entries in $\Upsilon \vec{\tilde{q}}$ do not contribute to the superpotential we can safely remove the last three of the $\vec{\tilde{\Pi}}$ and use it as our period vector $\vec{\Pi}$. Morally speaking, we just include the pairing matrix in the definition of the charge vector and then truncate the general period and charge vectors by removing the components that do not contribute to the superpotential. With these period and charge vectors the superpotential reduces to the one in \eqref{tensionW}.

For completeness, we present here the different $\textbf{a}_0$'s and $P_{(n)}$'s associated to several infinite distance points that enter the nilpotent orbit expansion of the period vector $\vec{\Pi}$ that we have just defined :
\begin{itemize}
{\item[{\bf (K.I)}]{\bf One K\"ahler modulus going to infinity: $t^1 \rightarrow \infty$.}\\
\begin{equation}
\textbf{a}_0^t=( 1, \underbrace{0, i T^d}_{h_{-}^{1,1} \ \mathrm{entries}}, -\oh \CK_{ade}T^d T^e, -\frac{i}{6} \CK_{def}T^d T^e T^f, N^K ), \qquad 
P_{(n)}=P_{a=1}
\end{equation}
where sum over repeated indices is understood and the indices $d,e,f=2,3...,h_-^{1,1}$.
}
{\item[{\bf (K.III)}]{\bf All the K\"ahler moduli going to infinity: $t^1\propto t^2 ... \propto t^{h_{-}^{1,1} }\rightarrow \infty$.}\\
\begin{equation}
\textbf{a}_0^t= \left( 1,0,0,0, i N^K  \right) , \qquad 
P_{(n)}= \sum_{a=1}^{{h_{-}^{1,1}}}P_a 
\end{equation}
}
{\item[{\bf (CS.I)}]{\bf One complex structure moduli going to infinity $n^1 \rightarrow \infty \ 1$. }\\
\begin{equation}
\textbf{a}_0^t=( 1, i T^a, -\oh \CK_{abc} T^aT^b, -\frac{i}{6} \CK_{abc} T^aT^bT^c, \underbrace{0, i N^M}_{h_+^{1,2} \ \mathrm{entries}} ) , \qquad P_{(n)}=P_{K=1} 
\end{equation}
where  the index $M$ runs over $2,3...,h_+^{1,2}$.
}
{\item[{\bf (CS.III)}]{\bf All complex structure moduli going to infinity $n^{1} \propto n^{2} ... n^{h_+^{1,2}} \rightarrow \infty$.}\\
\begin{equation}
\textbf{a}_0^t= ( 1, i T^a, -\oh \CK_{abc} T^aT^b, -\frac{i}{6} \CK_{abc} T^aT^bT^c, 0 ), \qquad P_{(n)}=\sum_{K=1}^{h_+^{1,2}}P_{K=1} 
\end{equation}
}
\end{itemize}
Note that all these ${\bf a}_0$ can be obtained from the one associated to  $Z^\mathcal{I} \rightarrow \infty, \ \forall \ \mathcal{I}$ by using the fact that all the monodromy generators commute. That is, all these can be calculated from
\begin{equation}
{\bf a}_0=\exp \left(\sum_\flat i \zeta^\flat P_\flat \right) \left(\begin{array}{c} 1 \\ 0\\ 0 \\ 0 \\ 0\end{array} \right)\, ,
\end{equation}
where $\flat$ runs over all the moduli that do not diverge

\section{Monodromy and towers}
\label{ap:moretowers}

In this appendix we complete the study of monodromies and infinite towers by considering the cases involving complex
structure moduli:

\begin{itemize}
{\item[{\bf (CS.I)}] {\bf One complex structure moduli going to infinity: $n^1 \to \infty$.}\\
Here, the conditions for a tensionless domain wall were $h_1=0$ if $K_Q$ is a polynomial of degree less than two in $n^1$ or no restrictions otherwise. For the former case all monodromies but the one about the singular point act non-trivially (i.e. only $(P_{K=1})^{t} \, \vec{q}=0$) and respect the tensionless condition, that is, if we begin with $h_1=0$ the action of the monodromy transformations never generate a non-zero value for that flux, that would render the wall tension non-vanishing. For the  latter  case every monodromy generates an infinite tensionless tower. In particular, if the infinite towers does not contain any D8-branes (i.e. $m=0$) the tadpole cancelation conditions \eqref{RRtadpole} are not modified across the wall.
}
{\item[{\bf (CS.II)}] {\bf Several complex structure moduli going to infinity along a path $n^1 \propto n^2... \propto n^J \rightarrow \infty, \  1< J  <h_+^{1,2}$.}\\
The tensionless  conditions for domain walls are $h_i=0$, for all $i=1...n$ if $K_Q$ is a polynomial of degree less than two in $n^i$ or no restrictions otherwise. For the first case all monodromies but the ones associated to $N^1... N^J$ act non-trivially (i.e. only $(P_{K=i})^{t} \, \vec{q}=0$) and they maintain the tensionless condition. For the second case all monodromies generate an infinite tower that is tensionless. As in case (CS.I), if the infinite tower does not contain any D8-brane (i.e. $m=0$ )the contribution to the tadpole cancelation conditions \eqref{RRtadpole} does not change across the wall.
}
{\item[{\bf (CS.III)}] {\bf All complex structure moduli going to infinity: $n^{1} \propto n^{2} ... \propto n^{h_+^{1,2}} \rightarrow \infty$.}\\
All the domain walls are tensionless in this case. If they contain some D4, D6 or D8-brane (i.e. $e_a, q^a, m\neq0$, respectively) there is always some monodromy generator $P_a$ that will act non-trivially and if they contain some NS5-brane (i.e. $h^I \neq 0$) the corresponding monodromy, $P_I$ will act non-trivially and generate an orbit. This is again a case in which we can identify a tower that is generated by the monodromies around the singular point or by the ones about the non-singular ones.
}
\end{itemize}

\end{document}